\documentclass[showpacs,superscriptaddress,twocolumn,10pt]{revtex4-1}
\makeatletter
\def\@biblabel#1{#1.}
\renewcommand\@bibitem[2][]{%
	\item\if@filesw \immediate\write\@auxout
	{\string\bibcite{#2}{\the\value{\@listctr}}}\fi\ignorespaces}
\makeatother

\usepackage{amssymb,amsmath, amsthm}

\usepackage{textcomp}
\usepackage{graphicx}
\usepackage[font=small]{caption}
\usepackage{overpic}
\usepackage{ulem} 

\usepackage[usenames]{color}
\usepackage{epsfig}
\usepackage{braket}
\usepackage[title]{appendix}

\usepackage{hyperref}

\usepackage{float}

\usepackage{xcolor}

\def\comment#1{}
\def\bra#1{\mathinner{\langle{#1}|}}
\def\ket#1{\mathinner{|{#1}\rangle}}

\def\comment#1{}

\begin{document}

\title{Precision gravimetry via harnessing interaction-induced resonances in optical lattices}

\author{Hassan Manshouri}
\email[]{h.manshouri@ph.iut.ac.ir}
\affiliation{Department of Physics, Isfahan University of Technology, Isfahan 84156-83111, Iran}
\affiliation{Quantum Technology Research Group, Isfahan University of Technology, Isfahan 84156-83111, Iran}
	
\author{Moslem Zarei}
\affiliation{Department of Physics, Isfahan University of Technology, Isfahan 84156-83111, Iran}
\affiliation{Quantum Technology Research Group, Isfahan University of Technology, Isfahan 84156-83111, Iran}

\author{Mehdi Abdi}
\affiliation{Department of Physics, Isfahan University of Technology, Isfahan 84156-83111, Iran}

\author{Yasser Omar}
\affiliation{Instituto Superior Técnico, Universidade de Lisboa, Portugal}
\affiliation{PQI -- Portuguese Quantum Institute, Portugal}
\affiliation{Physics of Information and Quantum Technologies Group, Centro de Física e Engenharia de Materiais Avançados (CeFEMA), Portugal}
\affiliation{LaPMET -- Laboratory of Physics for Materials and Emerging Technologies, Portugal}

\author{Sougato Bose}
\affiliation{Department of Physics and Astronomy, University College London, Gower Street, WC1E6BT, London, United Kingdom}

\author{Abolfazl Bayat}
\email[]{abolfazl.bayat@uestc.edu.cn}
\affiliation{Institute of Fundamental and Frontier Sciences, University of Electronic Science and Technology of China, Chengdu 611731, China}
\affiliation{Key Laboratory of Quantum Physics and Photonic Quantum Information, Ministry of Education, University of Electronic Science and Technology of China, Chengdu 611731, China}
\affiliation{Shimmer Center, Tianfu Jiangxi Laboratory, Chengdu 641419, China}

\date{\today}




\begin{abstract}

By confining a Bose-Einstein condensate in a vertical lattice subjected to a gravitational potential, we analyze the quantum Fisher information to determine its scaling with respect to time, system size and particle number. Our results reveal that in the localized phase, on-site interactions $U$ amplify the quantum Fisher information by a factor with respect to resonance condition $U=mh$ where $U$ is factor of gradient field amplitude $h$.
This precision enhancement can be employed in gravitational acceleration measurements with a finite number of particles trapped in optical lattices.

\end{abstract}

\maketitle
\section{Introduction}
Quantum
properties of physical
systems potentially make them a powerful resource for  sensing gravitational~\cite{TINO2013,Badurina_2020,baynham2025,Cepollaro2023}, electric~\cite{Block2021,Qiu_2022,Chen_2017,Iwasaki2017,Morales2024,Michl_2019,Yuan_2023,Liu_2022,liu2024} and magnetic fields~\cite{Budker_2007,Griffith2010,Yuantian2023,Rondin_2014,Hong_2013,Taylor_2008,Zhao2011,Quan2023,Yu2024,kurzyna2025}---surpassing the sensitivity limits of classical methods~\cite{Giovannetti_2006,Giovannetti_2011,Giovannetti_2004,Gross_2010,Poli2011,paris2009quantum,Uesli2025,Mihailescu2025,Braun_2018,RouhbakhshNabati2025,ghosh2025}. In particular, significant attention has been dedicated for the development of quantum gravimeters \cite{McGuirk2002,Fixler2007,Fattori2003,Szigeti2020,deAngelis_2009,Clade_2005,ferrari2006,Tino_2019,Canuel_2020,wu2023,Adams2021,Lellouch2022,Stray2022,Qvarfort_2018,Abend2016}.
Atomic interferometry, a technique based on the quantum interference of ultracold atoms, is widely used in precision gravitational physics. The applications include Raman interferometry for measuring Earth's gravitational acceleration $g$~\cite{peters1999} and its gradient~\cite{McGuirk2002}, determining the gravitational constant~\cite{Fixler2007,Fattori2003}, introducing a potential redefinition of the kilogram~\cite{Merlet_2008}, probing the quantum nature of gravity~\cite{bose2017}, and enabling geophysical studies~\cite{deAngelis_2009}.
A central aspect of atomic interferometry is the spatial delocalization of a large ensemble of atoms into two distinct vertical paths, which subsequently recombine to generate a measurable interference pattern~\cite{berman1997atom,Kasevich1991,Badurina_2020,Margalit_2021,Charriere2012,Pino_2018}. 
Bose-Einstein condensates (BECs) serves as an ideal atomic source for such atomic interferometers. While atomic interactions in dense BEC clouds can be beneficial to generate spin squeezing \cite{Gross_2010,Riedel2010,Szigeti2020,Malia2020,Perlin2020,Gietka2023}, they generally have detrimental effects on the precision of atom interferometry measurements \cite{Depasquale2013quantum,Yao2022,Simsarian2000,Jamison2011,Jamison2014,Jannin2015,Burchianti2020}.  
This raises a key question: can alternative many-body physics setups be used to create spatial superposition states in particles while leveraging the advantageous aspects of atomic interactions to enhance gravimetry?

Ultracold atoms confined in optical lattices, created by interfering laser beams \cite{Bloch2005,Bloch2007}, have a wide range of quantum applications such as analyzing strongly correlated phases \cite{Greiner_2002,Kinoshita2004,Fallani2007,Chin_2006,Mehboudi_2015,Muller2025}, determination of forces with micrometer resolution \cite{Anderson1998, Dimopoulos2003,Roati2004, Carusotto2005,Wolf2007, Ivanov2008}, quantum information processing \cite{Anderlini_2007,Bloch2008, Mandel2003, Mandel_2003,Agarwal2025} and realizing the quantum walk in the tilted lattice \cite{Sarkar_2020}. 
The high level of control and measurement precision in ultracold atoms enables detailed studies of ground-state properties \cite{Clement2009,Gemelke2009} or out-of-equilibrium dynamics \cite{Greiner_2002,Will2010} in quantum many-body systems. 
A key advantage of optical lattices is the high controllability by means of external fields and electromagnetic radiation.  
By adjusting the depth of the optical potential, one can adjust the ratio between interaction and kinetic energies in the system, inducing a quantum phase transition from a superfluid to a Mott insulator phase~\cite{Greiner_2002,GREINER2003,Sachdev_2002,Buyskikh_2019}.
Similarly, magnetic and optical Feshbach resonances allow the dynamics of repulsive and attractive interatomic potentials to be precisely controlled~\cite{Inouye1998,Theis2004,Chin2010}.
Such systems can also be served as quantum sensors. While recent studies have investigated the sensing properties in Wannier-Stark systems~\cite{he2023stark,Yousefjani2023,Sarkar2025,Manshouri_2025,yousefjani2026exponentially,Li2025nonequilibrium}, a comprehensive analysis in optical lattices is still missing. 
 
In this paper, we investigate the Quantum Fisher Information (QFI), as a figure of merit for sensing gravitational field, for an ultracold gas trapped in an optical lattice. In particular, we focus on QFI scaling with respect to  time, system size, and particle number across various range of on-site interaction. Our results demonstrate that in the localized phase, resonant conditions---where the on-site interaction strength is an integer multiple of the gradient potential---lead to significant enhancements in the QFI.
This precision enhancement allows for higher sensitivity in measuring the gravitational field
with the same number of trapped particles.

The structure of the paper is as following. In Sec. \ref{sec:est} a brief review of estimation theory is provided. We study the Stark Bose-Hubbard model, considering Bosons in an optical lattice in Sec. \ref{sec:Model}. In Sec. \ref{sec:QFI} we study the scaling of the QFI with respect to time, probe size and the number of trapped atoms. The scaling at resonance condition is investigated in Sec. \ref{sec:QFI-res}, while Sec. \ref{sec:table} provides quantitative estimates of this enhancement. 
Finally, Section \ref{sec:con} summarizes our principal conclusions and discusses their implications for quantum metrology with ultracold atomic systems.

\section{Estimation Theory}\label{sec:est}

In this section, we briefly review the theoretical foundations of estimation theory. 
The precision of a probe for sensing an unknown parameter $h$, is quantified by the standard deviation of the estimator $\hat{h}$ which is constructed by processing the measurement outcomes. 
A set of general quantum measurements $\{\hat{\Pi}_x\}$ consists of positive operator valued measure (POVM) $\hat{\Pi}_x$  with outcome $x$.
For an optimal estimator which is unbiased and efficient, the precision $\delta h$ is asymptotically bounded by Cram\'er-Rao inequality $\delta h \geq 1/ \sqrt{\mathcal{M}\mathcal{F}_{C}}$, where $\mathcal{M}$ is the number of independent measurements and $\mathcal{F}_{C}$ is classical Fisher information (CFI) \cite{fisher1922mathematical,paris2009quantum,liu2019quantum}. 
The ultimate quantum limit for the estimation of $h$ is obtained with maximization of the CFI over all possible POVMs, yielding the quantum Fisher information (QFI) which is independent of measurement basis \cite{paris2009quantum,MONTENEGRO2025}. Rewriting the Cram\'er-Rao inequality the precision is bounded by
\begin{equation}\label{eq:CRb}
	\delta h \ge \frac{1}{\sqrt{\mathcal{M}\mathcal{F}_C} } \ge \frac{1}{\sqrt{\mathcal{M}\mathcal{F}_Q}}~.
\end{equation}
For a probe with the pure quantum state $\rho(h){=}\ket{\Psi(h)}\bra{\Psi(h)}$, the QFI takes the explicit form of \cite{paris2009quantum} 
\begin{equation}\label{eq:QFI}
	\mathcal{F}_Q(h)= 4 [\braket{\partial_h \Psi|\partial_h \Psi}-\braket{\partial_h \Psi|\Psi}\braket{\Psi|\partial_h \Psi}]~,
\end{equation}
where $h$ is encoded in the quantum state $\ket{\Psi(h)}$. for mixed states the QFI is derived from the more general form involving the density matrix of the probe $\rho(h)$ and its logarithmic derivative \cite{liu2019quantum,paris2009quantum}.

\section{The Bose-Hubbard model}\label{sec:Model}
In optical lattice systems, ultracold particles are typically realized as a BEC, where interatomic interactions play a significant role in the system dynamics.
A BEC of dilute gases of bosonic species is realized as the Bose-Hubbard model when confined in periodic optical potentials \cite{GREINER2003,Greiner_2002}.
When subjected to a gradient field, these systems are described by the Stark Bose-Hubbard model \cite{Alberti_2010,Thommen_2004}.

A system of two-body interacting bosons can be described by the Hamiltonian \cite{Dalfovo1999}
\begin{eqnarray}\label{BEChamiltonian}
	\hat{H}_0&=& \frac{-\hbar^2}{2 m} \int \hat{\Psi}^\dagger(\vec{r}) \nabla^2 \hat{\Psi}(\vec{r}) d^3 r
	+ \int \hat{\Psi}^\dagger(\vec{r}) V_{\text{ext}}(\vec{r}) \hat{\Psi}(\vec{r}) d^3 r \nonumber \\
	&& + \frac{1}{2} \int \hat{\Psi}^\dagger(\vec{r})  \hat{\Psi}^\dagger(\vec{r'}) V(\vec{r'}-\vec{r}) \hat{\Psi}(\vec{r'})  \hat{\Psi}(\vec{r}) d^3 r d^3 r',
\end{eqnarray}
where $\hat{\Psi}(\vec{r})$ and $\hat{\Psi}^\dagger(\vec{r})$ are the bosonic field operators that annihilate and create a particle at the position $\vec{r}$, respectively. In the above Hamiltonian, the first term is the kinetic term of bosons with mass $m$, the second term is the interaction of particles with the external potential $V_{\text{ext}}$ which corresponds to the combination of the periodic potential of the lattice laser beam and the gravitational potential, and the last term describes the two-body interaction between bosons \cite{dalibard_2013,Dalfovo1999}. 
In a dilute and cold gas, when only the binary collisions at low energy are relevant the two-body interaction is $V(\vec{r}'-\vec{r})= u \delta(\vec{r}'-\vec{r})$ where $u$ is a single effective interaction parameter.

\begin{figure}[t]
	\includegraphics[width=.49\linewidth]{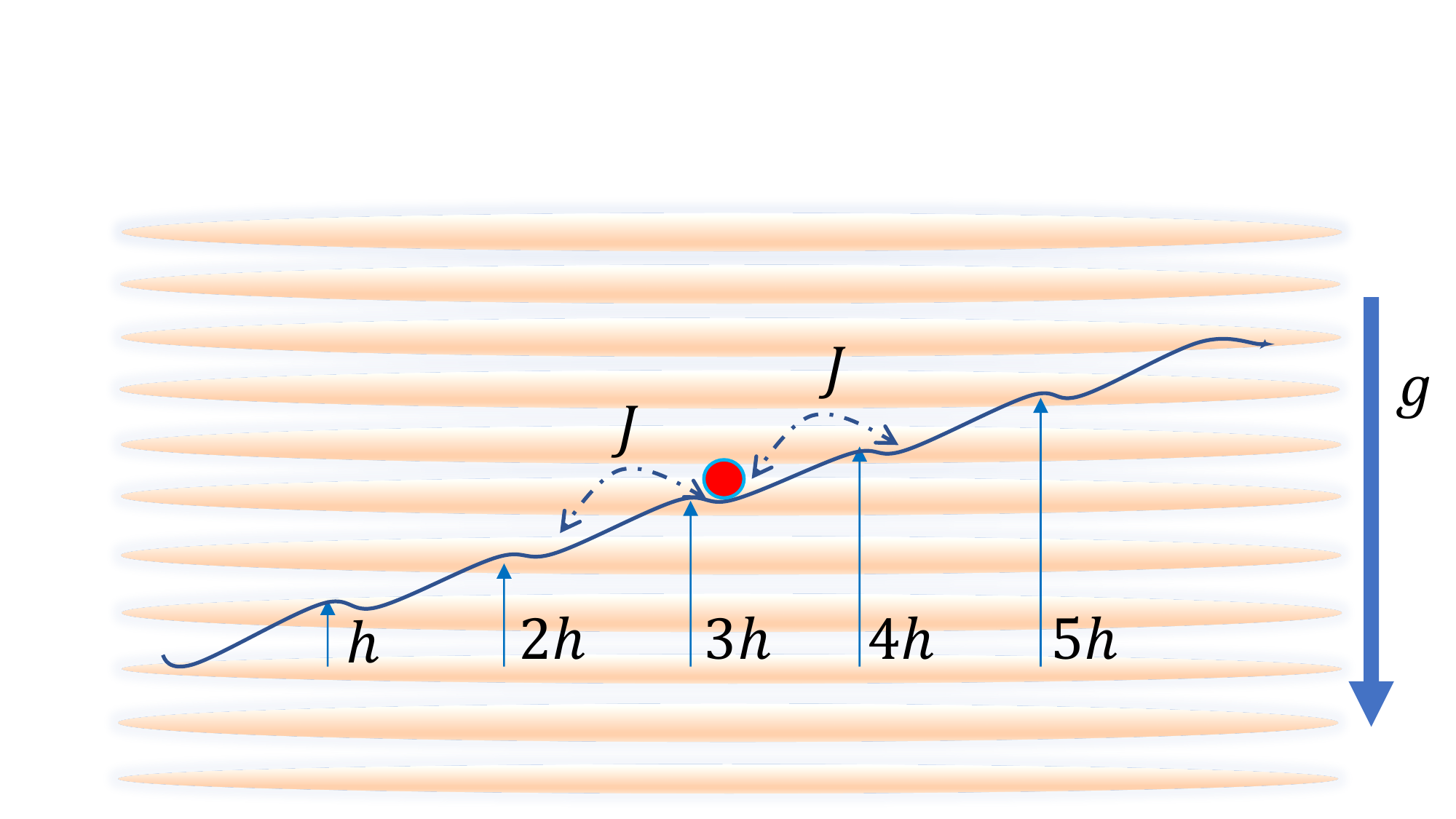}
	\put(-120,60){\color{black}(a)}
	\includegraphics[width=.49\linewidth]{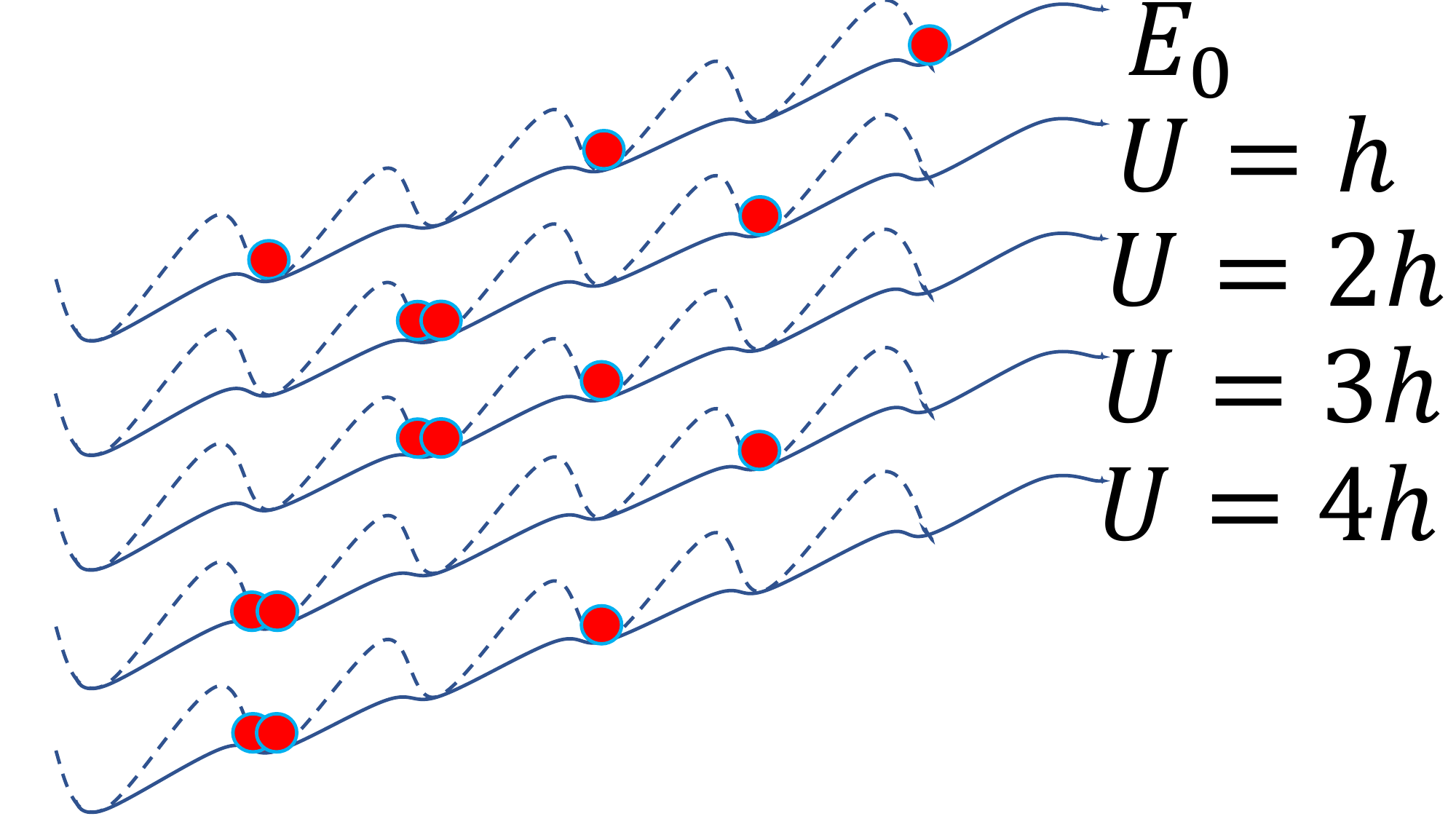}
	\put(-120,60){\color{black}(b)}
	\caption{(a) Schematic of a lattice with the tunneling rate $J$ and gradient field $h$. (b) Starting with the configuration $\ket{...0101010...}$ with the energy $E_0$, the corresponding configurations with the on-site interaction energy at resonance $U=mh$ are drawn where $m=1,2,3,4$. {color{red} (The lattice depth is too small. Enlarge it a bit.)}}
	\label{fig:stair}
\end{figure}

 We focus on a lattice in the vertical direction $z$, namely $\vec{r}=z \hat{z}$. In this situation one can write the lattice potential as
$V_{\rm ext}(z)=V_0 \sin^2(\frac{2 \pi}{\lambda_L} z)+h z$, where $h = mg\lambda_L/2$ represents the external gravitational potential strength and the potential intensity $V_0$ satisfies the condition $V_0 \gg E_R$ in the tight binding limit, with $E_R$ being the recoiling energy. 
The many-body lattice model of bosons can be derived from the second-quantization of the bosonic field operators in Eq.~\eqref{BEChamiltonian}. Considering only the Wannier function of the first band $w_0(z)$ \cite{Kohn_1959}, one can expand the field operator in terms of Wannier functions as
\begin{equation}
	\hat{\Psi}(z)= \sum_{l=-\infty}^{\infty} w_0(z-z_l) \hat{b}_l~,
\end{equation}
where $\hat{b}_l$ is the annihilation operator of a boson on the site $l$. The resulting Hamiltonian is the first-band Bose-Hubbard Hamiltonian with the nearest-neighbor tunneling \cite{Krutitsky_2016,Moseley_2008,dalibard_2013}

\begin{eqnarray}\label{eq:H}
	\hat{H}= &-&J \sum_l \left(\hat{b}^\dagger_l \hat{b}_{l+1} + \hat{b}^\dagger_{l+1} \hat{b}_{l}\right) +\frac{U}{2} \sum_{l} \hat{n}_l (\hat{n}_l-1)  \nonumber \\ 
    &+& h \sum_{l} l~\hat{n}_l~,
\end{eqnarray}
where the number operator is defined as $\hat{n}_l = \hat{b}^\dagger_l \hat{b}_l$, and   one can find the exchange coupling $J$ as \cite{Zwerger,Krutitsky_2016}
	\begin{equation}\label{J}
		\tilde{J}=\frac{4}{\sqrt{\pi}}(\tilde{V_0})^{3/4}e^{-2(\tilde{V_0})^{1/2}},
	\end{equation}
in which $\tilde{J}=\frac{J}{E_R}$ and $\tilde{V_0}=\frac{V_0}{E_R}$. Also, $U$ reads as
\begin{equation}
	U= u \int_{-\infty}^{\infty} |w_0(z)|^4 dz~. 
\end{equation}
The schematic picture of the system is given in Fig. \ref{fig:stair}(a). Note that due to gradient energy $h$ the tunneling between neighboring sites is suppressed, resulting in Stark-Wannier localization~\cite{Wannier_1960,Manshouri_2025}. However, the presence of on-site interaction $U$ may compensate for such energy mismatching allowing for particles to tunnel while changing the particle occupancy of certain sites, see Fig. \ref{fig:stair}(b). 

By considering the unitary time evolution of an initial state, $\ket{\Psi(t)} {=} e^{-iHt}\ket{\Psi(0)}$, with the Hamiltonian given by Eq.~\eqref{eq:H}, one can determine the ultimate precision of the gradient field $h$ \cite{Yuan_2015}. This is achieved by computing the time evolution of the QFI for the state $\ket{\Psi(t)}$ using Eq.~\eqref{eq:QFI}.
The initial state of the system is generated by symmetrically filling a lattice of size $L$ with $N$ particles, each separated by one vacant site. For example, for a lattice of size $L{=}5$ the initial state with $N=1$ particle would be $\ket{00100}$ and the initial state for $N=2$ particles would be $\ket{01010}$, and so on. In this situation, the size of the Hilbert space becomes $(L+N-1)!/N!(L-1)!$.

\begin{figure}[b]
	\includegraphics[width=.97\linewidth]{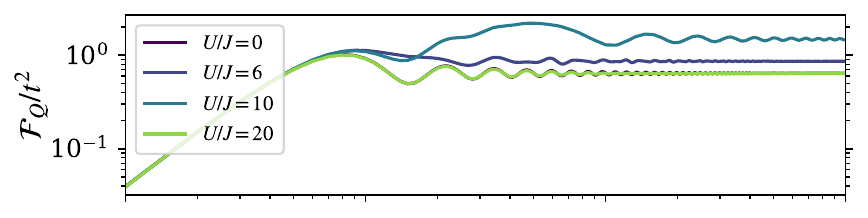}
	\put(-20,10){\color{black}(a)}\\
	\includegraphics[width=.97\linewidth]{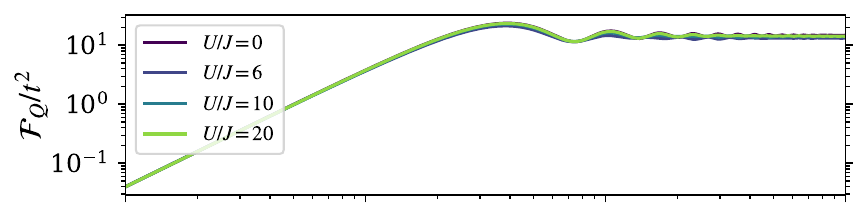}
	\put(-20,10){\color{black}(b)}\\
	\hspace*{.02cm} 
	\includegraphics[width=.98\linewidth]{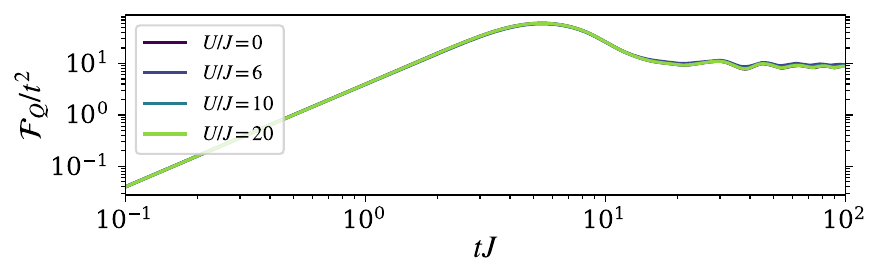}
	\put(-22,28){\color{black}(c)}\\
	\caption{The time evolution of the normalized QFI at different phases for $N=2$ particles in an $L=11$ lattice. $\mathcal{F}_{Q}/t^2$ with respect to time is plotted for different $U/J$ at (a)  $h/J=5$ in the localized phase, (b) $h/J=1$ at the transition point, and (c) $h/J=0.1$ in the extended phase. The QFI in the long time limit scales as $\mathcal{F}_{Q} \sim t^2$ in all plots.}
	\label{fig:BH_QFIt}
\end{figure}
\section{Gradient Field Sensing with Bose-Hubbard model}\label{sec:QFI}
In this section, we use the non-equilibrium quantum state of a Bose-Hubbard model for sensing the gradient field $h$.  
In Figs.~\ref{fig:BH_QFIt}(a)-(c) we plot the normalized QFI, namely $\mathcal{F}_{Q}/t^2$, as a function of time for three different choices of $h/J$ corresponding to localized,  critical and extended phases, respectively. 
The exact determination of the phases will be clarified below. As  Fig. \ref{fig:BH_QFIt} shows, in each phase, the normalized QFI for different choices of on-site interaction $U$ is depicted. Two observations are clearly evident. First, After a transient time the QFI scales quadratically with time which is expected due to unitary evolution~\cite{Boixo2007Generalized}. Second, the on-site interaction $U$ hardly affects  the evolution of the QFI for the choices of $h/J$ in the extended phase and the transition point, see Figs.~\ref{fig:BH_QFIt}(b) and (c). In fact, the on-site interaction $U$ can only weakly affect the QFI in the localized phase, as shown in Fig.~\ref{fig:BH_QFIt}(a). We will investigate the role of $U$ in the sensing capacity of our probe in more details in the next section. 

\begin{figure}[t]
	\includegraphics[width=.49\linewidth]{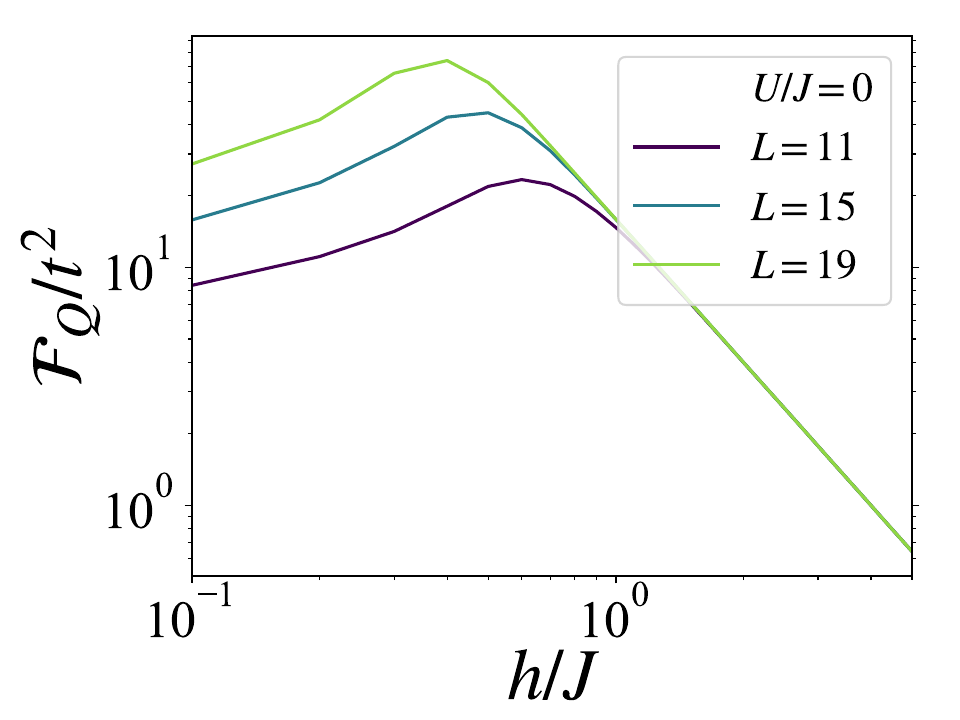}
	\put(-95,22){\color{black}(a)}
	\includegraphics[width=.49\linewidth]{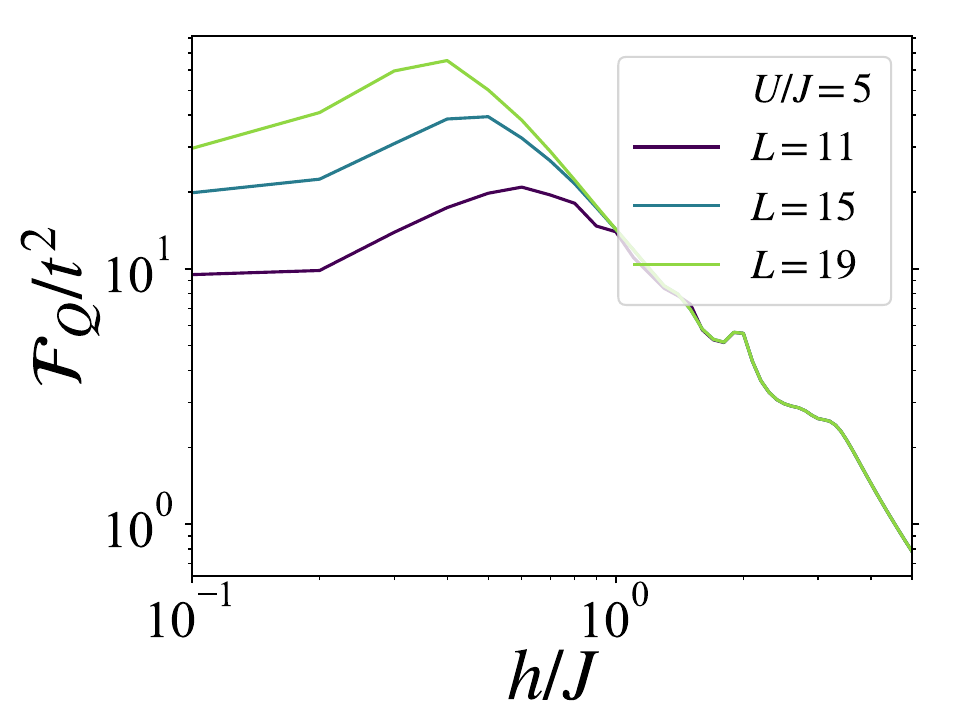}
	\put(-95,22){\color{black}(b)}\\
	\includegraphics[width=.49\linewidth]{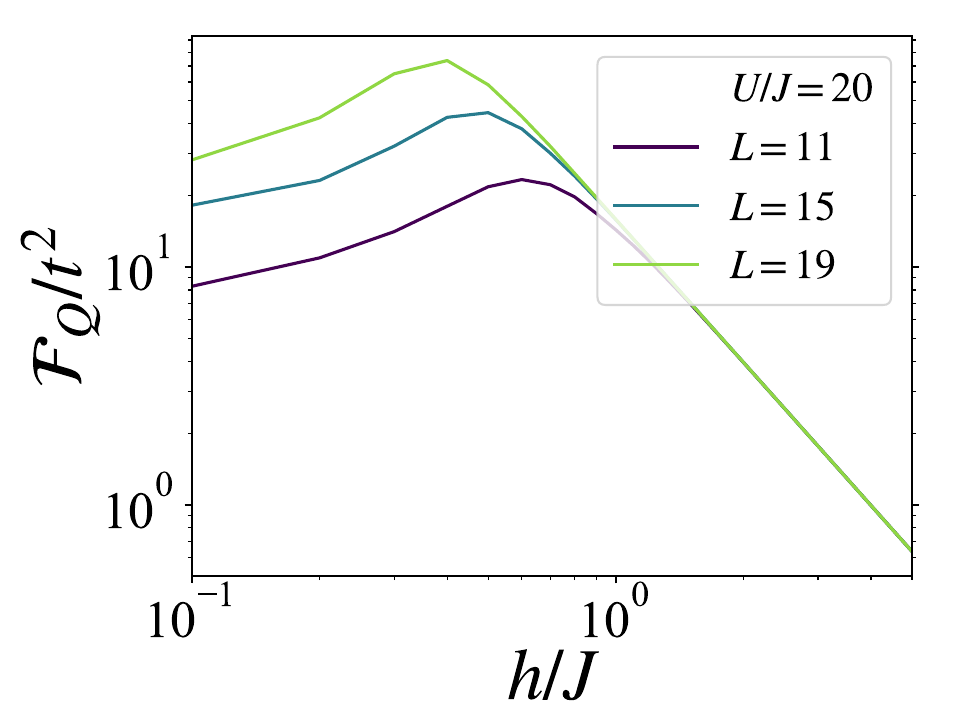}
	\put(-95,22){\color{black}(c)}
	\begin{overpic}[width=0.49\linewidth]{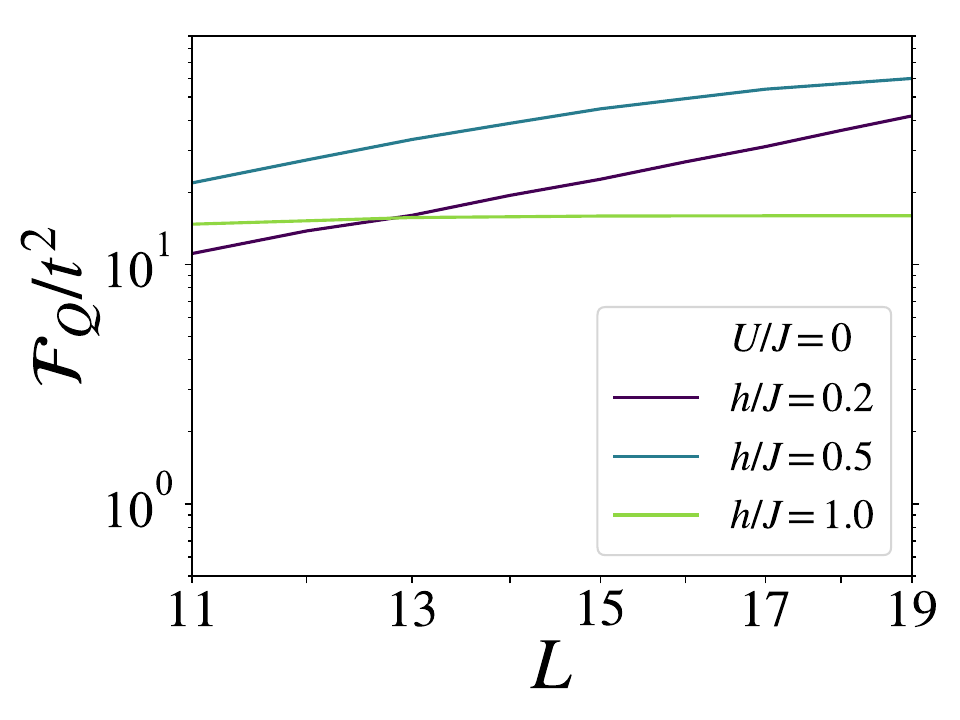}
		\put(20,15.1){\includegraphics[width=.2\linewidth]{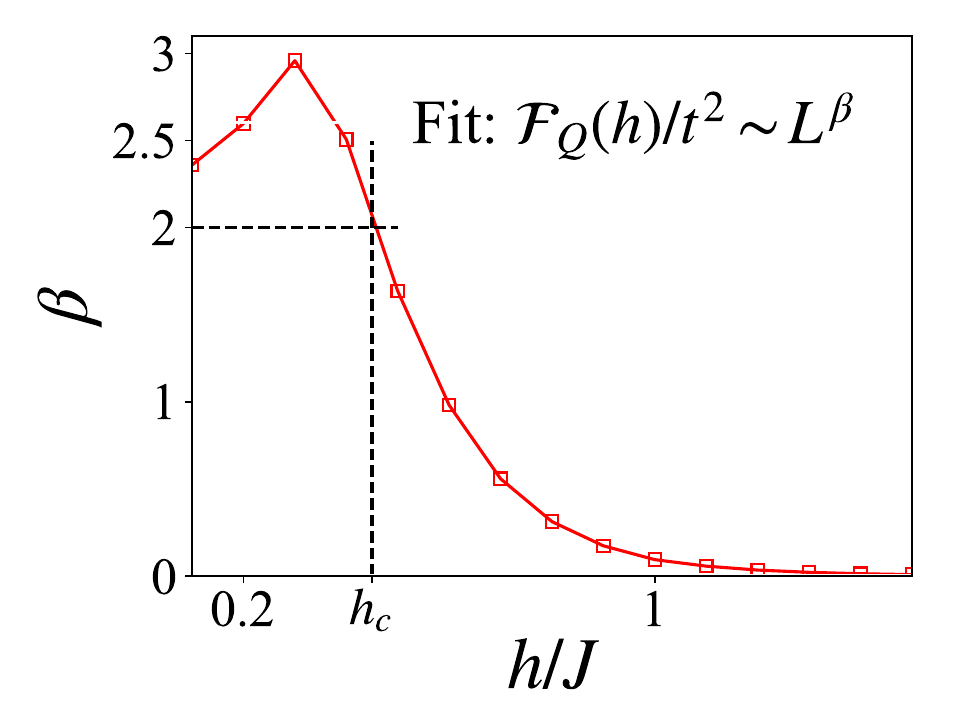}}
	\end{overpic}
	\put(-95,78){\color{black}(d)}
	\caption{In the long time limit, the normalized QFI for $N=2$ and different system size $L$ as a function of $h/J$ is plotted at (a) the non-interacting regime ($U/J=0$), (b) intermediate interactions ($U/J=5$), and (c) the strongly interacting limit ($U/J=20$). (d) The normalized QFI as a function of $L$ is plotted. The inset shows the scaling exponent $\beta$ with respect to gradient field $h/J$ for $L{=}20$, in which at the transition point $h_c/J$ exhibits $\beta=2$. 
	}
	\label{fig:BH_QFIT2L}
\end{figure}

Since in the long time limit the QFI scales quadratically with time, we now focus on extracting the scaling with respect to system size and particle numbers. To identify the QFI scaling with respect to system, in Figs.~\ref{fig:BH_QFIT2L}(a)-(c), we plot the normalized QFI, $\mathcal{F}_{Q}/t^2$ at long time limit, as a function of $h/J$ for three choices of on-site interaction $U/J$  keeping the particle numbers fixed to $N{=}2$. As the figures show, by increasing $h/J$ the QFI increases size-dependently in the extended phase (i.e. small values of $h/J$) until it reaches its peak at the  transition point $h_c/J$ after which it decays size-independently in the localized phase.

For all interaction strengths $U/J$, the normalized QFI scales as $\mathcal{F}_{Q}/ t^2 \sim L^\beta$. This scaling is demonstrated in Fig.~\ref{fig:BH_QFIT2L}(d) for the extended phase with $\beta > 2$, transition point with $\beta=2$ and the localized phase with $\beta=0$.  In the inset of Fig.~\ref{fig:BH_QFIT2L}(d), we plot the scaling exponent $\beta$ as a function of $h/J$. The exponent $\beta$  shows super-linear scaling of the QFI in the extended phase clearly identifying quantum-enhanced sensitivity. Interestingly, exactly at the transition point $h_c/J$ the exponent reaches  $\beta=2$ which is identical to a single-particle system in Ref.~\cite{Manshouri_2025}.

\begin{figure}[t]
	\includegraphics[width=.49\linewidth]{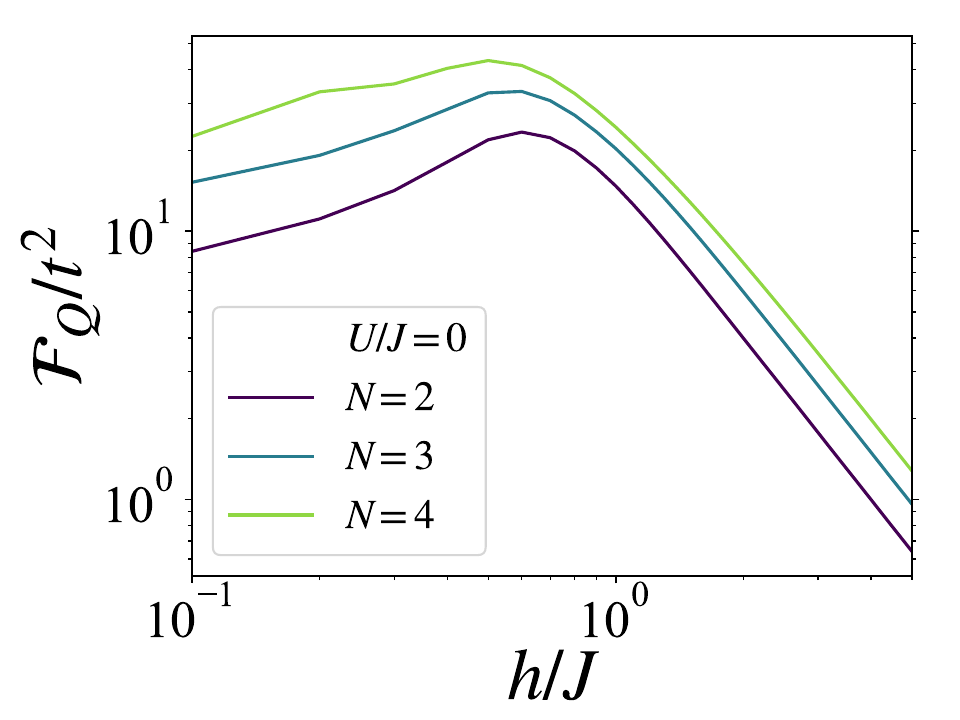}
	\put(-95,77){\color{black}(a)}
	\includegraphics[width=.49\linewidth]{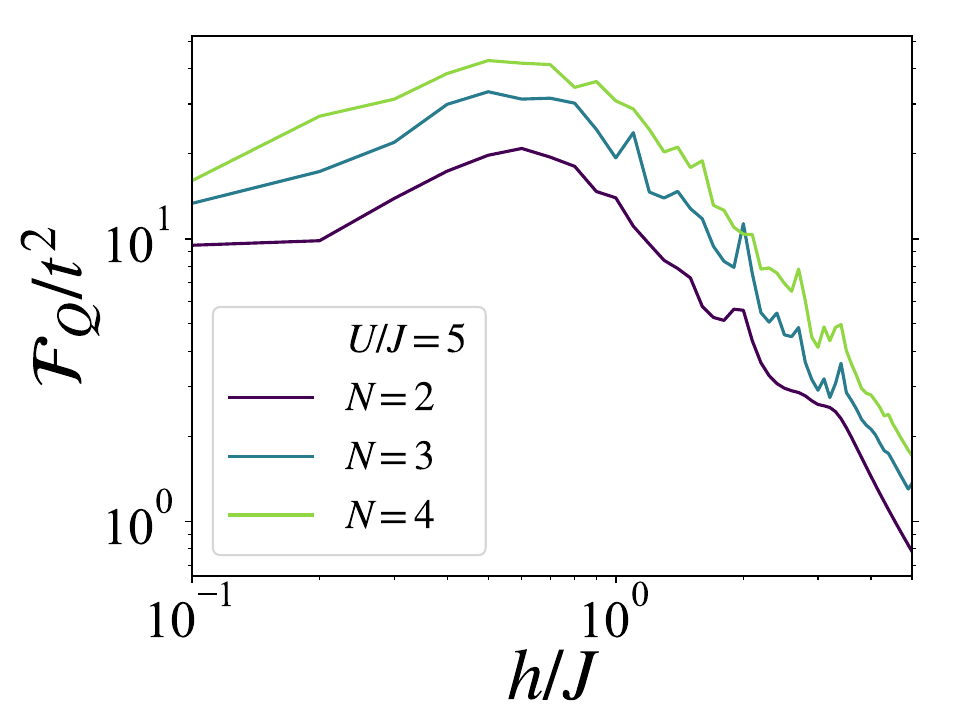}
	\put(-95,77){\color{black}(b)}\\
	\includegraphics[width=.49\linewidth]{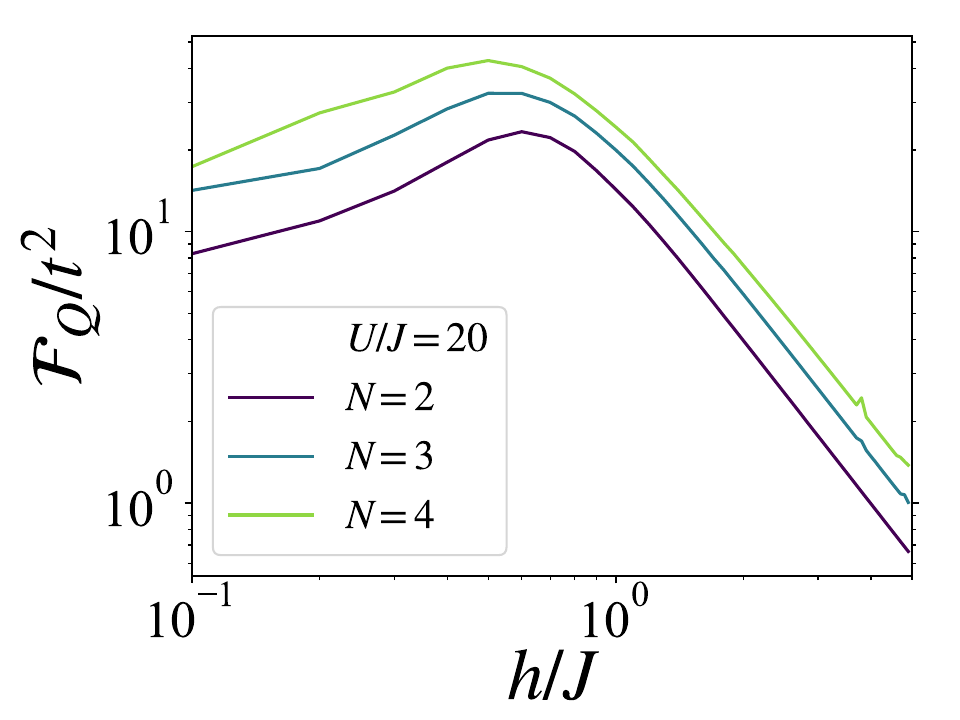}
	\put(-95,77){\color{black}(c)}
	\includegraphics[width=.49\linewidth]{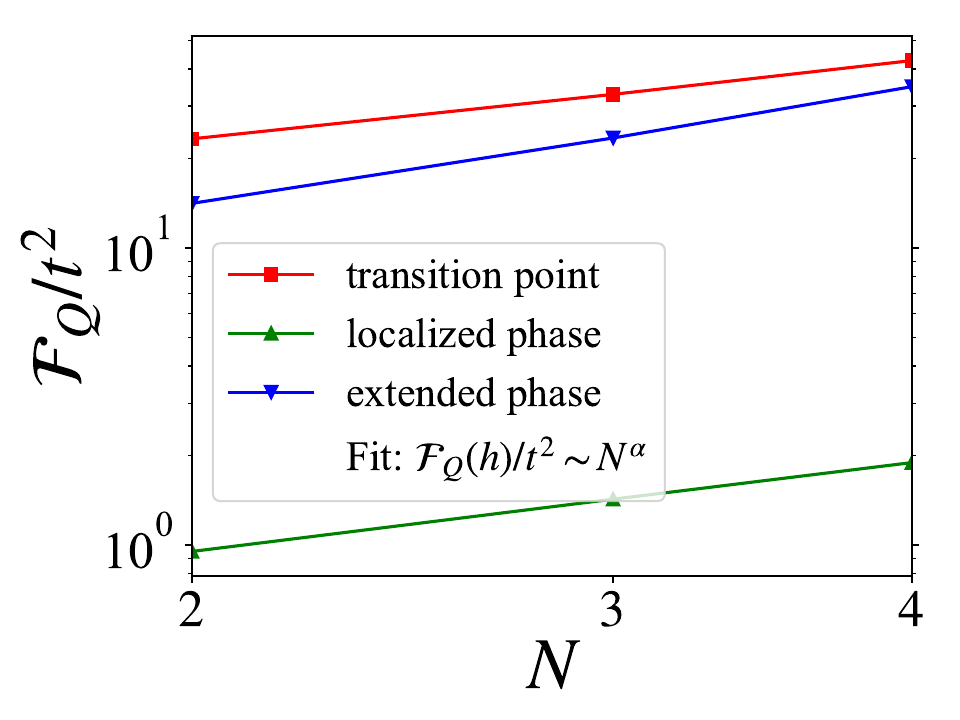}
	\put(-95,77){\color{black}(d)}\\
	\caption{Particle-number scaling of the normalized quantum Fisher information $\mathcal{F}_Q/t^2$ in the long-time limit. $\mathcal{F}_{Q}/ t^2$ as a function of $h/J$ for $L=11$ of different $N$ is figured in (a) non-interacting regime ($U=0$), (b) intermediate regime ($U=5$), and strongly interacting limit ($U/J= 20$). (d) The QFI scales as $\mathcal{F}_{Q}/ t^2 \sim N^\alpha$, in which for all phases $\alpha=1$, while in the extended phase becomes larger than the localized phase and the transition point value.}
	\label{fig:BH_QFIT2N}
\end{figure}

To investigate the scaling of the QFI with respect to particle number $N$, we fix the system size to $L{=}11$ and investigate the behavior of the long time normalized QFI as a function of $h/J$ for various particle numbers. The results are depicted in Figs.~\ref{fig:BH_QFIT2N}(a)-(c), where the system shows size-dependent scaling across the entire phase diagram for different values of on-site interaction $U/J$. Therefore, the QFI scaling can be written as $\mathcal{F}_{Q}/ t^2 \sim N^\alpha$. The fluctuations in Fig.~\ref{fig:BH_QFIT2N}(b), where $U/J=5$, are due to resonances which will be discussed in Sec.~\ref{sec:QFI-res}. 
In Fig.~\ref{fig:BH_QFIT2N}(d) we explicitly compare the normalized QFI scaling  for different regimes, namely the localized phase, the transition point and the extended phase in the non-interacting regime (i.e. $U=0$). Notably, the QFI scaling exponent $\alpha$ generally reads $\alpha\simeq 1$, while becomes slightly larger in the extended phase.



\section{QFI at resonance conditions}\label{sec:QFI-res}
\begin{figure}[t]
	\includegraphics[width=.49\linewidth]{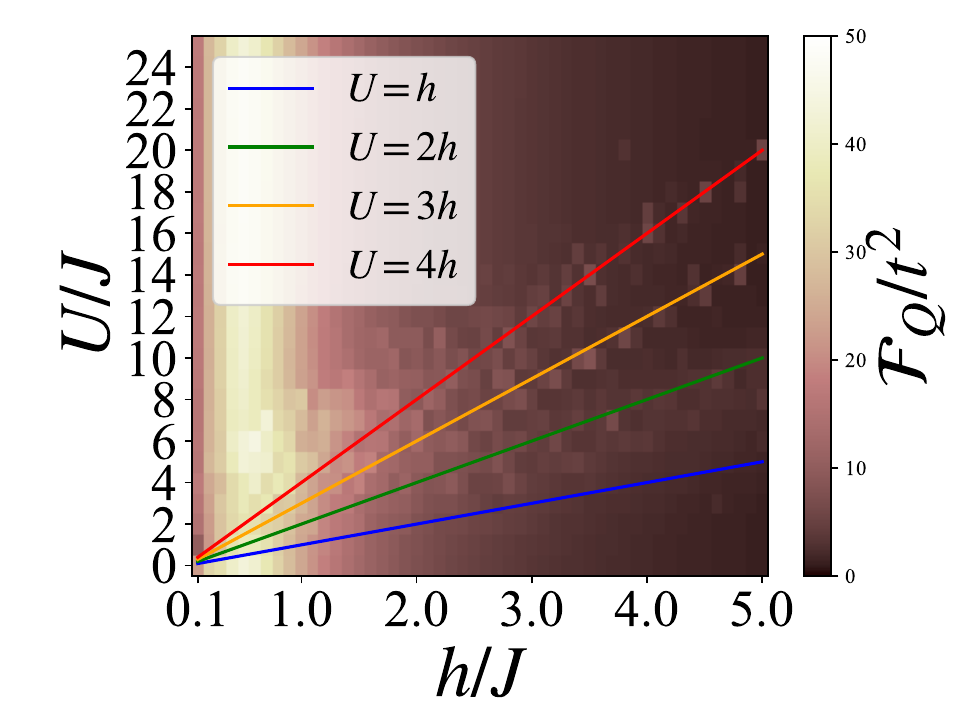}
	\put(-35,77){\color{white}(a)}
	\includegraphics[width=.49\linewidth]{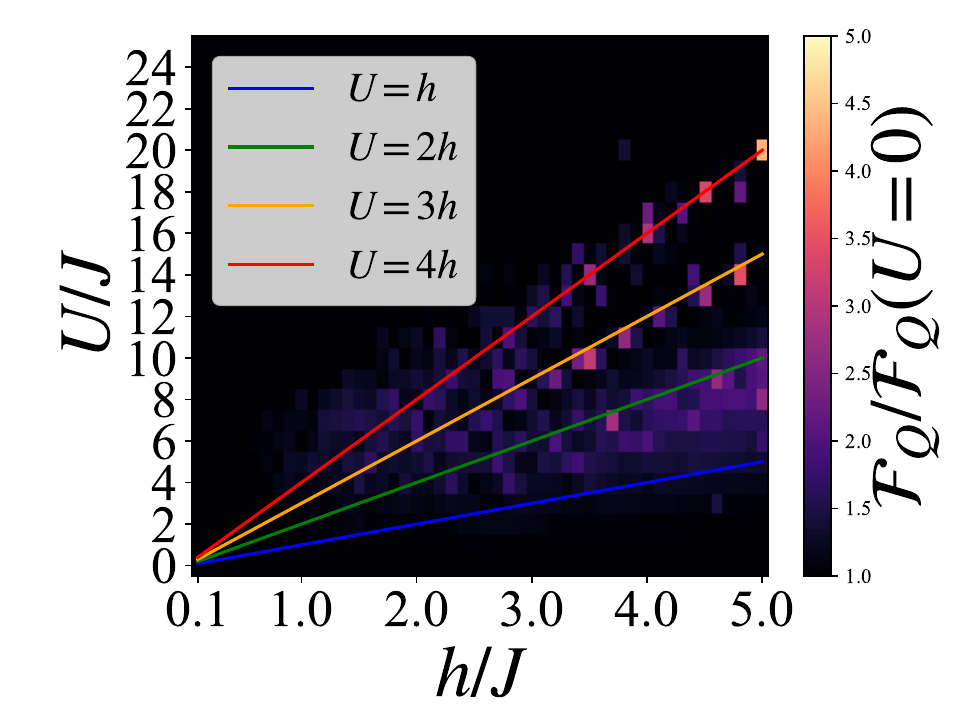}
	\put(-35,77){\color{white}(b)}
	\caption{(a) $\mathcal{F}_{Q}$ is plotted as a function of various on-site interaction energy and gradient field for $L=11$ and $N=4$. (b) Shows the normalized QFI for different $U/J$ divided by the non-interacting normalized QFI, $\mathcal{F}_{Q}/t^2(U{=}0)$. The lines $U=mh$ are signified with the color lines blue (m=1), orange (m=2), green (m=3) and red (m=4).}
	\label{fig:BH_QFIT2U}
\end{figure}

The occupation of lattice sites by multiple bosons is constrained by on-site repulsive interactions ($U > 0$). As shown in Fig.~\ref{fig:BH_QFIt}, while the impact of $U$ on sensitivity is weak in the extended phase and at the transition point, its effect on sensitivity in the localized phase can be complex and significant. In this section, we systematically investigate the impact of on-site interaction $U$ on sensing capacity of the Bose-Hubbard model. In Fig.~\ref{fig:BH_QFIT2U}(a) we display the normalized QFI as a function of both $U$ and $h$ for a system of size $L=11$ and boson number $N=4$. Interestingly, the normalized QFI reveals oscillations with consistent local maxima along the lines of $U = mh$ for $m=1,2,3,4$, which are highlighted in the figure by colored lines.  To quantify the metrological advantage along these lines, Fig.~\ref{fig:BH_QFIT2U}(b) displays the normalized QFI enhancement ratio $\mathcal{F}_Q(U)/\mathcal{F}_Q(U{=}0)$ for different $U/J$. One can see at the vicinity of $U{=}mh$ the significant QFI amplification, while out of resonance reads $\mathcal{F}_{Q}(U{\neq} mh)/\mathcal{F}_{Q}(U{=}0){=}1$ indicating that the sensing performance reduces to the non-interacting values.

The above fluctuations are associated with resonances between different particle configurations in which individual particles can move along the lattice despite the gradient field energy differences at the expense of having   double (or multiple) occupancies at certain lattice sites. To elucidate the underlying mechanism, starting with the initial state $\ket{...0101010...}$ in Fig.~\ref{fig:stair}(b), the corresponding configurations for $U{=}mh$ are illustrated conserving the initial energy $E_0$. The tunneling of bosons to occupied sites is suppressed by the repulsive on-site interaction $U$, except near these resonant points where energy matches $U{=}mh$. 

\begin{figure}[t]
	\includegraphics[width=.49\linewidth]{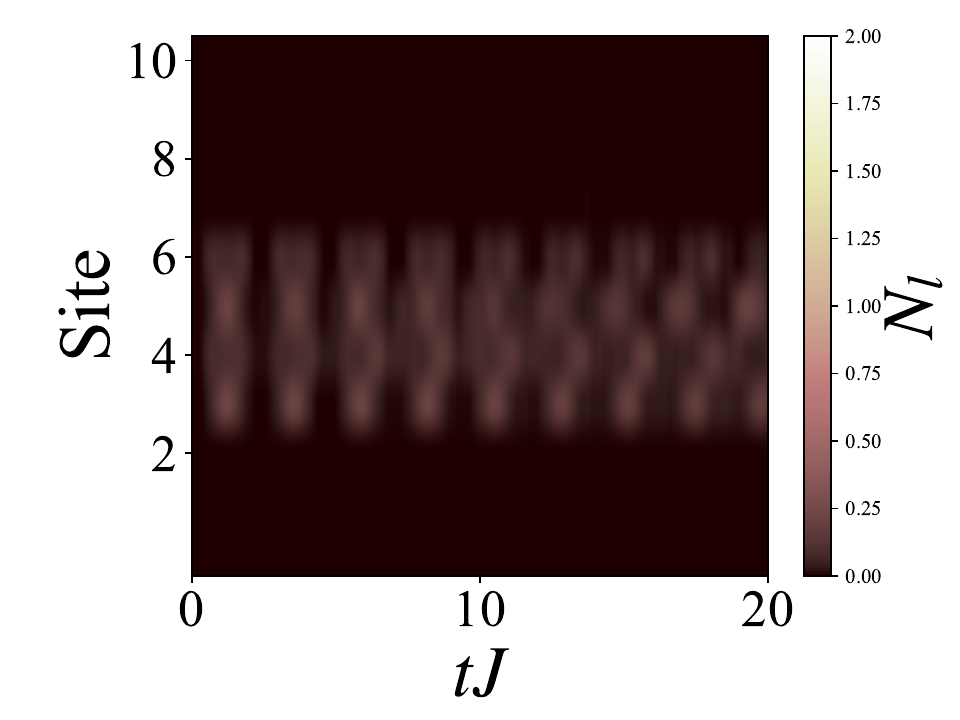}
	\put(-95,77){\color{white}(a)}
	\includegraphics[width=.49\linewidth]{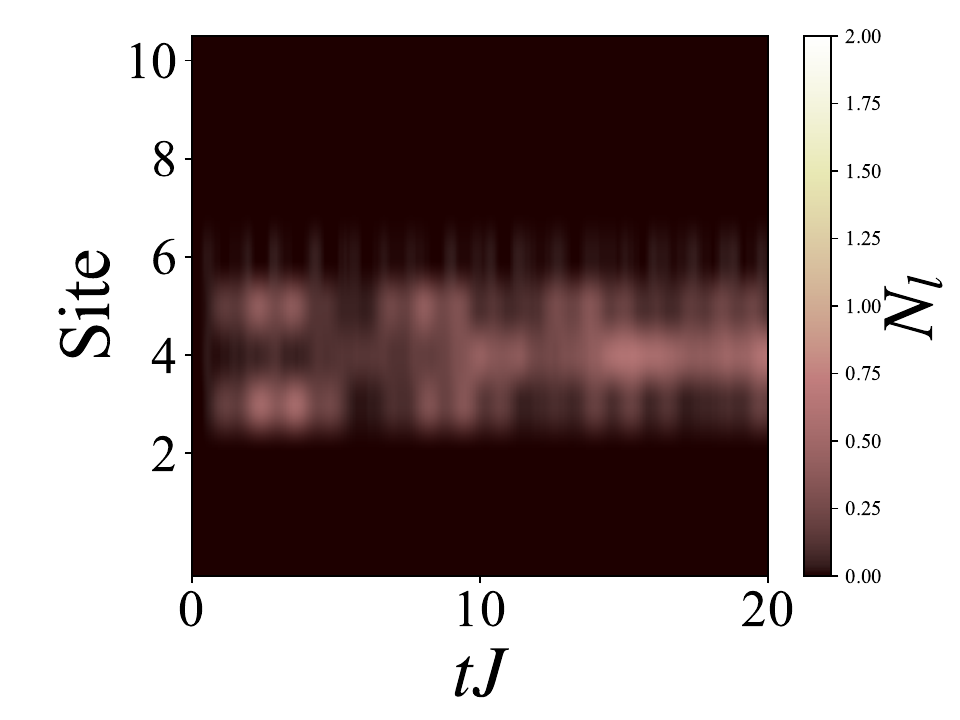}
	\put(-95,77){\color{white}(b)}\\
	\includegraphics[width=.49\linewidth]{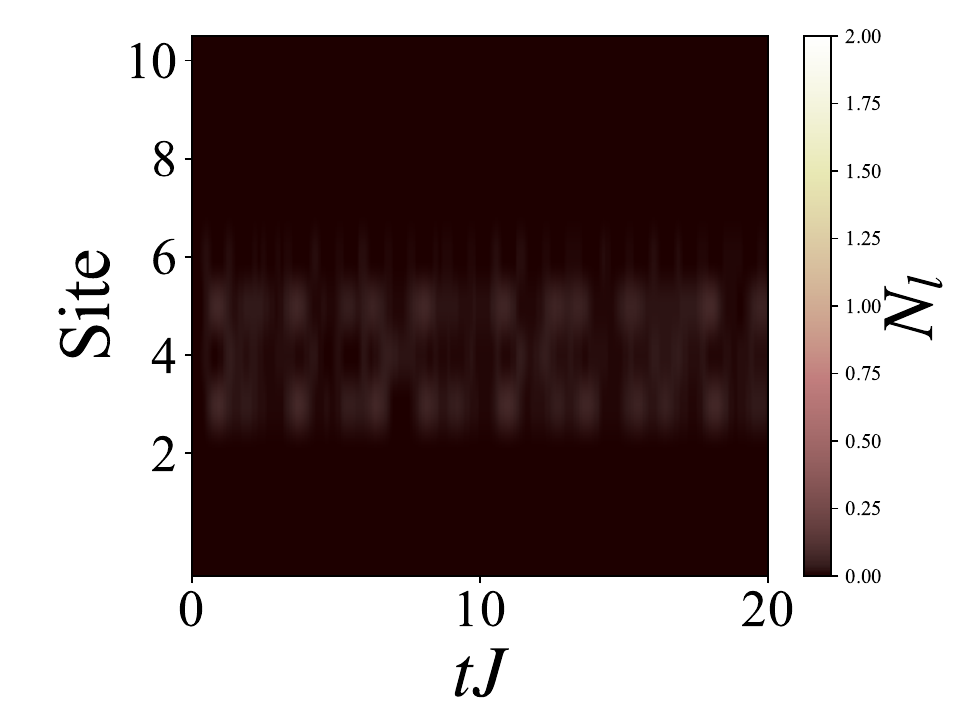}
	\put(-95,77){\color{white}(c)}
	\includegraphics[width=.49\linewidth]{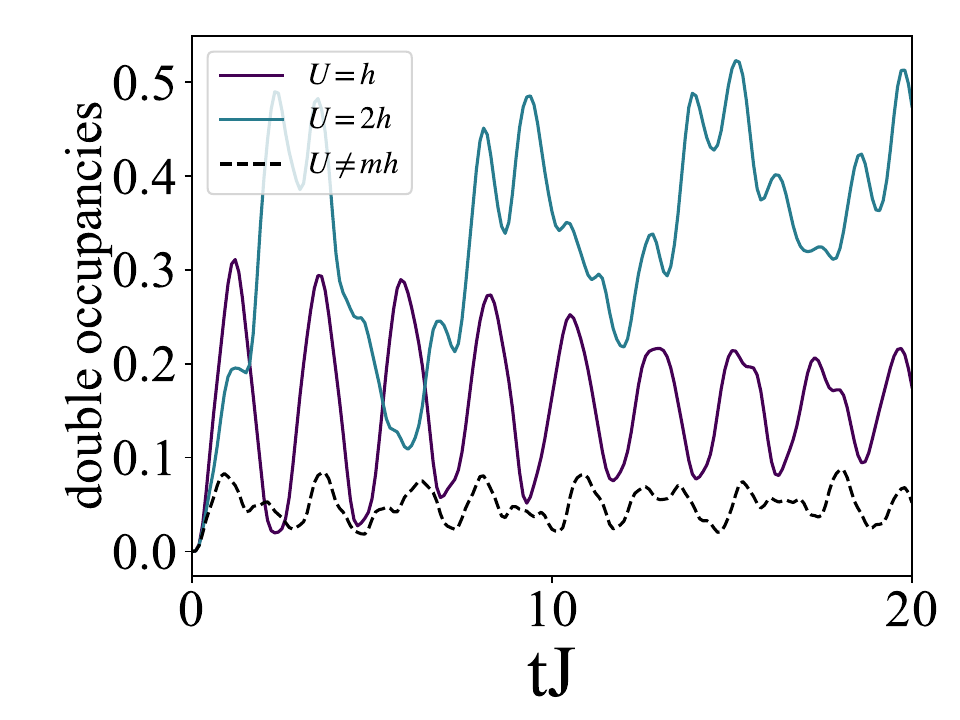}
	\put(-20,77){\color{black}(d)}\\
	\caption{Site occupations $N_l$ as a function of time for $L=11$ and $N=3$ is plotted for on-site interaction energies and gradient fields (a) $U{=}h$ ($U/J{=}4$, $h/J{=}4$), (b) $U{=}2h$ ($U/J{=}8$, $h/J{=}4$) and (c) $U{\neq}mh$ ($U/J{=}10$, $h/J{=}4$). In (d) the corresponding double occupancies in (a) to (c) are figured.}
	\label{fig:BH-BD}
\end{figure}

In order to see the impact of such resonances, one can investigate the expectation value of the following operator 
\begin{equation}
	N_l= \langle \Psi(t) | \hat{n}_l (\hat{n}_l - 1) | \Psi(t) \rangle.
\end{equation}	
The expectation will be zero if the number of particles in the site $l$ is either zero or one and becomes nonzero when there are more than one particles in the site. Fig.~\ref{fig:BH-BD}(a) and (b) display $N_l$ in a system of size $L=11$ with $N=3$ particles at resonance conditions of $m{=}1$ and $m{=}2$, respectively. Consistent with the particle configurations shown in Fig.~\ref{fig:stair}(b) of $U{=}h$ and $U{=}2h$, particles mostly occupy lower lattice sites. This spatial asymmetry directly correlates with the enhanced QFI observed at resonance. 
The contrast with off-resonant behavior is evident in Fig.~\ref{fig:BH-BD}(c), where non-resonant interaction strengths ($U \neq mh$) allow much weaker probability of multiple occupancies and correspondingly no QFI enhancement.
Another way of looking at the resonance condition is to investigate the probability of having any multiple occupancies across the chain, quantified by $ \sum_l N_l$, during the dynamics. In Fig.~\ref{fig:BH-BD}(d), we plot $ \sum_l N_l$ as a function of time where it clearly exhibits dramatic enhancement under resonant conditions ($U{=}mh$) compared to off-resonant cases. This not only indicates the role of resonances in the dynamics but also provides the reason behind the QFI enhancement as the wave function changes dramatically at the resonance conditions.  

Based on the above analysis one can summarize the sensing  results of the Bose-Hubbard model as
\begin{eqnarray}\label{eq:QFI_scaling}
	\mathcal{F}_{Q} & \sim & t^2 L^\beta N^\alpha,  \;\;\;\;\;\;\;\;\;\;\;\;\;\;\;\; \text{   for } h\le h_c~, \nonumber\\
	\mathcal{F}_{Q} & \sim& A_{r}(U,h) h^{-2} t^2 N^\alpha,  \;\; \text{   for } h> h_c~,
\end{eqnarray} 
where the coefficient $A_r(U,h)$ accounts for the resonance enhancement coefficient at $U=mh$. In appendix~\ref{appendixB} we show that  the coefficient $A_r(U,h)$ is independent of system size $L$ and particle number $N$ and solely depends on on-site interaction $U$ and gradient field $h$. The coefficient peaks at the resonance condition where $U=mh$.

\section{Optical lattice ultimate precision}\label{sec:table}

The optimal experimental platform for our proposal is an optical lattice. By considering an optical lattice in the vertical direction the gravitational field $g$ directly implies a gradient field $h=mg\lambda_L/2$ between adjacent sites. We can use the non-equilibrium dynamics of our Bose-Hubbard model for estimating $g$. 
The ultimate sensitivity of determining $g$ is governed by the QFI via the Cram\'er Rao bound in the form of $\delta g /g \ge 1/g\sqrt{\mathcal{M} \mathcal{F}_Q}$~. In Table \ref{table} for different particle numbers $N$ with the same system size $L{=}11$~, the gravitational sensitivity, $\delta g /g$ is calculated numerically under both off-resonance and resonance conditions. Because the QFI is enhanced by a factor of $A_r(U{=}mh) \simeq 4$ at resonance, the gravitational sensitivity improve by a factor of $\sqrt{A_r(U{=}mh)} \simeq 2$.
\begin{table}[t]
	\begin{tabular}{|c|c|c|c|}
		\hline 
		$N$ &2 & 3 &  4\\ \hline
		~~$\frac{\delta g}{g} (U{\neq}mh)$~~ & ~~$ 7.90\times10^{-6}$~~&~ ~$  6.45\times10^{-6}$~~&~~ $5.59\times10^{-6}$ ~~\\ \hline
		~~$\frac{\delta g}{g} (U{=}mh)$~~ & ~~$ 4.11\times10^{-6}$~~&~ ~$ 3.20\times10^{-6}$~~&~~ $2.74\times10^{-6}$ ~~\\ \hline	
	\end{tabular}
	

	\caption{The estimated  gravitational sensitivity $\delta g/g$ at the resonance and off-resonance conditions.
	the gravitational sensitivity $\delta g /g$ for different values of particle number $N$ with the same system size $L=11$, are calculated in the off-resonance ($U{\neq}mh$) and resonance ($U=4h$) conditions for the coherent time $t= 1$s~. 
}
\label{table}
\end{table}
The analysis reveals that under identical experimental constraints, operating at resonant conditions ($U = mh$) substantially reduces the gravitational sensitivity $\delta g/g$, compared to off-resonant operation.

\section{conclusion}\label{sec:con}

We study the scaling of the QFI with respect to time, probe size and the number of bosonic atoms in different on-site interaction strength regimes. Our analysis shows that operating at resonance condition, where the interaction strength in Stark Bose-Hubbard model relates to the gradient field as $U= mh$, significantly enhances measurement sensitivity by amplifying the QFI by a factor of $A_r(U,m)$~. In contrast, under off-resonance conditions $U\neq mh$, the interaction strength $U$ has no significant impact on the many-body Bloch oscillations in optical lattices and the corresponding QFI. 
This resonance enhancement of the QFI can be particularly advantageous for optical lattice gravimetry, as it allows for higher gravitational field sensitivity
with the same number of trapped particles.

\vspace{0.5cm}
\begin{acknowledgments}
AB acknowledges support from the National Natural Science Foundation of China (grants No. W2541020, No. 12274059,
No. 12574528 and No. 1251101297). HM and MZ thank the Center for International Scientific Studies  Collaborations (CISSC), "Ministry of Science, Research and Technology" of Iran. 
YO thanks the support from FCT -- Funda\c{c}\~{a}o para a Ci\^{e}ncia e a Tecnologia (Portugal), namely through project UID/04540/2025 and contract LA/P/0095/2020.

\end{acknowledgments}

\begingroup 
\makeatletter
\let\ps@plain\ps@empty
\makeatother
\endgroup

\bibliography{reference-opticalLattice}

\begin{thebibliography}{117}%
\makeatletter
\providecommand \@ifxundefined [1]{%
 \@ifx{#1\undefined}
}%
\providecommand \@ifnum [1]{%
 \ifnum #1\expandafter \@firstoftwo
 \else \expandafter \@secondoftwo
 \fi
}%
\providecommand \@ifx [1]{%
 \ifx #1\expandafter \@firstoftwo
 \else \expandafter \@secondoftwo
 \fi
}%
\providecommand \natexlab [1]{#1}%
\providecommand \enquote  [1]{``#1''}%
\providecommand \bibnamefont  [1]{#1}%
\providecommand \bibfnamefont [1]{#1}%
\providecommand \citenamefont [1]{#1}%
\providecommand \href@noop [0]{\@secondoftwo}%
\providecommand \href [0]{\begingroup \@sanitize@url \@href}%
\providecommand \@href[1]{\@@startlink{#1}\@@href}%
\providecommand \@@href[1]{\endgroup#1\@@endlink}%
\providecommand \@sanitize@url [0]{\catcode `\\12\catcode `\$12\catcode
  `\&12\catcode `\#12\catcode `\^12\catcode `\_12\catcode `\%12\relax}%
\providecommand \@@startlink[1]{}%
\providecommand \@@endlink[0]{}%
\providecommand \url  [0]{\begingroup\@sanitize@url \@url }%
\providecommand \@url [1]{\endgroup\@href {#1}{\urlprefix }}%
\providecommand \urlprefix  [0]{URL }%
\providecommand \Eprint [0]{\href }%
\providecommand \doibase [0]{http://dx.doi.org/}%
\providecommand \selectlanguage [0]{\@gobble}%
\providecommand \bibinfo  [0]{\@secondoftwo}%
\providecommand \bibfield  [0]{\@secondoftwo}%
\providecommand \translation [1]{[#1]}%
\providecommand \BibitemOpen [0]{}%
\providecommand \bibitemStop [0]{}%
\providecommand \bibitemNoStop [0]{.\EOS\space}%
\providecommand \EOS [0]{\spacefactor3000\relax}%
\providecommand \BibitemShut  [1]{\csname bibitem#1\endcsname}%
\let\auto@bib@innerbib\@empty
\bibitem [{\citenamefont {Tino}\ \emph {et~al.}(2013)\citenamefont {Tino},
  \citenamefont {Sorrentino}, \citenamefont {Aguilera}, \citenamefont
  {Battelier}, \citenamefont {Bertoldi}, \citenamefont {Bodart}, \citenamefont
  {Bongs}, \citenamefont {Bouyer}, \citenamefont {Braxmaier}, \citenamefont
  {Cacciapuoti}, \citenamefont {Gaaloul}, \citenamefont {Gürlebeck},
  \citenamefont {Hauth}, \citenamefont {Herrmann}, \citenamefont {Krutzik},
  \citenamefont {Kubelka}, \citenamefont {Landragin}, \citenamefont {Milke},
  \citenamefont {Peters}, \citenamefont {Rasel}, \citenamefont {Rocco},
  \citenamefont {Schubert}, \citenamefont {Schuldt}, \citenamefont
  {Sengstock},\ and\ \citenamefont {Wicht}}]{TINO2013}%
  \BibitemOpen
  \bibfield  {author} {\bibinfo {author} {\bibfnamefont {G.}~\bibnamefont
  {Tino}}, \bibinfo {author} {\bibfnamefont {F.}~\bibnamefont {Sorrentino}},
  \bibinfo {author} {\bibfnamefont {D.}~\bibnamefont {Aguilera}}, \bibinfo
  {author} {\bibfnamefont {B.}~\bibnamefont {Battelier}}, \bibinfo {author}
  {\bibfnamefont {A.}~\bibnamefont {Bertoldi}}, \bibinfo {author}
  {\bibfnamefont {Q.}~\bibnamefont {Bodart}}, \bibinfo {author} {\bibfnamefont
  {K.}~\bibnamefont {Bongs}}, \bibinfo {author} {\bibfnamefont
  {P.}~\bibnamefont {Bouyer}}, \bibinfo {author} {\bibfnamefont
  {C.}~\bibnamefont {Braxmaier}}, \bibinfo {author} {\bibfnamefont
  {L.}~\bibnamefont {Cacciapuoti}}, \bibinfo {author} {\bibfnamefont
  {N.}~\bibnamefont {Gaaloul}}, \bibinfo {author} {\bibfnamefont
  {N.}~\bibnamefont {Gürlebeck}}, \bibinfo {author} {\bibfnamefont
  {M.}~\bibnamefont {Hauth}}, \bibinfo {author} {\bibfnamefont
  {S.}~\bibnamefont {Herrmann}}, \bibinfo {author} {\bibfnamefont
  {M.}~\bibnamefont {Krutzik}}, \bibinfo {author} {\bibfnamefont
  {A.}~\bibnamefont {Kubelka}}, \bibinfo {author} {\bibfnamefont
  {A.}~\bibnamefont {Landragin}}, \bibinfo {author} {\bibfnamefont
  {A.}~\bibnamefont {Milke}}, \bibinfo {author} {\bibfnamefont
  {A.}~\bibnamefont {Peters}}, \bibinfo {author} {\bibfnamefont
  {E.}~\bibnamefont {Rasel}}, \bibinfo {author} {\bibfnamefont
  {E.}~\bibnamefont {Rocco}}, \bibinfo {author} {\bibfnamefont
  {C.}~\bibnamefont {Schubert}}, \bibinfo {author} {\bibfnamefont
  {T.}~\bibnamefont {Schuldt}}, \bibinfo {author} {\bibfnamefont
  {K.}~\bibnamefont {Sengstock}}, \ and\ \bibinfo {author} {\bibfnamefont
  {A.}~\bibnamefont {Wicht}},\ }\href {\doibase
  https://doi.org/10.1016/j.nuclphysbps.2013.09.023} {\bibfield  {journal}
  {\bibinfo  {journal} {Nuclear Physics B - Proceedings Supplements}\ }\textbf
  {\bibinfo {volume} {243-244}},\ \bibinfo {pages} {203} (\bibinfo {year}
  {2013})},\ \bibinfo {note} {proceedings of the IV International Conference on
  Particle and Fundamental Physics in Space}\BibitemShut {NoStop}%
\bibitem [{\citenamefont {Badurina}\ \emph {et~al.}(2020)\citenamefont
  {Badurina}, \citenamefont {Bentine}, \citenamefont {Blas}, \citenamefont
  {Bongs}, \citenamefont {Bortoletto}, \citenamefont {Bowcock}, \citenamefont
  {Bridges}, \citenamefont {Bowden}, \citenamefont {Buchmueller}, \citenamefont
  {Burrage}, \citenamefont {Coleman}, \citenamefont {Elertas}, \citenamefont
  {Ellis}, \citenamefont {Foot}, \citenamefont {Gibson}, \citenamefont
  {Haehnelt}, \citenamefont {Harte}, \citenamefont {Hedges}, \citenamefont
  {Hobson}, \citenamefont {Holynski}, \citenamefont {Jones}, \citenamefont
  {Langlois}, \citenamefont {Lellouch}, \citenamefont {Lewicki}, \citenamefont
  {Maiolino}, \citenamefont {Majewski}, \citenamefont {Malik}, \citenamefont
  {March-Russell}, \citenamefont {McCabe}, \citenamefont {Newbold},
  \citenamefont {Sauer}, \citenamefont {Schneider}, \citenamefont {Shipsey},
  \citenamefont {Singh}, \citenamefont {Uchida}, \citenamefont {Valenzuela},
  \citenamefont {van~der Grinten}, \citenamefont {Vaskonen}, \citenamefont
  {Vossebeld}, \citenamefont {Weatherill},\ and\ \citenamefont
  {Wilmut}}]{Badurina_2020}%
  \BibitemOpen
  \bibfield  {author} {\bibinfo {author} {\bibfnamefont {L.}~\bibnamefont
  {Badurina}}, \bibinfo {author} {\bibfnamefont {E.}~\bibnamefont {Bentine}},
  \bibinfo {author} {\bibfnamefont {D.}~\bibnamefont {Blas}}, \bibinfo {author}
  {\bibfnamefont {K.}~\bibnamefont {Bongs}}, \bibinfo {author} {\bibfnamefont
  {D.}~\bibnamefont {Bortoletto}}, \bibinfo {author} {\bibfnamefont
  {T.}~\bibnamefont {Bowcock}}, \bibinfo {author} {\bibfnamefont
  {K.}~\bibnamefont {Bridges}}, \bibinfo {author} {\bibfnamefont
  {W.}~\bibnamefont {Bowden}}, \bibinfo {author} {\bibfnamefont
  {O.}~\bibnamefont {Buchmueller}}, \bibinfo {author} {\bibfnamefont
  {C.}~\bibnamefont {Burrage}}, \bibinfo {author} {\bibfnamefont
  {J.}~\bibnamefont {Coleman}}, \bibinfo {author} {\bibfnamefont
  {G.}~\bibnamefont {Elertas}}, \bibinfo {author} {\bibfnamefont
  {J.}~\bibnamefont {Ellis}}, \bibinfo {author} {\bibfnamefont
  {C.}~\bibnamefont {Foot}}, \bibinfo {author} {\bibfnamefont {V.}~\bibnamefont
  {Gibson}}, \bibinfo {author} {\bibfnamefont {M.}~\bibnamefont {Haehnelt}},
  \bibinfo {author} {\bibfnamefont {T.}~\bibnamefont {Harte}}, \bibinfo
  {author} {\bibfnamefont {S.}~\bibnamefont {Hedges}}, \bibinfo {author}
  {\bibfnamefont {R.}~\bibnamefont {Hobson}}, \bibinfo {author} {\bibfnamefont
  {M.}~\bibnamefont {Holynski}}, \bibinfo {author} {\bibfnamefont
  {T.}~\bibnamefont {Jones}}, \bibinfo {author} {\bibfnamefont
  {M.}~\bibnamefont {Langlois}}, \bibinfo {author} {\bibfnamefont
  {S.}~\bibnamefont {Lellouch}}, \bibinfo {author} {\bibfnamefont
  {M.}~\bibnamefont {Lewicki}}, \bibinfo {author} {\bibfnamefont
  {R.}~\bibnamefont {Maiolino}}, \bibinfo {author} {\bibfnamefont
  {P.}~\bibnamefont {Majewski}}, \bibinfo {author} {\bibfnamefont
  {S.}~\bibnamefont {Malik}}, \bibinfo {author} {\bibfnamefont
  {J.}~\bibnamefont {March-Russell}}, \bibinfo {author} {\bibfnamefont
  {C.}~\bibnamefont {McCabe}}, \bibinfo {author} {\bibfnamefont
  {D.}~\bibnamefont {Newbold}}, \bibinfo {author} {\bibfnamefont
  {B.}~\bibnamefont {Sauer}}, \bibinfo {author} {\bibfnamefont
  {U.}~\bibnamefont {Schneider}}, \bibinfo {author} {\bibfnamefont
  {I.}~\bibnamefont {Shipsey}}, \bibinfo {author} {\bibfnamefont
  {Y.}~\bibnamefont {Singh}}, \bibinfo {author} {\bibfnamefont
  {M.}~\bibnamefont {Uchida}}, \bibinfo {author} {\bibfnamefont
  {T.}~\bibnamefont {Valenzuela}}, \bibinfo {author} {\bibfnamefont
  {M.}~\bibnamefont {van~der Grinten}}, \bibinfo {author} {\bibfnamefont
  {V.}~\bibnamefont {Vaskonen}}, \bibinfo {author} {\bibfnamefont
  {J.}~\bibnamefont {Vossebeld}}, \bibinfo {author} {\bibfnamefont
  {D.}~\bibnamefont {Weatherill}}, \ and\ \bibinfo {author} {\bibfnamefont
  {I.}~\bibnamefont {Wilmut}},\ }\href {\doibase 10.1088/1475-7516/2020/05/011}
  {\bibfield  {journal} {\bibinfo  {journal} {Journal of Cosmology and
  Astroparticle Physics}\ }\textbf {\bibinfo {volume} {2020}},\ \bibinfo
  {pages} {011} (\bibinfo {year} {2020})}\BibitemShut {NoStop}%
\bibitem [{\citenamefont {Baynham}\ \emph {et~al.}(2025)\citenamefont
  {Baynham}, \citenamefont {Hobson}, \citenamefont {Buchmueller}, \citenamefont
  {Evans}, \citenamefont {Hawkins}, \citenamefont {Iannizzotto-Venezze},
  \citenamefont {Josset}, \citenamefont {Lee}, \citenamefont {Pasatembou},
  \citenamefont {Sauer}, \citenamefont {Tarbutt}, \citenamefont {Walker},
  \citenamefont {Ennis}, \citenamefont {Chauhan}, \citenamefont {Brzakalik},
  \citenamefont {Dey}, \citenamefont {Hedges}, \citenamefont {Stray},
  \citenamefont {Langlois}, \citenamefont {Bongs}, \citenamefont {Hird},
  \citenamefont {Lellouch}, \citenamefont {Holynski}, \citenamefont {Bostwick},
  \citenamefont {Chen}, \citenamefont {Eyler}, \citenamefont {Gibson},
  \citenamefont {Harte}, \citenamefont {Hsu}, \citenamefont {Karzazi},
  \citenamefont {Lu}, \citenamefont {Millward}, \citenamefont {Mitchell},
  \citenamefont {Mouelle}, \citenamefont {Panchumarthi}, \citenamefont
  {Scheper}, \citenamefont {Schneider}, \citenamefont {Su}, \citenamefont
  {Tang}, \citenamefont {Tkalcec}, \citenamefont {Zeuner}, \citenamefont
  {Zhang}, \citenamefont {Zhi}, \citenamefont {Badurina}, \citenamefont
  {Beniwal}, \citenamefont {Blas}, \citenamefont {Carlton}, \citenamefont
  {Ellis}, \citenamefont {McCabe}, \citenamefont {Parish}, \citenamefont
  {Govardhan}, \citenamefont {Vaskonen}, \citenamefont {Bowcock}, \citenamefont
  {Bridges}, \citenamefont {Carroll}, \citenamefont {Coleman}, \citenamefont
  {Elertas}, \citenamefont {Hindley}, \citenamefont {Metelko}, \citenamefont
  {Throssell}, \citenamefont {Tinsley}, \citenamefont {Bentine}, \citenamefont
  {Booth}, \citenamefont {Bortoletto}, \citenamefont {Callaghan}, \citenamefont
  {Foot}, \citenamefont {Gomez-Monedero}, \citenamefont {Hughes}, \citenamefont
  {James}, \citenamefont {Leese}, \citenamefont {Lowe}, \citenamefont
  {March-Russell}, \citenamefont {Sander}, \citenamefont {Schelfhout},
  \citenamefont {Shipsey}, \citenamefont {Weatherill}, \citenamefont {Wood},
  \citenamefont {Bason}, \citenamefont {Hussain}, \citenamefont {Labiad},
  \citenamefont {Marchant}, \citenamefont {Thornton}, \citenamefont
  {Valenzuela}, \citenamefont {Balashov}, \citenamefont {Majewski},
  \citenamefont {Newbold}, \citenamefont {van~der Grinten}, \citenamefont
  {Pan}, \citenamefont {Tam}, \citenamefont {Wilmut}, \citenamefont {Clarke},\
  and\ \citenamefont {Vick}}]{baynham2025}%
  \BibitemOpen
  \bibfield  {author} {\bibinfo {author} {\bibfnamefont {C.~F.~A.}\
  \bibnamefont {Baynham}}, \bibinfo {author} {\bibfnamefont {R.}~\bibnamefont
  {Hobson}}, \bibinfo {author} {\bibfnamefont {O.}~\bibnamefont {Buchmueller}},
  \bibinfo {author} {\bibfnamefont {D.}~\bibnamefont {Evans}}, \bibinfo
  {author} {\bibfnamefont {L.}~\bibnamefont {Hawkins}}, \bibinfo {author}
  {\bibfnamefont {L.}~\bibnamefont {Iannizzotto-Venezze}}, \bibinfo {author}
  {\bibfnamefont {A.}~\bibnamefont {Josset}}, \bibinfo {author} {\bibfnamefont
  {D.}~\bibnamefont {Lee}}, \bibinfo {author} {\bibfnamefont {E.}~\bibnamefont
  {Pasatembou}}, \bibinfo {author} {\bibfnamefont {B.~E.}\ \bibnamefont
  {Sauer}}, \bibinfo {author} {\bibfnamefont {M.~R.}\ \bibnamefont {Tarbutt}},
  \bibinfo {author} {\bibfnamefont {T.}~\bibnamefont {Walker}}, \bibinfo
  {author} {\bibfnamefont {O.}~\bibnamefont {Ennis}}, \bibinfo {author}
  {\bibfnamefont {U.}~\bibnamefont {Chauhan}}, \bibinfo {author} {\bibfnamefont
  {A.}~\bibnamefont {Brzakalik}}, \bibinfo {author} {\bibfnamefont
  {S.}~\bibnamefont {Dey}}, \bibinfo {author} {\bibfnamefont {S.}~\bibnamefont
  {Hedges}}, \bibinfo {author} {\bibfnamefont {B.}~\bibnamefont {Stray}},
  \bibinfo {author} {\bibfnamefont {M.}~\bibnamefont {Langlois}}, \bibinfo
  {author} {\bibfnamefont {K.}~\bibnamefont {Bongs}}, \bibinfo {author}
  {\bibfnamefont {T.}~\bibnamefont {Hird}}, \bibinfo {author} {\bibfnamefont
  {S.}~\bibnamefont {Lellouch}}, \bibinfo {author} {\bibfnamefont
  {M.}~\bibnamefont {Holynski}}, \bibinfo {author} {\bibfnamefont
  {B.}~\bibnamefont {Bostwick}}, \bibinfo {author} {\bibfnamefont
  {J.}~\bibnamefont {Chen}}, \bibinfo {author} {\bibfnamefont {Z.}~\bibnamefont
  {Eyler}}, \bibinfo {author} {\bibfnamefont {V.}~\bibnamefont {Gibson}},
  \bibinfo {author} {\bibfnamefont {T.~L.}\ \bibnamefont {Harte}}, \bibinfo
  {author} {\bibfnamefont {C.~C.}\ \bibnamefont {Hsu}}, \bibinfo {author}
  {\bibfnamefont {M.}~\bibnamefont {Karzazi}}, \bibinfo {author} {\bibfnamefont
  {C.}~\bibnamefont {Lu}}, \bibinfo {author} {\bibfnamefont {B.}~\bibnamefont
  {Millward}}, \bibinfo {author} {\bibfnamefont {J.}~\bibnamefont {Mitchell}},
  \bibinfo {author} {\bibfnamefont {N.}~\bibnamefont {Mouelle}}, \bibinfo
  {author} {\bibfnamefont {B.}~\bibnamefont {Panchumarthi}}, \bibinfo {author}
  {\bibfnamefont {J.}~\bibnamefont {Scheper}}, \bibinfo {author} {\bibfnamefont
  {U.}~\bibnamefont {Schneider}}, \bibinfo {author} {\bibfnamefont
  {X.}~\bibnamefont {Su}}, \bibinfo {author} {\bibfnamefont {Y.}~\bibnamefont
  {Tang}}, \bibinfo {author} {\bibfnamefont {K.}~\bibnamefont {Tkalcec}},
  \bibinfo {author} {\bibfnamefont {M.}~\bibnamefont {Zeuner}}, \bibinfo
  {author} {\bibfnamefont {S.}~\bibnamefont {Zhang}}, \bibinfo {author}
  {\bibfnamefont {Y.}~\bibnamefont {Zhi}}, \bibinfo {author} {\bibfnamefont
  {L.}~\bibnamefont {Badurina}}, \bibinfo {author} {\bibfnamefont
  {A.}~\bibnamefont {Beniwal}}, \bibinfo {author} {\bibfnamefont
  {D.}~\bibnamefont {Blas}}, \bibinfo {author} {\bibfnamefont {J.}~\bibnamefont
  {Carlton}}, \bibinfo {author} {\bibfnamefont {J.}~\bibnamefont {Ellis}},
  \bibinfo {author} {\bibfnamefont {C.}~\bibnamefont {McCabe}}, \bibinfo
  {author} {\bibfnamefont {G.}~\bibnamefont {Parish}}, \bibinfo {author}
  {\bibfnamefont {D.~P.}\ \bibnamefont {Govardhan}}, \bibinfo {author}
  {\bibfnamefont {V.}~\bibnamefont {Vaskonen}}, \bibinfo {author}
  {\bibfnamefont {T.}~\bibnamefont {Bowcock}}, \bibinfo {author} {\bibfnamefont
  {K.}~\bibnamefont {Bridges}}, \bibinfo {author} {\bibfnamefont
  {A.}~\bibnamefont {Carroll}}, \bibinfo {author} {\bibfnamefont
  {J.}~\bibnamefont {Coleman}}, \bibinfo {author} {\bibfnamefont
  {G.}~\bibnamefont {Elertas}}, \bibinfo {author} {\bibfnamefont
  {S.}~\bibnamefont {Hindley}}, \bibinfo {author} {\bibfnamefont
  {C.}~\bibnamefont {Metelko}}, \bibinfo {author} {\bibfnamefont
  {H.}~\bibnamefont {Throssell}}, \bibinfo {author} {\bibfnamefont {J.~N.}\
  \bibnamefont {Tinsley}}, \bibinfo {author} {\bibfnamefont {E.}~\bibnamefont
  {Bentine}}, \bibinfo {author} {\bibfnamefont {M.}~\bibnamefont {Booth}},
  \bibinfo {author} {\bibfnamefont {D.}~\bibnamefont {Bortoletto}}, \bibinfo
  {author} {\bibfnamefont {N.}~\bibnamefont {Callaghan}}, \bibinfo {author}
  {\bibfnamefont {C.}~\bibnamefont {Foot}}, \bibinfo {author} {\bibfnamefont
  {C.}~\bibnamefont {Gomez-Monedero}}, \bibinfo {author} {\bibfnamefont
  {K.}~\bibnamefont {Hughes}}, \bibinfo {author} {\bibfnamefont
  {A.}~\bibnamefont {James}}, \bibinfo {author} {\bibfnamefont
  {T.}~\bibnamefont {Leese}}, \bibinfo {author} {\bibfnamefont
  {A.}~\bibnamefont {Lowe}}, \bibinfo {author} {\bibfnamefont {J.}~\bibnamefont
  {March-Russell}}, \bibinfo {author} {\bibfnamefont {J.}~\bibnamefont
  {Sander}}, \bibinfo {author} {\bibfnamefont {J.}~\bibnamefont {Schelfhout}},
  \bibinfo {author} {\bibfnamefont {I.}~\bibnamefont {Shipsey}}, \bibinfo
  {author} {\bibfnamefont {D.}~\bibnamefont {Weatherill}}, \bibinfo {author}
  {\bibfnamefont {D.}~\bibnamefont {Wood}}, \bibinfo {author} {\bibfnamefont
  {M.~G.}\ \bibnamefont {Bason}}, \bibinfo {author} {\bibfnamefont
  {K.}~\bibnamefont {Hussain}}, \bibinfo {author} {\bibfnamefont
  {H.}~\bibnamefont {Labiad}}, \bibinfo {author} {\bibfnamefont {A.~L.}\
  \bibnamefont {Marchant}}, \bibinfo {author} {\bibfnamefont {T.~C.}\
  \bibnamefont {Thornton}}, \bibinfo {author} {\bibfnamefont {T.}~\bibnamefont
  {Valenzuela}}, \bibinfo {author} {\bibfnamefont {S.~N.}\ \bibnamefont
  {Balashov}}, \bibinfo {author} {\bibfnamefont {P.}~\bibnamefont {Majewski}},
  \bibinfo {author} {\bibfnamefont {D.}~\bibnamefont {Newbold}}, \bibinfo
  {author} {\bibfnamefont {M.~G.~D.}\ \bibnamefont {van~der Grinten}}, \bibinfo
  {author} {\bibfnamefont {Z.}~\bibnamefont {Pan}}, \bibinfo {author}
  {\bibfnamefont {Z.}~\bibnamefont {Tam}}, \bibinfo {author} {\bibfnamefont
  {I.}~\bibnamefont {Wilmut}}, \bibinfo {author} {\bibfnamefont
  {K.}~\bibnamefont {Clarke}}, \ and\ \bibinfo {author} {\bibfnamefont
  {A.}~\bibnamefont {Vick}},\ }\href {https://arxiv.org/abs/2504.09158} {\
  (\bibinfo {year} {2025})},\ \Eprint {http://arxiv.org/abs/2504.09158}
  {arXiv:2504.09158 [hep-ex]} \BibitemShut {NoStop}%
\bibitem [{\citenamefont {Cepollaro}\ \emph {et~al.}(2023)\citenamefont
  {Cepollaro}, \citenamefont {Giacomini},\ and\ \citenamefont
  {Paris}}]{Cepollaro2023}%
  \BibitemOpen
  \bibfield  {author} {\bibinfo {author} {\bibfnamefont {C.}~\bibnamefont
  {Cepollaro}}, \bibinfo {author} {\bibfnamefont {F.}~\bibnamefont
  {Giacomini}}, \ and\ \bibinfo {author} {\bibfnamefont {M.~G.}\ \bibnamefont
  {Paris}},\ }\href {\doibase 10.22331/q-2023-03-13-946} {\bibfield  {journal}
  {\bibinfo  {journal} {{Quantum}}\ }\textbf {\bibinfo {volume} {7}},\ \bibinfo
  {pages} {946} (\bibinfo {year} {2023})}\BibitemShut {NoStop}%
\bibitem [{\citenamefont {Block}\ \emph {et~al.}(2021)\citenamefont {Block},
  \citenamefont {Kobrin}, \citenamefont {Jarmola}, \citenamefont {Hsieh},
  \citenamefont {Zu}, \citenamefont {Figueroa}, \citenamefont {Acosta},
  \citenamefont {Minguzzi}, \citenamefont {Maze}, \citenamefont {Budker},\ and\
  \citenamefont {Yao}}]{Block2021}%
  \BibitemOpen
  \bibfield  {author} {\bibinfo {author} {\bibfnamefont {M.}~\bibnamefont
  {Block}}, \bibinfo {author} {\bibfnamefont {B.}~\bibnamefont {Kobrin}},
  \bibinfo {author} {\bibfnamefont {A.}~\bibnamefont {Jarmola}}, \bibinfo
  {author} {\bibfnamefont {S.}~\bibnamefont {Hsieh}}, \bibinfo {author}
  {\bibfnamefont {C.}~\bibnamefont {Zu}}, \bibinfo {author} {\bibfnamefont
  {N.}~\bibnamefont {Figueroa}}, \bibinfo {author} {\bibfnamefont
  {V.}~\bibnamefont {Acosta}}, \bibinfo {author} {\bibfnamefont
  {J.}~\bibnamefont {Minguzzi}}, \bibinfo {author} {\bibfnamefont
  {J.}~\bibnamefont {Maze}}, \bibinfo {author} {\bibfnamefont {D.}~\bibnamefont
  {Budker}}, \ and\ \bibinfo {author} {\bibfnamefont {N.}~\bibnamefont {Yao}},\
  }\href {\doibase 10.1103/PhysRevApplied.16.024024} {\bibfield  {journal}
  {\bibinfo  {journal} {Phys. Rev. Appl.}\ }\textbf {\bibinfo {volume} {16}},\
  \bibinfo {pages} {024024} (\bibinfo {year} {2021})}\BibitemShut {NoStop}%
\bibitem [{\citenamefont {Qiu}\ \emph {et~al.}(2022)\citenamefont {Qiu},
  \citenamefont {Hamo}, \citenamefont {Vool}, \citenamefont {Zhou},\ and\
  \citenamefont {Yacoby}}]{Qiu_2022}%
  \BibitemOpen
  \bibfield  {author} {\bibinfo {author} {\bibfnamefont {Z.}~\bibnamefont
  {Qiu}}, \bibinfo {author} {\bibfnamefont {A.}~\bibnamefont {Hamo}}, \bibinfo
  {author} {\bibfnamefont {U.}~\bibnamefont {Vool}}, \bibinfo {author}
  {\bibfnamefont {T.~X.}\ \bibnamefont {Zhou}}, \ and\ \bibinfo {author}
  {\bibfnamefont {A.}~\bibnamefont {Yacoby}},\ }\href {\doibase
  10.1038/s41534-022-00622-3} {\bibfield  {journal} {\bibinfo  {journal} {npj
  Quantum Information}\ }\textbf {\bibinfo {volume} {8}} (\bibinfo {year}
  {2022}),\ 10.1038/s41534-022-00622-3}\BibitemShut {NoStop}%
\bibitem [{\citenamefont {Chen}\ \emph {et~al.}(2017)\citenamefont {Chen},
  \citenamefont {Clevenson}, \citenamefont {Johnson}, \citenamefont {Pham},
  \citenamefont {Englund}, \citenamefont {Hemmer},\ and\ \citenamefont
  {Braje}}]{Chen_2017}%
  \BibitemOpen
  \bibfield  {author} {\bibinfo {author} {\bibfnamefont {E.~H.}\ \bibnamefont
  {Chen}}, \bibinfo {author} {\bibfnamefont {H.~A.}\ \bibnamefont {Clevenson}},
  \bibinfo {author} {\bibfnamefont {K.~A.}\ \bibnamefont {Johnson}}, \bibinfo
  {author} {\bibfnamefont {L.~M.}\ \bibnamefont {Pham}}, \bibinfo {author}
  {\bibfnamefont {D.~R.}\ \bibnamefont {Englund}}, \bibinfo {author}
  {\bibfnamefont {P.~R.}\ \bibnamefont {Hemmer}}, \ and\ \bibinfo {author}
  {\bibfnamefont {D.~A.}\ \bibnamefont {Braje}},\ }\href {\doibase
  10.1103/physreva.95.053417} {\bibfield  {journal} {\bibinfo  {journal}
  {Physical Review A}\ }\textbf {\bibinfo {volume} {95}} (\bibinfo {year}
  {2017}),\ 10.1103/physreva.95.053417}\BibitemShut {NoStop}%
\bibitem [{\citenamefont {Iwasaki}\ \emph {et~al.}(2017)\citenamefont
  {Iwasaki}, \citenamefont {Naruki}, \citenamefont {Tahara}, \citenamefont
  {Makino}, \citenamefont {Kato}, \citenamefont {Ogura}, \citenamefont
  {Takeuchi}, \citenamefont {Yamasaki},\ and\ \citenamefont
  {Hatano}}]{Iwasaki2017}%
  \BibitemOpen
  \bibfield  {author} {\bibinfo {author} {\bibfnamefont {T.}~\bibnamefont
  {Iwasaki}}, \bibinfo {author} {\bibfnamefont {W.}~\bibnamefont {Naruki}},
  \bibinfo {author} {\bibfnamefont {K.}~\bibnamefont {Tahara}}, \bibinfo
  {author} {\bibfnamefont {T.}~\bibnamefont {Makino}}, \bibinfo {author}
  {\bibfnamefont {H.}~\bibnamefont {Kato}}, \bibinfo {author} {\bibfnamefont
  {M.}~\bibnamefont {Ogura}}, \bibinfo {author} {\bibfnamefont
  {D.}~\bibnamefont {Takeuchi}}, \bibinfo {author} {\bibfnamefont
  {S.}~\bibnamefont {Yamasaki}}, \ and\ \bibinfo {author} {\bibfnamefont
  {M.}~\bibnamefont {Hatano}},\ }\href {\doibase 10.1021/acsnano.6b04460}
  {\bibfield  {journal} {\bibinfo  {journal} {ACS Nano}\ }\textbf {\bibinfo
  {volume} {11}} (\bibinfo {year} {2017}),\
  10.1021/acsnano.6b04460}\BibitemShut {NoStop}%
\bibitem [{\citenamefont {L\'opez-Morales}\ \emph {et~al.}(2024)\citenamefont
  {L\'opez-Morales}, \citenamefont {Zajac}, \citenamefont {Flick},
  \citenamefont {Meriles},\ and\ \citenamefont {Dreyer}}]{Morales2024}%
  \BibitemOpen
  \bibfield  {author} {\bibinfo {author} {\bibfnamefont {G.~I.}\ \bibnamefont
  {L\'opez-Morales}}, \bibinfo {author} {\bibfnamefont {J.~M.}\ \bibnamefont
  {Zajac}}, \bibinfo {author} {\bibfnamefont {J.}~\bibnamefont {Flick}},
  \bibinfo {author} {\bibfnamefont {C.~A.}\ \bibnamefont {Meriles}}, \ and\
  \bibinfo {author} {\bibfnamefont {C.~E.}\ \bibnamefont {Dreyer}},\ }\href
  {\doibase 10.1103/PhysRevB.110.245127} {\bibfield  {journal} {\bibinfo
  {journal} {Phys. Rev. B}\ }\textbf {\bibinfo {volume} {110}},\ \bibinfo
  {pages} {245127} (\bibinfo {year} {2024})}\BibitemShut {NoStop}%
\bibitem [{\citenamefont {Michl}\ \emph {et~al.}(2019)\citenamefont {Michl},
  \citenamefont {Steiner}, \citenamefont {Denisenko}, \citenamefont {Bülau},
  \citenamefont {Zimmermann}, \citenamefont {Nakamura}, \citenamefont {Sumiya},
  \citenamefont {Onoda}, \citenamefont {Neumann}, \citenamefont {Isoya},\ and\
  \citenamefont {Wrachtrup}}]{Michl_2019}%
  \BibitemOpen
  \bibfield  {author} {\bibinfo {author} {\bibfnamefont {J.}~\bibnamefont
  {Michl}}, \bibinfo {author} {\bibfnamefont {J.}~\bibnamefont {Steiner}},
  \bibinfo {author} {\bibfnamefont {A.}~\bibnamefont {Denisenko}}, \bibinfo
  {author} {\bibfnamefont {A.}~\bibnamefont {Bülau}}, \bibinfo {author}
  {\bibfnamefont {A.}~\bibnamefont {Zimmermann}}, \bibinfo {author}
  {\bibfnamefont {K.}~\bibnamefont {Nakamura}}, \bibinfo {author}
  {\bibfnamefont {H.}~\bibnamefont {Sumiya}}, \bibinfo {author} {\bibfnamefont
  {S.}~\bibnamefont {Onoda}}, \bibinfo {author} {\bibfnamefont
  {P.}~\bibnamefont {Neumann}}, \bibinfo {author} {\bibfnamefont
  {J.}~\bibnamefont {Isoya}}, \ and\ \bibinfo {author} {\bibfnamefont
  {J.}~\bibnamefont {Wrachtrup}},\ }\href {\doibase
  10.1021/acs.nanolett.9b00900} {\bibfield  {journal} {\bibinfo  {journal}
  {Nano Letters}\ }\textbf {\bibinfo {volume} {19}},\ \bibinfo {pages}
  {4904–4910} (\bibinfo {year} {2019})}\BibitemShut {NoStop}%
\bibitem [{\citenamefont {Yuan}\ \emph {et~al.}(2023)\citenamefont {Yuan},
  \citenamefont {Yang}, \citenamefont {Jing}, \citenamefont {Zhang},
  \citenamefont {Jiao}, \citenamefont {Li}, \citenamefont {Zhang},
  \citenamefont {Xiao},\ and\ \citenamefont {Jia}}]{Yuan_2023}%
  \BibitemOpen
  \bibfield  {author} {\bibinfo {author} {\bibfnamefont {J.}~\bibnamefont
  {Yuan}}, \bibinfo {author} {\bibfnamefont {W.}~\bibnamefont {Yang}}, \bibinfo
  {author} {\bibfnamefont {M.}~\bibnamefont {Jing}}, \bibinfo {author}
  {\bibfnamefont {H.}~\bibnamefont {Zhang}}, \bibinfo {author} {\bibfnamefont
  {Y.}~\bibnamefont {Jiao}}, \bibinfo {author} {\bibfnamefont {W.}~\bibnamefont
  {Li}}, \bibinfo {author} {\bibfnamefont {L.}~\bibnamefont {Zhang}}, \bibinfo
  {author} {\bibfnamefont {L.}~\bibnamefont {Xiao}}, \ and\ \bibinfo {author}
  {\bibfnamefont {S.}~\bibnamefont {Jia}},\ }\href {\doibase
  10.1088/1361-6633/acf22f} {\bibfield  {journal} {\bibinfo  {journal} {Reports
  on Progress in Physics}\ }\textbf {\bibinfo {volume} {86}},\ \bibinfo {pages}
  {106001} (\bibinfo {year} {2023})}\BibitemShut {NoStop}%
\bibitem [{\citenamefont {Liu}\ \emph {et~al.}(2022)\citenamefont {Liu},
  \citenamefont {Zhang}, \citenamefont {Liu}, \citenamefont {Zhang},
  \citenamefont {Zhu}, \citenamefont {Gao}, \citenamefont {Guo}, \citenamefont
  {Ding},\ and\ \citenamefont {Shi}}]{Liu_2022}%
  \BibitemOpen
  \bibfield  {author} {\bibinfo {author} {\bibfnamefont {B.}~\bibnamefont
  {Liu}}, \bibinfo {author} {\bibfnamefont {L.-H.}\ \bibnamefont {Zhang}},
  \bibinfo {author} {\bibfnamefont {Z.-K.}\ \bibnamefont {Liu}}, \bibinfo
  {author} {\bibfnamefont {Z.-Y.}\ \bibnamefont {Zhang}}, \bibinfo {author}
  {\bibfnamefont {Z.-H.}\ \bibnamefont {Zhu}}, \bibinfo {author} {\bibfnamefont
  {W.}~\bibnamefont {Gao}}, \bibinfo {author} {\bibfnamefont {G.-C.}\
  \bibnamefont {Guo}}, \bibinfo {author} {\bibfnamefont {D.-S.}\ \bibnamefont
  {Ding}}, \ and\ \bibinfo {author} {\bibfnamefont {B.-S.}\ \bibnamefont
  {Shi}},\ }\href {\doibase 10.1103/physrevapplied.18.014045} {\bibfield
  {journal} {\bibinfo  {journal} {Physical Review Applied}\ }\textbf {\bibinfo
  {volume} {18}} (\bibinfo {year} {2022}),\
  10.1103/physrevapplied.18.014045}\BibitemShut {NoStop}%
\bibitem [{\citenamefont {Liu}\ \emph {et~al.}(2024)\citenamefont {Liu},
  \citenamefont {Zhang}, \citenamefont {Liu}, \citenamefont {Wang},
  \citenamefont {Ma}, \citenamefont {Han}, \citenamefont {Zhang}, \citenamefont
  {Shao}, \citenamefont {Zhang}, \citenamefont {Li}, \citenamefont {Chen},
  \citenamefont {Ding},\ and\ \citenamefont {Shi}}]{liu2024}%
  \BibitemOpen
  \bibfield  {author} {\bibinfo {author} {\bibfnamefont {B.}~\bibnamefont
  {Liu}}, \bibinfo {author} {\bibfnamefont {L.-H.}\ \bibnamefont {Zhang}},
  \bibinfo {author} {\bibfnamefont {Z.-K.}\ \bibnamefont {Liu}}, \bibinfo
  {author} {\bibfnamefont {Q.-F.}\ \bibnamefont {Wang}}, \bibinfo {author}
  {\bibfnamefont {Y.}~\bibnamefont {Ma}}, \bibinfo {author} {\bibfnamefont
  {T.-Y.}\ \bibnamefont {Han}}, \bibinfo {author} {\bibfnamefont {Z.-Y.}\
  \bibnamefont {Zhang}}, \bibinfo {author} {\bibfnamefont {S.-Y.}\ \bibnamefont
  {Shao}}, \bibinfo {author} {\bibfnamefont {J.}~\bibnamefont {Zhang}},
  \bibinfo {author} {\bibfnamefont {Q.}~\bibnamefont {Li}}, \bibinfo {author}
  {\bibfnamefont {H.-C.}\ \bibnamefont {Chen}}, \bibinfo {author}
  {\bibfnamefont {D.-S.}\ \bibnamefont {Ding}}, \ and\ \bibinfo {author}
  {\bibfnamefont {B.-S.}\ \bibnamefont {Shi}},\ }\href
  {https://arxiv.org/abs/2404.06915} {\  (\bibinfo {year} {2024})},\ \Eprint
  {http://arxiv.org/abs/2404.06915} {arXiv:2404.06915 [physics.atom-ph]}
  \BibitemShut {NoStop}%
\bibitem [{\citenamefont {Budker}\ and\ \citenamefont
  {Romalis}(2007)}]{Budker_2007}%
  \BibitemOpen
  \bibfield  {author} {\bibinfo {author} {\bibfnamefont {D.}~\bibnamefont
  {Budker}}\ and\ \bibinfo {author} {\bibfnamefont {M.}~\bibnamefont
  {Romalis}},\ }\href {\doibase 10.1038/nphys566} {\bibfield  {journal}
  {\bibinfo  {journal} {Nature Physics}\ }\textbf {\bibinfo {volume} {3}},\
  \bibinfo {pages} {227–234} (\bibinfo {year} {2007})}\BibitemShut {NoStop}%
\bibitem [{\citenamefont {Griffith}\ \emph {et~al.}(2010)\citenamefont
  {Griffith}, \citenamefont {Knappe},\ and\ \citenamefont
  {Kitching}}]{Griffith2010}%
  \BibitemOpen
  \bibfield  {author} {\bibinfo {author} {\bibfnamefont {W.~C.}\ \bibnamefont
  {Griffith}}, \bibinfo {author} {\bibfnamefont {S.}~\bibnamefont {Knappe}}, \
  and\ \bibinfo {author} {\bibfnamefont {J.}~\bibnamefont {Kitching}},\ }\href
  {\doibase 10.1364/OE.18.027167} {\bibfield  {journal} {\bibinfo  {journal}
  {Opt. Express}\ }\textbf {\bibinfo {volume} {18}},\ \bibinfo {pages} {27167}
  (\bibinfo {year} {2010})}\BibitemShut {NoStop}%
\bibitem [{\citenamefont {lu}\ \emph {et~al.}(2023)\citenamefont {lu},
  \citenamefont {Zhao}, \citenamefont {Zhu}, \citenamefont {Liu}, \citenamefont
  {Zhuang}, \citenamefont {Guangyou},\ and\ \citenamefont
  {Zhang}}]{Yuantian2023}%
  \BibitemOpen
  \bibfield  {author} {\bibinfo {author} {\bibfnamefont {Y.}~\bibnamefont
  {lu}}, \bibinfo {author} {\bibfnamefont {T.}~\bibnamefont {Zhao}}, \bibinfo
  {author} {\bibfnamefont {W.}~\bibnamefont {Zhu}}, \bibinfo {author}
  {\bibfnamefont {L.}~\bibnamefont {Liu}}, \bibinfo {author} {\bibfnamefont
  {X.}~\bibnamefont {Zhuang}}, \bibinfo {author} {\bibfnamefont
  {F.}~\bibnamefont {Guangyou}}, \ and\ \bibinfo {author} {\bibfnamefont
  {X.}~\bibnamefont {Zhang}},\ }\href {\doibase 10.3390/s23115318} {\bibfield
  {journal} {\bibinfo  {journal} {Sensors}\ }\textbf {\bibinfo {volume} {23}},\
  \bibinfo {pages} {5318} (\bibinfo {year} {2023})}\BibitemShut {NoStop}%
\bibitem [{\citenamefont {Rondin}\ \emph {et~al.}(2014)\citenamefont {Rondin},
  \citenamefont {Tetienne}, \citenamefont {Hingant}, \citenamefont {Roch},
  \citenamefont {Maletinsky},\ and\ \citenamefont {Jacques}}]{Rondin_2014}%
  \BibitemOpen
  \bibfield  {author} {\bibinfo {author} {\bibfnamefont {L.}~\bibnamefont
  {Rondin}}, \bibinfo {author} {\bibfnamefont {J.-P.}\ \bibnamefont
  {Tetienne}}, \bibinfo {author} {\bibfnamefont {T.}~\bibnamefont {Hingant}},
  \bibinfo {author} {\bibfnamefont {J.-F.}\ \bibnamefont {Roch}}, \bibinfo
  {author} {\bibfnamefont {P.}~\bibnamefont {Maletinsky}}, \ and\ \bibinfo
  {author} {\bibfnamefont {V.}~\bibnamefont {Jacques}},\ }\href {\doibase
  10.1088/0034-4885/77/5/056503} {\bibfield  {journal} {\bibinfo  {journal}
  {Reports on Progress in Physics}\ }\textbf {\bibinfo {volume} {77}},\
  \bibinfo {pages} {056503} (\bibinfo {year} {2014})}\BibitemShut {NoStop}%
\bibitem [{\citenamefont {Hong}\ \emph {et~al.}(2013)\citenamefont {Hong},
  \citenamefont {Grinolds}, \citenamefont {Pham}, \citenamefont {Le~Sage},
  \citenamefont {Luan}, \citenamefont {Walsworth},\ and\ \citenamefont
  {Yacoby}}]{Hong_2013}%
  \BibitemOpen
  \bibfield  {author} {\bibinfo {author} {\bibfnamefont {S.}~\bibnamefont
  {Hong}}, \bibinfo {author} {\bibfnamefont {M.~S.}\ \bibnamefont {Grinolds}},
  \bibinfo {author} {\bibfnamefont {L.~M.}\ \bibnamefont {Pham}}, \bibinfo
  {author} {\bibfnamefont {D.}~\bibnamefont {Le~Sage}}, \bibinfo {author}
  {\bibfnamefont {L.}~\bibnamefont {Luan}}, \bibinfo {author} {\bibfnamefont
  {R.~L.}\ \bibnamefont {Walsworth}}, \ and\ \bibinfo {author} {\bibfnamefont
  {A.}~\bibnamefont {Yacoby}},\ }\href {\doibase 10.1557/mrs.2013.23}
  {\bibfield  {journal} {\bibinfo  {journal} {MRS Bulletin}\ }\textbf {\bibinfo
  {volume} {38}},\ \bibinfo {pages} {155–161} (\bibinfo {year}
  {2013})}\BibitemShut {NoStop}%
\bibitem [{\citenamefont {Taylor}\ \emph {et~al.}(2008)\citenamefont {Taylor},
  \citenamefont {Cappellaro}, \citenamefont {Childress}, \citenamefont {Jiang},
  \citenamefont {Budker}, \citenamefont {Hemmer}, \citenamefont {Yacoby},
  \citenamefont {Walsworth},\ and\ \citenamefont {Lukin}}]{Taylor_2008}%
  \BibitemOpen
  \bibfield  {author} {\bibinfo {author} {\bibfnamefont {J.~M.}\ \bibnamefont
  {Taylor}}, \bibinfo {author} {\bibfnamefont {P.}~\bibnamefont {Cappellaro}},
  \bibinfo {author} {\bibfnamefont {L.}~\bibnamefont {Childress}}, \bibinfo
  {author} {\bibfnamefont {L.}~\bibnamefont {Jiang}}, \bibinfo {author}
  {\bibfnamefont {D.}~\bibnamefont {Budker}}, \bibinfo {author} {\bibfnamefont
  {P.~R.}\ \bibnamefont {Hemmer}}, \bibinfo {author} {\bibfnamefont
  {A.}~\bibnamefont {Yacoby}}, \bibinfo {author} {\bibfnamefont
  {R.}~\bibnamefont {Walsworth}}, \ and\ \bibinfo {author} {\bibfnamefont
  {M.~D.}\ \bibnamefont {Lukin}},\ }\href {\doibase 10.1038/nphys1075}
  {\bibfield  {journal} {\bibinfo  {journal} {Nature Physics}\ }\textbf
  {\bibinfo {volume} {4}},\ \bibinfo {pages} {810–816} (\bibinfo {year}
  {2008})}\BibitemShut {NoStop}%
\bibitem [{\citenamefont {Zhao}\ \emph {et~al.}(2011)\citenamefont {Zhao},
  \citenamefont {Hu}, \citenamefont {Ho}, \citenamefont {Wan},\ and\
  \citenamefont {Liu}}]{Zhao2011}%
  \BibitemOpen
  \bibfield  {author} {\bibinfo {author} {\bibfnamefont {N.}~\bibnamefont
  {Zhao}}, \bibinfo {author} {\bibfnamefont {J.-L.}\ \bibnamefont {Hu}},
  \bibinfo {author} {\bibfnamefont {S.-W.}\ \bibnamefont {Ho}}, \bibinfo
  {author} {\bibfnamefont {J.}~\bibnamefont {Wan}}, \ and\ \bibinfo {author}
  {\bibfnamefont {R.-B.}\ \bibnamefont {Liu}},\ }\href {\doibase
  10.1038/nnano.2011.22} {\bibfield  {journal} {\bibinfo  {journal} {Nature
  nanotechnology}\ }\textbf {\bibinfo {volume} {6}},\ \bibinfo {pages} {242}
  (\bibinfo {year} {2011})}\BibitemShut {NoStop}%
\bibitem [{\citenamefont {{Lin}}\ \emph {et~al.}(2023)\citenamefont {{Lin}},
  \citenamefont {{Yi}},\ and\ \citenamefont {{Xue}}}]{Quan2023}%
  \BibitemOpen
  \bibfield  {author} {\bibinfo {author} {\bibfnamefont {Q.}~\bibnamefont
  {{Lin}}}, \bibinfo {author} {\bibfnamefont {W.}~\bibnamefont {{Yi}}}, \ and\
  \bibinfo {author} {\bibfnamefont {P.}~\bibnamefont {{Xue}}},\ }\href
  {\doibase 10.1038/s41467-023-42045-4} {\bibfield  {journal} {\bibinfo
  {journal} {Nature Communications}\ }\textbf {\bibinfo {volume} {14}},\
  \bibinfo {eid} {6283} (\bibinfo {year} {2023})},\ \Eprint
  {http://arxiv.org/abs/2212.00387} {arXiv:2212.00387 [cond-mat.mes-hall]}
  \BibitemShut {NoStop}%
\bibitem [{\citenamefont {Yu}\ \emph {et~al.}(2024)\citenamefont {Yu},
  \citenamefont {Zhang}, \citenamefont {Xu}, \citenamefont {Chen},
  \citenamefont {Zheng}, \citenamefont {Li}, \citenamefont {Liu}, \citenamefont
  {Pan}, \citenamefont {Budker},\ and\ \citenamefont {Liu}}]{Yu2024}%
  \BibitemOpen
  \bibfield  {author} {\bibinfo {author} {\bibfnamefont {Y.-H.}\ \bibnamefont
  {Yu}}, \bibinfo {author} {\bibfnamefont {R.-Z.}\ \bibnamefont {Zhang}},
  \bibinfo {author} {\bibfnamefont {Y.}~\bibnamefont {Xu}}, \bibinfo {author}
  {\bibfnamefont {X.-Q.}\ \bibnamefont {Chen}}, \bibinfo {author}
  {\bibfnamefont {H.}~\bibnamefont {Zheng}}, \bibinfo {author} {\bibfnamefont
  {Q.}~\bibnamefont {Li}}, \bibinfo {author} {\bibfnamefont {R.-B.}\
  \bibnamefont {Liu}}, \bibinfo {author} {\bibfnamefont {X.-Y.}\ \bibnamefont
  {Pan}}, \bibinfo {author} {\bibfnamefont {D.}~\bibnamefont {Budker}}, \ and\
  \bibinfo {author} {\bibfnamefont {G.-Q.}\ \bibnamefont {Liu}},\ }\href
  {\doibase 10.1103/PhysRevApplied.21.044051} {\bibfield  {journal} {\bibinfo
  {journal} {Phys. Rev. Appl.}\ }\textbf {\bibinfo {volume} {21}},\ \bibinfo
  {pages} {044051} (\bibinfo {year} {2024})}\BibitemShut {NoStop}%
\bibitem [{\citenamefont {Kurzyna}\ \emph {et~al.}(2025)\citenamefont
  {Kurzyna}, \citenamefont {Niewelt}, \citenamefont {Mazelanik}, \citenamefont
  {Wasilewski}, \citenamefont {Demkowicz-Dobrza{\'n}ski},\ and\ \citenamefont
  {Parniak}}]{kurzyna2025}%
  \BibitemOpen
  \bibfield  {author} {\bibinfo {author} {\bibfnamefont {S.}~\bibnamefont
  {Kurzyna}}, \bibinfo {author} {\bibfnamefont {B.}~\bibnamefont {Niewelt}},
  \bibinfo {author} {\bibfnamefont {M.}~\bibnamefont {Mazelanik}}, \bibinfo
  {author} {\bibfnamefont {W.}~\bibnamefont {Wasilewski}}, \bibinfo {author}
  {\bibfnamefont {R.}~\bibnamefont {Demkowicz-Dobrza{\'n}ski}}, \ and\ \bibinfo
  {author} {\bibfnamefont {M.}~\bibnamefont {Parniak}},\ }\href
  {https://arxiv.org/abs/2505.01506} {\  (\bibinfo {year} {2025})},\ \Eprint
  {http://arxiv.org/abs/2505.01506} {arXiv:2505.01506 [quant-ph]} \BibitemShut
  {NoStop}%
\bibitem [{\citenamefont {Giovannetti}\ \emph {et~al.}(2006)\citenamefont
  {Giovannetti}, \citenamefont {Lloyd},\ and\ \citenamefont
  {Maccone}}]{Giovannetti_2006}%
  \BibitemOpen
  \bibfield  {author} {\bibinfo {author} {\bibfnamefont {V.}~\bibnamefont
  {Giovannetti}}, \bibinfo {author} {\bibfnamefont {S.}~\bibnamefont {Lloyd}},
  \ and\ \bibinfo {author} {\bibfnamefont {L.}~\bibnamefont {Maccone}},\ }\href
  {\doibase 10.1103/physrevlett.96.010401} {\bibfield  {journal} {\bibinfo
  {journal} {Physical Review Letters}\ }\textbf {\bibinfo {volume} {96}}
  (\bibinfo {year} {2006}),\ 10.1103/physrevlett.96.010401}\BibitemShut
  {NoStop}%
\bibitem [{\citenamefont {Giovannetti}\ \emph {et~al.}(2011)\citenamefont
  {Giovannetti}, \citenamefont {Lloyd},\ and\ \citenamefont
  {Maccone}}]{Giovannetti_2011}%
  \BibitemOpen
  \bibfield  {author} {\bibinfo {author} {\bibfnamefont {V.}~\bibnamefont
  {Giovannetti}}, \bibinfo {author} {\bibfnamefont {S.}~\bibnamefont {Lloyd}},
  \ and\ \bibinfo {author} {\bibfnamefont {L.}~\bibnamefont {Maccone}},\ }\href
  {\doibase 10.1038/nphoton.2011.35} {\bibfield  {journal} {\bibinfo  {journal}
  {Nature Photonics}\ }\textbf {\bibinfo {volume} {5}},\ \bibinfo {pages}
  {222–229} (\bibinfo {year} {2011})}\BibitemShut {NoStop}%
\bibitem [{\citenamefont {Giovannetti}\ \emph {et~al.}(2004)\citenamefont
  {Giovannetti}, \citenamefont {Lloyd},\ and\ \citenamefont
  {Maccone}}]{Giovannetti_2004}%
  \BibitemOpen
  \bibfield  {author} {\bibinfo {author} {\bibfnamefont {V.}~\bibnamefont
  {Giovannetti}}, \bibinfo {author} {\bibfnamefont {S.}~\bibnamefont {Lloyd}},
  \ and\ \bibinfo {author} {\bibfnamefont {L.}~\bibnamefont {Maccone}},\ }\href
  {\doibase 10.1126/science.1104149} {\bibfield  {journal} {\bibinfo  {journal}
  {Science}\ }\textbf {\bibinfo {volume} {306}},\ \bibinfo {pages}
  {1330–1336} (\bibinfo {year} {2004})}\BibitemShut {NoStop}%
\bibitem [{\citenamefont {Gross}\ \emph {et~al.}(2010)\citenamefont {Gross},
  \citenamefont {Zibold}, \citenamefont {Nicklas}, \citenamefont {Estève},\
  and\ \citenamefont {Oberthaler}}]{Gross_2010}%
  \BibitemOpen
  \bibfield  {author} {\bibinfo {author} {\bibfnamefont {C.}~\bibnamefont
  {Gross}}, \bibinfo {author} {\bibfnamefont {T.}~\bibnamefont {Zibold}},
  \bibinfo {author} {\bibfnamefont {E.}~\bibnamefont {Nicklas}}, \bibinfo
  {author} {\bibfnamefont {J.}~\bibnamefont {Estève}}, \ and\ \bibinfo
  {author} {\bibfnamefont {M.~K.}\ \bibnamefont {Oberthaler}},\ }\href
  {\doibase 10.1038/nature08919} {\bibfield  {journal} {\bibinfo  {journal}
  {Nature}\ }\textbf {\bibinfo {volume} {464}},\ \bibinfo {pages} {1165–1169}
  (\bibinfo {year} {2010})}\BibitemShut {NoStop}%
\bibitem [{\citenamefont {Poli}\ \emph {et~al.}(2011)\citenamefont {Poli},
  \citenamefont {Wang}, \citenamefont {Tarallo}, \citenamefont {Alberti},
  \citenamefont {Prevedelli},\ and\ \citenamefont {Tino}}]{Poli2011}%
  \BibitemOpen
  \bibfield  {author} {\bibinfo {author} {\bibfnamefont {N.}~\bibnamefont
  {Poli}}, \bibinfo {author} {\bibfnamefont {F.-Y.}\ \bibnamefont {Wang}},
  \bibinfo {author} {\bibfnamefont {M.~G.}\ \bibnamefont {Tarallo}}, \bibinfo
  {author} {\bibfnamefont {A.}~\bibnamefont {Alberti}}, \bibinfo {author}
  {\bibfnamefont {M.}~\bibnamefont {Prevedelli}}, \ and\ \bibinfo {author}
  {\bibfnamefont {G.~M.}\ \bibnamefont {Tino}},\ }\href {\doibase
  10.1103/PhysRevLett.106.038501} {\bibfield  {journal} {\bibinfo  {journal}
  {Phys. Rev. Lett.}\ }\textbf {\bibinfo {volume} {106}},\ \bibinfo {pages}
  {038501} (\bibinfo {year} {2011})}\BibitemShut {NoStop}%
\bibitem [{\citenamefont {Paris}(2009)}]{paris2009quantum}%
  \BibitemOpen
  \bibfield  {author} {\bibinfo {author} {\bibfnamefont {M.~G.}\ \bibnamefont
  {Paris}},\ }\href {\doibase 10.1142/S0219749909004839} {\bibfield  {journal}
  {\bibinfo  {journal} {International Journal of Quantum Information}\ }\textbf
  {\bibinfo {volume} {7}},\ \bibinfo {pages} {125} (\bibinfo {year}
  {2009})}\BibitemShut {NoStop}%
\bibitem [{\citenamefont {Alushi}\ \emph {et~al.}(2025)\citenamefont {Alushi},
  \citenamefont {Coppo}, \citenamefont {Brosco}, \citenamefont {\{Di
  Candia\}},\ and\ \citenamefont {Felicetti}}]{Uesli2025}%
  \BibitemOpen
  \bibfield  {author} {\bibinfo {author} {\bibfnamefont {U.}~\bibnamefont
  {Alushi}}, \bibinfo {author} {\bibfnamefont {A.}~\bibnamefont {Coppo}},
  \bibinfo {author} {\bibfnamefont {V.}~\bibnamefont {Brosco}}, \bibinfo
  {author} {\bibfnamefont {R.}~\bibnamefont {\{Di Candia\}}}, \ and\ \bibinfo
  {author} {\bibfnamefont {S.}~\bibnamefont {Felicetti}},\ }\href {\doibase
  10.1038/s42005-025-01975-9} {\bibfield  {journal} {\bibinfo  {journal}
  {Communications Physics}\ }\textbf {\bibinfo {volume} {8}} (\bibinfo {year}
  {2025}),\ 10.1038/s42005-025-01975-9},\ \bibinfo {note} {publisher Copyright:
  {\textcopyright} The Author(s) 2025.}\BibitemShut {Stop}%
\bibitem [{\citenamefont {Mihailescu}\ \emph {et~al.}(2025)\citenamefont
  {Mihailescu}, \citenamefont {Alushi}, \citenamefont {Candia}, \citenamefont
  {Felicetti},\ and\ \citenamefont {Gietka}}]{Mihailescu2025}%
  \BibitemOpen
  \bibfield  {author} {\bibinfo {author} {\bibfnamefont {G.}~\bibnamefont
  {Mihailescu}}, \bibinfo {author} {\bibfnamefont {U.}~\bibnamefont {Alushi}},
  \bibinfo {author} {\bibfnamefont {R.~D.}\ \bibnamefont {Candia}}, \bibinfo
  {author} {\bibfnamefont {S.}~\bibnamefont {Felicetti}}, \ and\ \bibinfo
  {author} {\bibfnamefont {K.}~\bibnamefont {Gietka}},\ }\href
  {https://arxiv.org/abs/2510.02035} {\  (\bibinfo {year} {2025})},\ \Eprint
  {http://arxiv.org/abs/2510.02035} {arXiv:2510.02035 [quant-ph]} \BibitemShut
  {NoStop}%
\bibitem [{\citenamefont {Braun}\ \emph {et~al.}(2018)\citenamefont {Braun},
  \citenamefont {Adesso}, \citenamefont {Benatti}, \citenamefont {Floreanini},
  \citenamefont {Marzolino}, \citenamefont {Mitchell},\ and\ \citenamefont
  {Pirandola}}]{Braun_2018}%
  \BibitemOpen
  \bibfield  {author} {\bibinfo {author} {\bibfnamefont {D.}~\bibnamefont
  {Braun}}, \bibinfo {author} {\bibfnamefont {G.}~\bibnamefont {Adesso}},
  \bibinfo {author} {\bibfnamefont {F.}~\bibnamefont {Benatti}}, \bibinfo
  {author} {\bibfnamefont {R.}~\bibnamefont {Floreanini}}, \bibinfo {author}
  {\bibfnamefont {U.}~\bibnamefont {Marzolino}}, \bibinfo {author}
  {\bibfnamefont {M.~W.}\ \bibnamefont {Mitchell}}, \ and\ \bibinfo {author}
  {\bibfnamefont {S.}~\bibnamefont {Pirandola}},\ }\href {\doibase
  10.1103/revmodphys.90.035006} {\bibfield  {journal} {\bibinfo  {journal}
  {Reviews of Modern Physics}\ }\textbf {\bibinfo {volume} {90}} (\bibinfo
  {year} {2018}),\ 10.1103/revmodphys.90.035006}\BibitemShut {NoStop}%
\bibitem [{\citenamefont {RouhbakhshNabati}\ \emph {et~al.}(2025)\citenamefont
  {RouhbakhshNabati}, \citenamefont {Braun},\ and\ \citenamefont
  {Schomerus}}]{RouhbakhshNabati2025}%
  \BibitemOpen
  \bibfield  {author} {\bibinfo {author} {\bibfnamefont {M.}~\bibnamefont
  {RouhbakhshNabati}}, \bibinfo {author} {\bibfnamefont {D.}~\bibnamefont
  {Braun}}, \ and\ \bibinfo {author} {\bibfnamefont {H.}~\bibnamefont
  {Schomerus}},\ }\href {\doibase 10.1103/h4sg-mq1m} {\bibfield  {journal}
  {\bibinfo  {journal} {Phys. Rev. Lett.}\ }\textbf {\bibinfo {volume} {135}},\
  \bibinfo {pages} {190202} (\bibinfo {year} {2025})}\BibitemShut {NoStop}%
\bibitem [{\citenamefont {Ghosh}\ \emph {et~al.}(2025)\citenamefont {Ghosh},
  \citenamefont {Kobus}, \citenamefont {Kurdzialek},\ and\ \citenamefont
  {Demkowicz-Dobrza{\'n}ski}}]{ghosh2025}%
  \BibitemOpen
  \bibfield  {author} {\bibinfo {author} {\bibfnamefont {S.}~\bibnamefont
  {Ghosh}}, \bibinfo {author} {\bibfnamefont {A.}~\bibnamefont {Kobus}},
  \bibinfo {author} {\bibfnamefont {S.}~\bibnamefont {Kurdzialek}}, \ and\
  \bibinfo {author} {\bibfnamefont {R.}~\bibnamefont
  {Demkowicz-Dobrza{\'n}ski}},\ }\href {https://arxiv.org/abs/2511.07211} {\
  (\bibinfo {year} {2025})},\ \Eprint {http://arxiv.org/abs/2511.07211}
  {arXiv:2511.07211 [quant-ph]} \BibitemShut {NoStop}%
\bibitem [{\citenamefont {McGuirk}\ \emph {et~al.}(2002)\citenamefont
  {McGuirk}, \citenamefont {Foster}, \citenamefont {Fixler}, \citenamefont
  {Snadden},\ and\ \citenamefont {Kasevich}}]{McGuirk2002}%
  \BibitemOpen
  \bibfield  {author} {\bibinfo {author} {\bibfnamefont {J.~M.}\ \bibnamefont
  {McGuirk}}, \bibinfo {author} {\bibfnamefont {G.~T.}\ \bibnamefont {Foster}},
  \bibinfo {author} {\bibfnamefont {J.~B.}\ \bibnamefont {Fixler}}, \bibinfo
  {author} {\bibfnamefont {M.~J.}\ \bibnamefont {Snadden}}, \ and\ \bibinfo
  {author} {\bibfnamefont {M.~A.}\ \bibnamefont {Kasevich}},\ }\href {\doibase
  10.1103/PhysRevA.65.033608} {\bibfield  {journal} {\bibinfo  {journal} {Phys.
  Rev. A}\ }\textbf {\bibinfo {volume} {65}},\ \bibinfo {pages} {033608}
  (\bibinfo {year} {2002})}\BibitemShut {NoStop}%
\bibitem [{\citenamefont {Fixler}\ \emph {et~al.}(2007)\citenamefont {Fixler},
  \citenamefont {Foster}, \citenamefont {McGuirk},\ and\ \citenamefont
  {Kasevich}}]{Fixler2007}%
  \BibitemOpen
  \bibfield  {author} {\bibinfo {author} {\bibfnamefont {J.~B.}\ \bibnamefont
  {Fixler}}, \bibinfo {author} {\bibfnamefont {G.~T.}\ \bibnamefont {Foster}},
  \bibinfo {author} {\bibfnamefont {J.~M.}\ \bibnamefont {McGuirk}}, \ and\
  \bibinfo {author} {\bibfnamefont {M.~A.}\ \bibnamefont {Kasevich}},\ }\href
  {\doibase 10.1126/science.1135459} {\bibfield  {journal} {\bibinfo  {journal}
  {Science}\ }\textbf {\bibinfo {volume} {315}},\ \bibinfo {pages} {74}
  (\bibinfo {year} {2007})},\ \Eprint
  {http://arxiv.org/abs/https://www.science.org/doi/pdf/10.1126/science.1135459}
  {https://www.science.org/doi/pdf/10.1126/science.1135459} \BibitemShut
  {NoStop}%
\bibitem [{\citenamefont {Fattori}\ \emph {et~al.}(2003)\citenamefont
  {Fattori}, \citenamefont {Lamporesi}, \citenamefont {Petelski}, \citenamefont
  {Stuhler},\ and\ \citenamefont {Tino}}]{Fattori2003}%
  \BibitemOpen
  \bibfield  {author} {\bibinfo {author} {\bibfnamefont {M.}~\bibnamefont
  {Fattori}}, \bibinfo {author} {\bibfnamefont {G.}~\bibnamefont {Lamporesi}},
  \bibinfo {author} {\bibfnamefont {T.}~\bibnamefont {Petelski}}, \bibinfo
  {author} {\bibfnamefont {J.}~\bibnamefont {Stuhler}}, \ and\ \bibinfo
  {author} {\bibfnamefont {G.}~\bibnamefont {Tino}},\ }\href {\doibase
  https://doi.org/10.1016/j.physleta.2003.07.011} {\bibfield  {journal}
  {\bibinfo  {journal} {Physics Letters A}\ }\textbf {\bibinfo {volume}
  {318}},\ \bibinfo {pages} {184} (\bibinfo {year} {2003})}\BibitemShut
  {NoStop}%
\bibitem [{\citenamefont {Szigeti}\ \emph {et~al.}(2020)\citenamefont
  {Szigeti}, \citenamefont {Nolan}, \citenamefont {Close},\ and\ \citenamefont
  {Haine}}]{Szigeti2020}%
  \BibitemOpen
  \bibfield  {author} {\bibinfo {author} {\bibfnamefont {S.~S.}\ \bibnamefont
  {Szigeti}}, \bibinfo {author} {\bibfnamefont {S.~P.}\ \bibnamefont {Nolan}},
  \bibinfo {author} {\bibfnamefont {J.~D.}\ \bibnamefont {Close}}, \ and\
  \bibinfo {author} {\bibfnamefont {S.~A.}\ \bibnamefont {Haine}},\ }\href
  {\doibase 10.1103/PhysRevLett.125.100402} {\bibfield  {journal} {\bibinfo
  {journal} {Phys. Rev. Lett.}\ }\textbf {\bibinfo {volume} {125}},\ \bibinfo
  {pages} {100402} (\bibinfo {year} {2020})}\BibitemShut {NoStop}%
\bibitem [{\citenamefont {de~Angelis}\ \emph {et~al.}(2008)\citenamefont
  {de~Angelis}, \citenamefont {Bertoldi}, \citenamefont {Cacciapuoti},
  \citenamefont {Giorgini}, \citenamefont {Lamporesi}, \citenamefont
  {Prevedelli}, \citenamefont {Saccorotti}, \citenamefont {Sorrentino},\ and\
  \citenamefont {Tino}}]{deAngelis_2009}%
  \BibitemOpen
  \bibfield  {author} {\bibinfo {author} {\bibfnamefont {M.}~\bibnamefont
  {de~Angelis}}, \bibinfo {author} {\bibfnamefont {A.}~\bibnamefont
  {Bertoldi}}, \bibinfo {author} {\bibfnamefont {L.}~\bibnamefont
  {Cacciapuoti}}, \bibinfo {author} {\bibfnamefont {A.}~\bibnamefont
  {Giorgini}}, \bibinfo {author} {\bibfnamefont {G.}~\bibnamefont {Lamporesi}},
  \bibinfo {author} {\bibfnamefont {M.}~\bibnamefont {Prevedelli}}, \bibinfo
  {author} {\bibfnamefont {G.}~\bibnamefont {Saccorotti}}, \bibinfo {author}
  {\bibfnamefont {F.}~\bibnamefont {Sorrentino}}, \ and\ \bibinfo {author}
  {\bibfnamefont {G.~M.}\ \bibnamefont {Tino}},\ }\href {\doibase
  10.1088/0957-0233/20/2/022001} {\bibfield  {journal} {\bibinfo  {journal}
  {Measurement Science and Technology}\ }\textbf {\bibinfo {volume} {20}},\
  \bibinfo {pages} {022001} (\bibinfo {year} {2008})}\BibitemShut {NoStop}%
\bibitem [{\citenamefont {Cladé}\ \emph {et~al.}(2005)\citenamefont {Cladé},
  \citenamefont {Guellati-Khélifa}, \citenamefont {Schwob}, \citenamefont
  {Nez}, \citenamefont {Julien},\ and\ \citenamefont {Biraben}}]{Clade_2005}%
  \BibitemOpen
  \bibfield  {author} {\bibinfo {author} {\bibfnamefont {P.}~\bibnamefont
  {Cladé}}, \bibinfo {author} {\bibfnamefont {S.}~\bibnamefont
  {Guellati-Khélifa}}, \bibinfo {author} {\bibfnamefont {C.}~\bibnamefont
  {Schwob}}, \bibinfo {author} {\bibfnamefont {F.}~\bibnamefont {Nez}},
  \bibinfo {author} {\bibfnamefont {L.}~\bibnamefont {Julien}}, \ and\ \bibinfo
  {author} {\bibfnamefont {F.}~\bibnamefont {Biraben}},\ }\href {\doibase
  10.1209/epl/i2005-10163-6} {\bibfield  {journal} {\bibinfo  {journal}
  {Europhysics Letters}\ }\textbf {\bibinfo {volume} {71}},\ \bibinfo {pages}
  {730} (\bibinfo {year} {2005})}\BibitemShut {NoStop}%
\bibitem [{\citenamefont {Ferrari}\ \emph {et~al.}(2006)\citenamefont
  {Ferrari}, \citenamefont {Poli}, \citenamefont {Sorrentino},\ and\
  \citenamefont {Tino}}]{ferrari2006}%
  \BibitemOpen
  \bibfield  {author} {\bibinfo {author} {\bibfnamefont {G.}~\bibnamefont
  {Ferrari}}, \bibinfo {author} {\bibfnamefont {N.}~\bibnamefont {Poli}},
  \bibinfo {author} {\bibfnamefont {F.}~\bibnamefont {Sorrentino}}, \ and\
  \bibinfo {author} {\bibfnamefont {G.~M.}\ \bibnamefont {Tino}},\ }\href
  {\doibase 10.1103/PhysRevLett.97.060402} {\bibfield  {journal} {\bibinfo
  {journal} {Phys. Rev. Lett.}\ }\textbf {\bibinfo {volume} {97}},\ \bibinfo
  {pages} {060402} (\bibinfo {year} {2006})}\BibitemShut {NoStop}%
\bibitem [{\citenamefont {Tino}\ \emph {et~al.}(2019)\citenamefont {Tino},
  \citenamefont {Bassi}, \citenamefont {Bianco}, \citenamefont {Bongs},
  \citenamefont {Bouyer}, \citenamefont {Cacciapuoti}, \citenamefont
  {Capozziello}, \citenamefont {Chen}, \citenamefont {Chiofalo}, \citenamefont
  {Derevianko}, \citenamefont {Ertmer}, \citenamefont {Gaaloul}, \citenamefont
  {Gill}, \citenamefont {Graham}, \citenamefont {Hogan}, \citenamefont {Iess},
  \citenamefont {Kasevich}, \citenamefont {Katori}, \citenamefont {Klempt},
  \citenamefont {Lu}, \citenamefont {Ma}, \citenamefont {Müller},
  \citenamefont {Newbury}, \citenamefont {Oates}, \citenamefont {Peters},
  \citenamefont {Poli}, \citenamefont {Rasel}, \citenamefont {Rosi},
  \citenamefont {Roura}, \citenamefont {Salomon}, \citenamefont {Schiller},
  \citenamefont {Schleich}, \citenamefont {Schlippert}, \citenamefont
  {Schreck}, \citenamefont {Schubert}, \citenamefont {Sorrentino},
  \citenamefont {Sterr}, \citenamefont {Thomsen}, \citenamefont {Vallone},
  \citenamefont {Vetrano}, \citenamefont {Villoresi}, \citenamefont {von
  Klitzing}, \citenamefont {Wilkowski}, \citenamefont {Wolf}, \citenamefont
  {Ye}, \citenamefont {Yu},\ and\ \citenamefont {Zhan}}]{Tino_2019}%
  \BibitemOpen
  \bibfield  {author} {\bibinfo {author} {\bibfnamefont {G.~M.}\ \bibnamefont
  {Tino}}, \bibinfo {author} {\bibfnamefont {A.}~\bibnamefont {Bassi}},
  \bibinfo {author} {\bibfnamefont {G.}~\bibnamefont {Bianco}}, \bibinfo
  {author} {\bibfnamefont {K.}~\bibnamefont {Bongs}}, \bibinfo {author}
  {\bibfnamefont {P.}~\bibnamefont {Bouyer}}, \bibinfo {author} {\bibfnamefont
  {L.}~\bibnamefont {Cacciapuoti}}, \bibinfo {author} {\bibfnamefont
  {S.}~\bibnamefont {Capozziello}}, \bibinfo {author} {\bibfnamefont
  {X.}~\bibnamefont {Chen}}, \bibinfo {author} {\bibfnamefont {M.~L.}\
  \bibnamefont {Chiofalo}}, \bibinfo {author} {\bibfnamefont {A.}~\bibnamefont
  {Derevianko}}, \bibinfo {author} {\bibfnamefont {W.}~\bibnamefont {Ertmer}},
  \bibinfo {author} {\bibfnamefont {N.}~\bibnamefont {Gaaloul}}, \bibinfo
  {author} {\bibfnamefont {P.}~\bibnamefont {Gill}}, \bibinfo {author}
  {\bibfnamefont {P.~W.}\ \bibnamefont {Graham}}, \bibinfo {author}
  {\bibfnamefont {J.~M.}\ \bibnamefont {Hogan}}, \bibinfo {author}
  {\bibfnamefont {L.}~\bibnamefont {Iess}}, \bibinfo {author} {\bibfnamefont
  {M.~A.}\ \bibnamefont {Kasevich}}, \bibinfo {author} {\bibfnamefont
  {H.}~\bibnamefont {Katori}}, \bibinfo {author} {\bibfnamefont
  {C.}~\bibnamefont {Klempt}}, \bibinfo {author} {\bibfnamefont
  {X.}~\bibnamefont {Lu}}, \bibinfo {author} {\bibfnamefont {L.-S.}\
  \bibnamefont {Ma}}, \bibinfo {author} {\bibfnamefont {H.}~\bibnamefont
  {Müller}}, \bibinfo {author} {\bibfnamefont {N.~R.}\ \bibnamefont
  {Newbury}}, \bibinfo {author} {\bibfnamefont {C.~W.}\ \bibnamefont {Oates}},
  \bibinfo {author} {\bibfnamefont {A.}~\bibnamefont {Peters}}, \bibinfo
  {author} {\bibfnamefont {N.}~\bibnamefont {Poli}}, \bibinfo {author}
  {\bibfnamefont {E.~M.}\ \bibnamefont {Rasel}}, \bibinfo {author}
  {\bibfnamefont {G.}~\bibnamefont {Rosi}}, \bibinfo {author} {\bibfnamefont
  {A.}~\bibnamefont {Roura}}, \bibinfo {author} {\bibfnamefont
  {C.}~\bibnamefont {Salomon}}, \bibinfo {author} {\bibfnamefont
  {S.}~\bibnamefont {Schiller}}, \bibinfo {author} {\bibfnamefont
  {W.}~\bibnamefont {Schleich}}, \bibinfo {author} {\bibfnamefont
  {D.}~\bibnamefont {Schlippert}}, \bibinfo {author} {\bibfnamefont
  {F.}~\bibnamefont {Schreck}}, \bibinfo {author} {\bibfnamefont
  {C.}~\bibnamefont {Schubert}}, \bibinfo {author} {\bibfnamefont
  {F.}~\bibnamefont {Sorrentino}}, \bibinfo {author} {\bibfnamefont
  {U.}~\bibnamefont {Sterr}}, \bibinfo {author} {\bibfnamefont {J.~W.}\
  \bibnamefont {Thomsen}}, \bibinfo {author} {\bibfnamefont {G.}~\bibnamefont
  {Vallone}}, \bibinfo {author} {\bibfnamefont {F.}~\bibnamefont {Vetrano}},
  \bibinfo {author} {\bibfnamefont {P.}~\bibnamefont {Villoresi}}, \bibinfo
  {author} {\bibfnamefont {W.}~\bibnamefont {von Klitzing}}, \bibinfo {author}
  {\bibfnamefont {D.}~\bibnamefont {Wilkowski}}, \bibinfo {author}
  {\bibfnamefont {P.}~\bibnamefont {Wolf}}, \bibinfo {author} {\bibfnamefont
  {J.}~\bibnamefont {Ye}}, \bibinfo {author} {\bibfnamefont {N.}~\bibnamefont
  {Yu}}, \ and\ \bibinfo {author} {\bibfnamefont {M.}~\bibnamefont {Zhan}},\
  }\href {\doibase 10.1140/epjd/e2019-100324-6} {\bibfield  {journal} {\bibinfo
   {journal} {The European Physical Journal D}\ }\textbf {\bibinfo {volume}
  {73}} (\bibinfo {year} {2019}),\ 10.1140/epjd/e2019-100324-6}\BibitemShut
  {NoStop}%
\bibitem [{\citenamefont {Canuel}\ \emph {et~al.}(2020)\citenamefont {Canuel},
  \citenamefont {Abend}, \citenamefont {Amaro-Seoane}, \citenamefont
  {Badaracco}, \citenamefont {Beaufils}, \citenamefont {Bertoldi},
  \citenamefont {Bongs}, \citenamefont {Bouyer}, \citenamefont {Braxmaier},
  \citenamefont {Chaibi}, \citenamefont {Christensen}, \citenamefont {Fitzek},
  \citenamefont {Flouris}, \citenamefont {Gaaloul}, \citenamefont {Gaffet},
  \citenamefont {Garrido~Alzar}, \citenamefont {Geiger}, \citenamefont
  {Guellati-Khelifa}, \citenamefont {Hammerer}, \citenamefont {Harms},
  \citenamefont {Hinderer}, \citenamefont {Holynski}, \citenamefont {Junca},
  \citenamefont {Katsanevas}, \citenamefont {Klempt}, \citenamefont
  {Kozanitis}, \citenamefont {Krutzik}, \citenamefont {Landragin},
  \citenamefont {Làzaro~Roche}, \citenamefont {Leykauf}, \citenamefont {Lien},
  \citenamefont {Loriani}, \citenamefont {Merlet}, \citenamefont {Merzougui},
  \citenamefont {Nofrarias}, \citenamefont {Papadakos}, \citenamefont
  {Pereira~dos Santos}, \citenamefont {Peters}, \citenamefont {Plexousakis},
  \citenamefont {Prevedelli}, \citenamefont {Rasel}, \citenamefont {Rogister},
  \citenamefont {Rosat}, \citenamefont {Roura}, \citenamefont {Sabulsky},
  \citenamefont {Schkolnik}, \citenamefont {Schlippert}, \citenamefont
  {Schubert}, \citenamefont {Sidorenkov}, \citenamefont {Siemß}, \citenamefont
  {Sopuerta}, \citenamefont {Sorrentino}, \citenamefont {Struckmann},
  \citenamefont {Tino}, \citenamefont {Tsagkatakis}, \citenamefont {Viceré},
  \citenamefont {von Klitzing}, \citenamefont {Woerner},\ and\ \citenamefont
  {Zou}}]{Canuel_2020}%
  \BibitemOpen
  \bibfield  {author} {\bibinfo {author} {\bibfnamefont {B.}~\bibnamefont
  {Canuel}}, \bibinfo {author} {\bibfnamefont {S.}~\bibnamefont {Abend}},
  \bibinfo {author} {\bibfnamefont {P.}~\bibnamefont {Amaro-Seoane}}, \bibinfo
  {author} {\bibfnamefont {F.}~\bibnamefont {Badaracco}}, \bibinfo {author}
  {\bibfnamefont {Q.}~\bibnamefont {Beaufils}}, \bibinfo {author}
  {\bibfnamefont {A.}~\bibnamefont {Bertoldi}}, \bibinfo {author}
  {\bibfnamefont {K.}~\bibnamefont {Bongs}}, \bibinfo {author} {\bibfnamefont
  {P.}~\bibnamefont {Bouyer}}, \bibinfo {author} {\bibfnamefont
  {C.}~\bibnamefont {Braxmaier}}, \bibinfo {author} {\bibfnamefont
  {W.}~\bibnamefont {Chaibi}}, \bibinfo {author} {\bibfnamefont
  {N.}~\bibnamefont {Christensen}}, \bibinfo {author} {\bibfnamefont
  {F.}~\bibnamefont {Fitzek}}, \bibinfo {author} {\bibfnamefont
  {G.}~\bibnamefont {Flouris}}, \bibinfo {author} {\bibfnamefont
  {N.}~\bibnamefont {Gaaloul}}, \bibinfo {author} {\bibfnamefont
  {S.}~\bibnamefont {Gaffet}}, \bibinfo {author} {\bibfnamefont {C.~L.}\
  \bibnamefont {Garrido~Alzar}}, \bibinfo {author} {\bibfnamefont
  {R.}~\bibnamefont {Geiger}}, \bibinfo {author} {\bibfnamefont
  {S.}~\bibnamefont {Guellati-Khelifa}}, \bibinfo {author} {\bibfnamefont
  {K.}~\bibnamefont {Hammerer}}, \bibinfo {author} {\bibfnamefont
  {J.}~\bibnamefont {Harms}}, \bibinfo {author} {\bibfnamefont
  {J.}~\bibnamefont {Hinderer}}, \bibinfo {author} {\bibfnamefont
  {M.}~\bibnamefont {Holynski}}, \bibinfo {author} {\bibfnamefont
  {J.}~\bibnamefont {Junca}}, \bibinfo {author} {\bibfnamefont
  {S.}~\bibnamefont {Katsanevas}}, \bibinfo {author} {\bibfnamefont
  {C.}~\bibnamefont {Klempt}}, \bibinfo {author} {\bibfnamefont
  {C.}~\bibnamefont {Kozanitis}}, \bibinfo {author} {\bibfnamefont
  {M.}~\bibnamefont {Krutzik}}, \bibinfo {author} {\bibfnamefont
  {A.}~\bibnamefont {Landragin}}, \bibinfo {author} {\bibfnamefont
  {I.}~\bibnamefont {Làzaro~Roche}}, \bibinfo {author} {\bibfnamefont
  {B.}~\bibnamefont {Leykauf}}, \bibinfo {author} {\bibfnamefont {Y.-H.}\
  \bibnamefont {Lien}}, \bibinfo {author} {\bibfnamefont {S.}~\bibnamefont
  {Loriani}}, \bibinfo {author} {\bibfnamefont {S.}~\bibnamefont {Merlet}},
  \bibinfo {author} {\bibfnamefont {M.}~\bibnamefont {Merzougui}}, \bibinfo
  {author} {\bibfnamefont {M.}~\bibnamefont {Nofrarias}}, \bibinfo {author}
  {\bibfnamefont {P.}~\bibnamefont {Papadakos}}, \bibinfo {author}
  {\bibfnamefont {F.}~\bibnamefont {Pereira~dos Santos}}, \bibinfo {author}
  {\bibfnamefont {A.}~\bibnamefont {Peters}}, \bibinfo {author} {\bibfnamefont
  {D.}~\bibnamefont {Plexousakis}}, \bibinfo {author} {\bibfnamefont
  {M.}~\bibnamefont {Prevedelli}}, \bibinfo {author} {\bibfnamefont {E.~M.}\
  \bibnamefont {Rasel}}, \bibinfo {author} {\bibfnamefont {Y.}~\bibnamefont
  {Rogister}}, \bibinfo {author} {\bibfnamefont {S.}~\bibnamefont {Rosat}},
  \bibinfo {author} {\bibfnamefont {A.}~\bibnamefont {Roura}}, \bibinfo
  {author} {\bibfnamefont {D.~O.}\ \bibnamefont {Sabulsky}}, \bibinfo {author}
  {\bibfnamefont {V.}~\bibnamefont {Schkolnik}}, \bibinfo {author}
  {\bibfnamefont {D.}~\bibnamefont {Schlippert}}, \bibinfo {author}
  {\bibfnamefont {C.}~\bibnamefont {Schubert}}, \bibinfo {author}
  {\bibfnamefont {L.}~\bibnamefont {Sidorenkov}}, \bibinfo {author}
  {\bibfnamefont {J.-N.}\ \bibnamefont {Siemß}}, \bibinfo {author}
  {\bibfnamefont {C.~F.}\ \bibnamefont {Sopuerta}}, \bibinfo {author}
  {\bibfnamefont {F.}~\bibnamefont {Sorrentino}}, \bibinfo {author}
  {\bibfnamefont {C.}~\bibnamefont {Struckmann}}, \bibinfo {author}
  {\bibfnamefont {G.~M.}\ \bibnamefont {Tino}}, \bibinfo {author}
  {\bibfnamefont {G.}~\bibnamefont {Tsagkatakis}}, \bibinfo {author}
  {\bibfnamefont {A.}~\bibnamefont {Viceré}}, \bibinfo {author} {\bibfnamefont
  {W.}~\bibnamefont {von Klitzing}}, \bibinfo {author} {\bibfnamefont
  {L.}~\bibnamefont {Woerner}}, \ and\ \bibinfo {author} {\bibfnamefont
  {X.}~\bibnamefont {Zou}},\ }\href {\doibase 10.1088/1361-6382/aba80e}
  {\bibfield  {journal} {\bibinfo  {journal} {Classical and Quantum Gravity}\
  }\textbf {\bibinfo {volume} {37}},\ \bibinfo {pages} {225017} (\bibinfo
  {year} {2020})}\BibitemShut {NoStop}%
\bibitem [{\citenamefont {Wu}\ \emph {et~al.}(2023)\citenamefont {Wu},
  \citenamefont {Toro\ifmmode~\check{s}\else \v{s}\fi{}}, \citenamefont
  {Bose},\ and\ \citenamefont {Mazumdar}}]{wu2023}%
  \BibitemOpen
  \bibfield  {author} {\bibinfo {author} {\bibfnamefont {M.-Z.}\ \bibnamefont
  {Wu}}, \bibinfo {author} {\bibfnamefont {M.}~\bibnamefont
  {Toro\ifmmode~\check{s}\else \v{s}\fi{}}}, \bibinfo {author} {\bibfnamefont
  {S.}~\bibnamefont {Bose}}, \ and\ \bibinfo {author} {\bibfnamefont
  {A.}~\bibnamefont {Mazumdar}},\ }\href {\doibase 10.1103/PhysRevD.107.104053}
  {\bibfield  {journal} {\bibinfo  {journal} {Phys. Rev. D}\ }\textbf {\bibinfo
  {volume} {107}},\ \bibinfo {pages} {104053} (\bibinfo {year}
  {2023})}\BibitemShut {NoStop}%
\bibitem [{\citenamefont {Adams}\ \emph {et~al.}(2021)\citenamefont {Adams},
  \citenamefont {Macrae}, \citenamefont {Entezami}, \citenamefont {Ridley},
  \citenamefont {Kubba}, \citenamefont {Lien}, \citenamefont {Kinge},\ and\
  \citenamefont {Bongs}}]{Adams2021}%
  \BibitemOpen
  \bibfield  {author} {\bibinfo {author} {\bibfnamefont {B.}~\bibnamefont
  {Adams}}, \bibinfo {author} {\bibfnamefont {C.}~\bibnamefont {Macrae}},
  \bibinfo {author} {\bibfnamefont {M.}~\bibnamefont {Entezami}}, \bibinfo
  {author} {\bibfnamefont {K.}~\bibnamefont {Ridley}}, \bibinfo {author}
  {\bibfnamefont {A.}~\bibnamefont {Kubba}}, \bibinfo {author} {\bibfnamefont
  {Y.-H.}\ \bibnamefont {Lien}}, \bibinfo {author} {\bibfnamefont
  {S.}~\bibnamefont {Kinge}}, \ and\ \bibinfo {author} {\bibfnamefont
  {K.}~\bibnamefont {Bongs}},\ }\href {\doibase
  10.1109/INERTIAL51137.2021.9430461} {\ ,\ \bibinfo {pages} {1} (\bibinfo
  {year} {2021})}\BibitemShut {NoStop}%
\bibitem [{\citenamefont {Lellouch}\ \emph {et~al.}(2022)\citenamefont
  {Lellouch}, \citenamefont {Bongs},\ and\ \citenamefont
  {Holynski}}]{Lellouch2022}%
  \BibitemOpen
  \bibfield  {author} {\bibinfo {author} {\bibfnamefont {S.}~\bibnamefont
  {Lellouch}}, \bibinfo {author} {\bibfnamefont {K.}~\bibnamefont {Bongs}}, \
  and\ \bibinfo {author} {\bibfnamefont {M.}~\bibnamefont {Holynski}},\ }\href
  {\doibase 10.1080/00107514.2023.2180860} {\bibfield  {journal} {\bibinfo
  {journal} {Contemporary Physics}\ }\textbf {\bibinfo {volume} {63}},\
  \bibinfo {pages} {138} (\bibinfo {year} {2022})},\ \Eprint
  {http://arxiv.org/abs/https://doi.org/10.1080/00107514.2023.2180860}
  {https://doi.org/10.1080/00107514.2023.2180860} \BibitemShut {NoStop}%
\bibitem [{\citenamefont {Stray}\ \emph {et~al.}(2022)\citenamefont {Stray},
  \citenamefont {Lamb}, \citenamefont {Kaushik}, \citenamefont {Vovrosh},
  \citenamefont {Rodgers}, \citenamefont {Winch}, \citenamefont {Hayati},
  \citenamefont {Boddice}, \citenamefont {Stabrawa}, \citenamefont {Niggebaum},
  \citenamefont {Langlois}, \citenamefont {Lien}, \citenamefont {Lellouch},
  \citenamefont {Roshanmanesh}, \citenamefont {Ridley}, \citenamefont
  {Villiers}, \citenamefont {Brown}, \citenamefont {Cross}, \citenamefont
  {Tuckwell},\ and\ \citenamefont {Holynski}}]{Stray2022}%
  \BibitemOpen
  \bibfield  {author} {\bibinfo {author} {\bibfnamefont {B.}~\bibnamefont
  {Stray}}, \bibinfo {author} {\bibfnamefont {A.}~\bibnamefont {Lamb}},
  \bibinfo {author} {\bibfnamefont {A.}~\bibnamefont {Kaushik}}, \bibinfo
  {author} {\bibfnamefont {J.}~\bibnamefont {Vovrosh}}, \bibinfo {author}
  {\bibfnamefont {A.}~\bibnamefont {Rodgers}}, \bibinfo {author} {\bibfnamefont
  {J.}~\bibnamefont {Winch}}, \bibinfo {author} {\bibfnamefont
  {F.}~\bibnamefont {Hayati}}, \bibinfo {author} {\bibfnamefont
  {D.}~\bibnamefont {Boddice}}, \bibinfo {author} {\bibfnamefont
  {A.}~\bibnamefont {Stabrawa}}, \bibinfo {author} {\bibfnamefont
  {A.}~\bibnamefont {Niggebaum}}, \bibinfo {author} {\bibfnamefont
  {M.}~\bibnamefont {Langlois}}, \bibinfo {author} {\bibfnamefont {Y.-H.}\
  \bibnamefont {Lien}}, \bibinfo {author} {\bibfnamefont {S.}~\bibnamefont
  {Lellouch}}, \bibinfo {author} {\bibfnamefont {S.}~\bibnamefont
  {Roshanmanesh}}, \bibinfo {author} {\bibfnamefont {K.}~\bibnamefont
  {Ridley}}, \bibinfo {author} {\bibfnamefont {G.}~\bibnamefont {Villiers}},
  \bibinfo {author} {\bibfnamefont {G.}~\bibnamefont {Brown}}, \bibinfo
  {author} {\bibfnamefont {T.}~\bibnamefont {Cross}}, \bibinfo {author}
  {\bibfnamefont {G.}~\bibnamefont {Tuckwell}}, \ and\ \bibinfo {author}
  {\bibfnamefont {M.}~\bibnamefont {Holynski}},\ }\href {\doibase
  10.1038/s41586-021-04315-3} {\bibfield  {journal} {\bibinfo  {journal}
  {Nature}\ }\textbf {\bibinfo {volume} {602}},\ \bibinfo {pages} {590}
  (\bibinfo {year} {2022})}\BibitemShut {NoStop}%
\bibitem [{\citenamefont {Qvarfort}\ \emph {et~al.}(2018)\citenamefont
  {Qvarfort}, \citenamefont {Serafini}, \citenamefont {Barker},\ and\
  \citenamefont {Bose}}]{Qvarfort_2018}%
  \BibitemOpen
  \bibfield  {author} {\bibinfo {author} {\bibfnamefont {S.}~\bibnamefont
  {Qvarfort}}, \bibinfo {author} {\bibfnamefont {A.}~\bibnamefont {Serafini}},
  \bibinfo {author} {\bibfnamefont {P.~F.}\ \bibnamefont {Barker}}, \ and\
  \bibinfo {author} {\bibfnamefont {S.}~\bibnamefont {Bose}},\ }\href {\doibase
  10.1038/s41467-018-06037-z} {\bibfield  {journal} {\bibinfo  {journal}
  {Nature Communications}\ }\textbf {\bibinfo {volume} {9}} (\bibinfo {year}
  {2018}),\ 10.1038/s41467-018-06037-z}\BibitemShut {NoStop}%
\bibitem [{\citenamefont {Abend}\ \emph {et~al.}(2016)\citenamefont {Abend},
  \citenamefont {Gebbe}, \citenamefont {Gersemann}, \citenamefont {Ahlers},
  \citenamefont {M\"untinga}, \citenamefont {Giese}, \citenamefont {Gaaloul},
  \citenamefont {Schubert}, \citenamefont {L\"ammerzahl}, \citenamefont
  {Ertmer}, \citenamefont {Schleich},\ and\ \citenamefont {Rasel}}]{Abend2016}%
  \BibitemOpen
  \bibfield  {author} {\bibinfo {author} {\bibfnamefont {S.}~\bibnamefont
  {Abend}}, \bibinfo {author} {\bibfnamefont {M.}~\bibnamefont {Gebbe}},
  \bibinfo {author} {\bibfnamefont {M.}~\bibnamefont {Gersemann}}, \bibinfo
  {author} {\bibfnamefont {H.}~\bibnamefont {Ahlers}}, \bibinfo {author}
  {\bibfnamefont {H.}~\bibnamefont {M\"untinga}}, \bibinfo {author}
  {\bibfnamefont {E.}~\bibnamefont {Giese}}, \bibinfo {author} {\bibfnamefont
  {N.}~\bibnamefont {Gaaloul}}, \bibinfo {author} {\bibfnamefont
  {C.}~\bibnamefont {Schubert}}, \bibinfo {author} {\bibfnamefont
  {C.}~\bibnamefont {L\"ammerzahl}}, \bibinfo {author} {\bibfnamefont
  {W.}~\bibnamefont {Ertmer}}, \bibinfo {author} {\bibfnamefont {W.~P.}\
  \bibnamefont {Schleich}}, \ and\ \bibinfo {author} {\bibfnamefont {E.~M.}\
  \bibnamefont {Rasel}},\ }\href {\doibase 10.1103/PhysRevLett.117.203003}
  {\bibfield  {journal} {\bibinfo  {journal} {Phys. Rev. Lett.}\ }\textbf
  {\bibinfo {volume} {117}},\ \bibinfo {pages} {203003} (\bibinfo {year}
  {2016})}\BibitemShut {NoStop}%
\bibitem [{\citenamefont {Peters}\ \emph {et~al.}(1999)\citenamefont {Peters},
  \citenamefont {Achim}, \citenamefont {Chung}, \citenamefont {Yeow},
  \citenamefont {Chu},\ and\ \citenamefont {Steven}}]{peters1999}%
  \BibitemOpen
  \bibfield  {author} {\bibinfo {author} {\bibnamefont {Peters}}, \bibinfo
  {author} {\bibnamefont {Achim}}, \bibinfo {author} {\bibnamefont {Chung}},
  \bibinfo {author} {\bibfnamefont {K.}~\bibnamefont {Yeow}}, \bibinfo {author}
  {\bibfnamefont {S.}~\bibnamefont {Chu}}, \ and\ \bibinfo {author}
  {\bibnamefont {Steven}},\ }\href {\doibase 10.1038/23655} {\bibfield
  {journal} {\bibinfo  {journal} {Nature}\ }\textbf {\bibinfo {volume} {400}},\
  \bibinfo {pages} {849} (\bibinfo {year} {1999})}\BibitemShut {NoStop}%
\bibitem [{\citenamefont {Merlet}\ \emph {et~al.}(2008)\citenamefont {Merlet},
  \citenamefont {Kopaev}, \citenamefont {Diament}, \citenamefont {Geneves},
  \citenamefont {Landragin},\ and\ \citenamefont {Pereira
  Dos~Santos}}]{Merlet_2008}%
  \BibitemOpen
  \bibfield  {author} {\bibinfo {author} {\bibfnamefont {S.}~\bibnamefont
  {Merlet}}, \bibinfo {author} {\bibfnamefont {A.}~\bibnamefont {Kopaev}},
  \bibinfo {author} {\bibfnamefont {M.}~\bibnamefont {Diament}}, \bibinfo
  {author} {\bibfnamefont {G.}~\bibnamefont {Geneves}}, \bibinfo {author}
  {\bibfnamefont {A.}~\bibnamefont {Landragin}}, \ and\ \bibinfo {author}
  {\bibfnamefont {F.}~\bibnamefont {Pereira Dos~Santos}},\ }\href {\doibase
  10.1088/0026-1394/45/3/002} {\bibfield  {journal} {\bibinfo  {journal}
  {Metrologia}\ }\textbf {\bibinfo {volume} {45}},\ \bibinfo {pages} {265}
  (\bibinfo {year} {2008})}\BibitemShut {NoStop}%
\bibitem [{\citenamefont {Bose}\ \emph {et~al.}(2017)\citenamefont {Bose},
  \citenamefont {Mazumdar}, \citenamefont {Morley}, \citenamefont {Ulbricht},
  \citenamefont {Toro\ifmmode~\check{s}\else \v{s}\fi{}}, \citenamefont
  {Paternostro}, \citenamefont {Geraci}, \citenamefont {Barker}, \citenamefont
  {Kim},\ and\ \citenamefont {Milburn}}]{bose2017}%
  \BibitemOpen
  \bibfield  {author} {\bibinfo {author} {\bibfnamefont {S.}~\bibnamefont
  {Bose}}, \bibinfo {author} {\bibfnamefont {A.}~\bibnamefont {Mazumdar}},
  \bibinfo {author} {\bibfnamefont {G.~W.}\ \bibnamefont {Morley}}, \bibinfo
  {author} {\bibfnamefont {H.}~\bibnamefont {Ulbricht}}, \bibinfo {author}
  {\bibfnamefont {M.}~\bibnamefont {Toro\ifmmode~\check{s}\else \v{s}\fi{}}},
  \bibinfo {author} {\bibfnamefont {M.}~\bibnamefont {Paternostro}}, \bibinfo
  {author} {\bibfnamefont {A.~A.}\ \bibnamefont {Geraci}}, \bibinfo {author}
  {\bibfnamefont {P.~F.}\ \bibnamefont {Barker}}, \bibinfo {author}
  {\bibfnamefont {M.~S.}\ \bibnamefont {Kim}}, \ and\ \bibinfo {author}
  {\bibfnamefont {G.}~\bibnamefont {Milburn}},\ }\href {\doibase
  10.1103/PhysRevLett.119.240401} {\bibfield  {journal} {\bibinfo  {journal}
  {Phys. Rev. Lett.}\ }\textbf {\bibinfo {volume} {119}},\ \bibinfo {pages}
  {240401} (\bibinfo {year} {2017})}\BibitemShut {NoStop}%
\bibitem [{\citenamefont {Berman}(1997)}]{berman1997atom}%
  \BibitemOpen
  \bibfield  {author} {\bibinfo {author} {\bibfnamefont {P.}~\bibnamefont
  {Berman}},\ }\href {https://books.google.com/books?id=D0fDPFAcaMkC} {\emph
  {\bibinfo {title} {Atom Interferometry}}}\ (\bibinfo  {publisher} {Academic
  Press},\ \bibinfo {year} {1997})\BibitemShut {NoStop}%
\bibitem [{\citenamefont {Kasevich}\ and\ \citenamefont
  {Chu}(1991)}]{Kasevich1991}%
  \BibitemOpen
  \bibfield  {author} {\bibinfo {author} {\bibfnamefont {M.}~\bibnamefont
  {Kasevich}}\ and\ \bibinfo {author} {\bibfnamefont {S.}~\bibnamefont {Chu}},\
  }\href {\doibase 10.1103/PhysRevLett.67.181} {\bibfield  {journal} {\bibinfo
  {journal} {Phys. Rev. Lett.}\ }\textbf {\bibinfo {volume} {67}},\ \bibinfo
  {pages} {181} (\bibinfo {year} {1991})}\BibitemShut {NoStop}%
\bibitem [{\citenamefont {Margalit}\ \emph {et~al.}(2021)\citenamefont
  {Margalit}, \citenamefont {Dobkowski}, \citenamefont {Zhou}, \citenamefont
  {Amit}, \citenamefont {Japha}, \citenamefont {Moukouri}, \citenamefont
  {Rohrlich}, \citenamefont {Mazumdar}, \citenamefont {Bose}, \citenamefont
  {Henkel},\ and\ \citenamefont {Folman}}]{Margalit_2021}%
  \BibitemOpen
  \bibfield  {author} {\bibinfo {author} {\bibfnamefont {Y.}~\bibnamefont
  {Margalit}}, \bibinfo {author} {\bibfnamefont {O.}~\bibnamefont {Dobkowski}},
  \bibinfo {author} {\bibfnamefont {Z.}~\bibnamefont {Zhou}}, \bibinfo {author}
  {\bibfnamefont {O.}~\bibnamefont {Amit}}, \bibinfo {author} {\bibfnamefont
  {Y.}~\bibnamefont {Japha}}, \bibinfo {author} {\bibfnamefont
  {S.}~\bibnamefont {Moukouri}}, \bibinfo {author} {\bibfnamefont
  {D.}~\bibnamefont {Rohrlich}}, \bibinfo {author} {\bibfnamefont
  {A.}~\bibnamefont {Mazumdar}}, \bibinfo {author} {\bibfnamefont
  {S.}~\bibnamefont {Bose}}, \bibinfo {author} {\bibfnamefont {C.}~\bibnamefont
  {Henkel}}, \ and\ \bibinfo {author} {\bibfnamefont {R.}~\bibnamefont
  {Folman}},\ }\href {\doibase 10.1126/sciadv.abg2879} {\bibfield  {journal}
  {\bibinfo  {journal} {Science Advances}\ }\textbf {\bibinfo {volume} {7}}
  (\bibinfo {year} {2021}),\ 10.1126/sciadv.abg2879}\BibitemShut {NoStop}%
\bibitem [{\citenamefont {Charri\`ere}\ \emph {et~al.}(2012)\citenamefont
  {Charri\`ere}, \citenamefont {Cadoret}, \citenamefont {Zahzam}, \citenamefont
  {Bidel},\ and\ \citenamefont {Bresson}}]{Charriere2012}%
  \BibitemOpen
  \bibfield  {author} {\bibinfo {author} {\bibfnamefont {R.}~\bibnamefont
  {Charri\`ere}}, \bibinfo {author} {\bibfnamefont {M.}~\bibnamefont
  {Cadoret}}, \bibinfo {author} {\bibfnamefont {N.}~\bibnamefont {Zahzam}},
  \bibinfo {author} {\bibfnamefont {Y.}~\bibnamefont {Bidel}}, \ and\ \bibinfo
  {author} {\bibfnamefont {A.}~\bibnamefont {Bresson}},\ }\href {\doibase
  10.1103/PhysRevA.85.013639} {\bibfield  {journal} {\bibinfo  {journal} {Phys.
  Rev. A}\ }\textbf {\bibinfo {volume} {85}},\ \bibinfo {pages} {013639}
  (\bibinfo {year} {2012})}\BibitemShut {NoStop}%
\bibitem [{\citenamefont {Pino}\ \emph {et~al.}(2018)\citenamefont {Pino},
  \citenamefont {Prat-Camps}, \citenamefont {Sinha}, \citenamefont
  {Venkatesh},\ and\ \citenamefont {Romero-Isart}}]{Pino_2018}%
  \BibitemOpen
  \bibfield  {author} {\bibinfo {author} {\bibfnamefont {H.}~\bibnamefont
  {Pino}}, \bibinfo {author} {\bibfnamefont {J.}~\bibnamefont {Prat-Camps}},
  \bibinfo {author} {\bibfnamefont {K.}~\bibnamefont {Sinha}}, \bibinfo
  {author} {\bibfnamefont {B.~P.}\ \bibnamefont {Venkatesh}}, \ and\ \bibinfo
  {author} {\bibfnamefont {O.}~\bibnamefont {Romero-Isart}},\ }\href {\doibase
  10.1088/2058-9565/aa9d15} {\bibfield  {journal} {\bibinfo  {journal} {Quantum
  Science and Technology}\ }\textbf {\bibinfo {volume} {3}},\ \bibinfo {pages}
  {025001} (\bibinfo {year} {2018})}\BibitemShut {NoStop}%
\bibitem [{\citenamefont {{Riedel}}\ \emph {et~al.}(2010)\citenamefont
  {{Riedel}}, \citenamefont {{B{\"o}hi}}, \citenamefont {{Li}}, \citenamefont
  {{H{\"a}nsch}}, \citenamefont {{Sinatra}},\ and\ \citenamefont
  {{Treutlein}}}]{Riedel2010}%
  \BibitemOpen
  \bibfield  {author} {\bibinfo {author} {\bibfnamefont {M.~F.}\ \bibnamefont
  {{Riedel}}}, \bibinfo {author} {\bibfnamefont {P.}~\bibnamefont
  {{B{\"o}hi}}}, \bibinfo {author} {\bibfnamefont {Y.}~\bibnamefont {{Li}}},
  \bibinfo {author} {\bibfnamefont {T.~W.}\ \bibnamefont {{H{\"a}nsch}}},
  \bibinfo {author} {\bibfnamefont {A.}~\bibnamefont {{Sinatra}}}, \ and\
  \bibinfo {author} {\bibfnamefont {P.}~\bibnamefont {{Treutlein}}},\ }\href
  {\doibase 10.1038/nature08988} {\bibfield  {journal} {\bibinfo  {journal}
  {\nat}\ }\textbf {\bibinfo {volume} {464}},\ \bibinfo {pages} {1170}
  (\bibinfo {year} {2010})},\ \Eprint {http://arxiv.org/abs/1003.1651}
  {arXiv:1003.1651 [quant-ph]} \BibitemShut {NoStop}%
\bibitem [{\citenamefont {Malia}\ \emph {et~al.}(2020)\citenamefont {Malia},
  \citenamefont {Mart\'{\i}nez-Rinc\'on}, \citenamefont {Wu}, \citenamefont
  {Hosten},\ and\ \citenamefont {Kasevich}}]{Malia2020}%
  \BibitemOpen
  \bibfield  {author} {\bibinfo {author} {\bibfnamefont {B.~K.}\ \bibnamefont
  {Malia}}, \bibinfo {author} {\bibfnamefont {J.}~\bibnamefont
  {Mart\'{\i}nez-Rinc\'on}}, \bibinfo {author} {\bibfnamefont {Y.}~\bibnamefont
  {Wu}}, \bibinfo {author} {\bibfnamefont {O.}~\bibnamefont {Hosten}}, \ and\
  \bibinfo {author} {\bibfnamefont {M.~A.}\ \bibnamefont {Kasevich}},\ }\href
  {\doibase 10.1103/PhysRevLett.125.043202} {\bibfield  {journal} {\bibinfo
  {journal} {Phys. Rev. Lett.}\ }\textbf {\bibinfo {volume} {125}},\ \bibinfo
  {pages} {043202} (\bibinfo {year} {2020})}\BibitemShut {NoStop}%
\bibitem [{\citenamefont {Perlin}\ \emph {et~al.}(2020)\citenamefont {Perlin},
  \citenamefont {Qu},\ and\ \citenamefont {Rey}}]{Perlin2020}%
  \BibitemOpen
  \bibfield  {author} {\bibinfo {author} {\bibfnamefont {M.~A.}\ \bibnamefont
  {Perlin}}, \bibinfo {author} {\bibfnamefont {C.}~\bibnamefont {Qu}}, \ and\
  \bibinfo {author} {\bibfnamefont {A.~M.}\ \bibnamefont {Rey}},\ }\href
  {\doibase 10.1103/PhysRevLett.125.223401} {\bibfield  {journal} {\bibinfo
  {journal} {Phys. Rev. Lett.}\ }\textbf {\bibinfo {volume} {125}},\ \bibinfo
  {pages} {223401} (\bibinfo {year} {2020})}\BibitemShut {NoStop}%
\bibitem [{\citenamefont {Gietka}\ and\ \citenamefont
  {Ritsch}(2023)}]{Gietka2023}%
  \BibitemOpen
  \bibfield  {author} {\bibinfo {author} {\bibfnamefont {K.}~\bibnamefont
  {Gietka}}\ and\ \bibinfo {author} {\bibfnamefont {H.}~\bibnamefont
  {Ritsch}},\ }\href {\doibase 10.1103/PhysRevLett.130.090802} {\bibfield
  {journal} {\bibinfo  {journal} {Phys. Rev. Lett.}\ }\textbf {\bibinfo
  {volume} {130}},\ \bibinfo {pages} {090802} (\bibinfo {year}
  {2023})}\BibitemShut {NoStop}%
\bibitem [{\citenamefont {De~Pasquale}\ \emph {et~al.}(2013)\citenamefont
  {De~Pasquale}, \citenamefont {Rossini}, \citenamefont {Facchi},\ and\
  \citenamefont {Giovannetti}}]{Depasquale2013quantum}%
  \BibitemOpen
  \bibfield  {author} {\bibinfo {author} {\bibfnamefont {A.}~\bibnamefont
  {De~Pasquale}}, \bibinfo {author} {\bibfnamefont {D.}~\bibnamefont
  {Rossini}}, \bibinfo {author} {\bibfnamefont {P.}~\bibnamefont {Facchi}}, \
  and\ \bibinfo {author} {\bibfnamefont {V.}~\bibnamefont {Giovannetti}},\
  }\href {\doibase 10.1103/PhysRevA.88.052117} {\bibfield  {journal} {\bibinfo
  {journal} {Phys. Rev. A}\ }\textbf {\bibinfo {volume} {88}},\ \bibinfo
  {pages} {052117} (\bibinfo {year} {2013})}\BibitemShut {NoStop}%
\bibitem [{\citenamefont {Yao}\ \emph {et~al.}(2022)\citenamefont {Yao},
  \citenamefont {Solaro}, \citenamefont {Carrez}, \citenamefont {Clad\'e},\
  and\ \citenamefont {Guellati-Khelifa}}]{Yao2022}%
  \BibitemOpen
  \bibfield  {author} {\bibinfo {author} {\bibfnamefont {Z.}~\bibnamefont
  {Yao}}, \bibinfo {author} {\bibfnamefont {C.}~\bibnamefont {Solaro}},
  \bibinfo {author} {\bibfnamefont {C.}~\bibnamefont {Carrez}}, \bibinfo
  {author} {\bibfnamefont {P.}~\bibnamefont {Clad\'e}}, \ and\ \bibinfo
  {author} {\bibfnamefont {S.}~\bibnamefont {Guellati-Khelifa}},\ }\href
  {\doibase 10.1103/PhysRevA.106.043312} {\bibfield  {journal} {\bibinfo
  {journal} {Phys. Rev. A}\ }\textbf {\bibinfo {volume} {106}},\ \bibinfo
  {pages} {043312} (\bibinfo {year} {2022})}\BibitemShut {NoStop}%
\bibitem [{\citenamefont {Simsarian}\ \emph {et~al.}(2000)\citenamefont
  {Simsarian}, \citenamefont {Denschlag}, \citenamefont {Edwards},
  \citenamefont {Clark}, \citenamefont {Deng}, \citenamefont {Hagley},
  \citenamefont {Helmerson}, \citenamefont {Rolston},\ and\ \citenamefont
  {Phillips}}]{Simsarian2000}%
  \BibitemOpen
  \bibfield  {author} {\bibinfo {author} {\bibfnamefont {J.~E.}\ \bibnamefont
  {Simsarian}}, \bibinfo {author} {\bibfnamefont {J.}~\bibnamefont
  {Denschlag}}, \bibinfo {author} {\bibfnamefont {M.}~\bibnamefont {Edwards}},
  \bibinfo {author} {\bibfnamefont {C.~W.}\ \bibnamefont {Clark}}, \bibinfo
  {author} {\bibfnamefont {L.}~\bibnamefont {Deng}}, \bibinfo {author}
  {\bibfnamefont {E.~W.}\ \bibnamefont {Hagley}}, \bibinfo {author}
  {\bibfnamefont {K.}~\bibnamefont {Helmerson}}, \bibinfo {author}
  {\bibfnamefont {S.~L.}\ \bibnamefont {Rolston}}, \ and\ \bibinfo {author}
  {\bibfnamefont {W.~D.}\ \bibnamefont {Phillips}},\ }\href {\doibase
  10.1103/PhysRevLett.85.2040} {\bibfield  {journal} {\bibinfo  {journal}
  {Phys. Rev. Lett.}\ }\textbf {\bibinfo {volume} {85}},\ \bibinfo {pages}
  {2040} (\bibinfo {year} {2000})}\BibitemShut {NoStop}%
\bibitem [{\citenamefont {Jamison}\ \emph {et~al.}(2011)\citenamefont
  {Jamison}, \citenamefont {Kutz},\ and\ \citenamefont {Gupta}}]{Jamison2011}%
  \BibitemOpen
  \bibfield  {author} {\bibinfo {author} {\bibfnamefont {A.~O.}\ \bibnamefont
  {Jamison}}, \bibinfo {author} {\bibfnamefont {J.~N.}\ \bibnamefont {Kutz}}, \
  and\ \bibinfo {author} {\bibfnamefont {S.}~\bibnamefont {Gupta}},\ }\href
  {\doibase 10.1103/PhysRevA.84.043643} {\bibfield  {journal} {\bibinfo
  {journal} {Phys. Rev. A}\ }\textbf {\bibinfo {volume} {84}},\ \bibinfo
  {pages} {043643} (\bibinfo {year} {2011})}\BibitemShut {NoStop}%
\bibitem [{\citenamefont {Jamison}\ \emph {et~al.}(2014)\citenamefont
  {Jamison}, \citenamefont {Plotkin-Swing},\ and\ \citenamefont
  {Gupta}}]{Jamison2014}%
  \BibitemOpen
  \bibfield  {author} {\bibinfo {author} {\bibfnamefont {A.~O.}\ \bibnamefont
  {Jamison}}, \bibinfo {author} {\bibfnamefont {B.}~\bibnamefont
  {Plotkin-Swing}}, \ and\ \bibinfo {author} {\bibfnamefont {S.}~\bibnamefont
  {Gupta}},\ }\href {\doibase 10.1103/PhysRevA.90.063606} {\bibfield  {journal}
  {\bibinfo  {journal} {Phys. Rev. A}\ }\textbf {\bibinfo {volume} {90}},\
  \bibinfo {pages} {063606} (\bibinfo {year} {2014})}\BibitemShut {NoStop}%
\bibitem [{\citenamefont {Jannin}\ \emph {et~al.}(2015)\citenamefont {Jannin},
  \citenamefont {Clad\'e},\ and\ \citenamefont
  {Guellati-Kh\'elifa}}]{Jannin2015}%
  \BibitemOpen
  \bibfield  {author} {\bibinfo {author} {\bibfnamefont {R.}~\bibnamefont
  {Jannin}}, \bibinfo {author} {\bibfnamefont {P.}~\bibnamefont {Clad\'e}}, \
  and\ \bibinfo {author} {\bibfnamefont {S.}~\bibnamefont
  {Guellati-Kh\'elifa}},\ }\href {\doibase 10.1103/PhysRevA.92.013616}
  {\bibfield  {journal} {\bibinfo  {journal} {Phys. Rev. A}\ }\textbf {\bibinfo
  {volume} {92}},\ \bibinfo {pages} {013616} (\bibinfo {year}
  {2015})}\BibitemShut {NoStop}%
\bibitem [{\citenamefont {Burchianti}\ \emph {et~al.}(2020)\citenamefont
  {Burchianti}, \citenamefont {D'Errico}, \citenamefont {Marconi},
  \citenamefont {Minardi}, \citenamefont {Fort},\ and\ \citenamefont
  {Modugno}}]{Burchianti2020}%
  \BibitemOpen
  \bibfield  {author} {\bibinfo {author} {\bibfnamefont {A.}~\bibnamefont
  {Burchianti}}, \bibinfo {author} {\bibfnamefont {C.}~\bibnamefont
  {D'Errico}}, \bibinfo {author} {\bibfnamefont {L.}~\bibnamefont {Marconi}},
  \bibinfo {author} {\bibfnamefont {F.}~\bibnamefont {Minardi}}, \bibinfo
  {author} {\bibfnamefont {C.}~\bibnamefont {Fort}}, \ and\ \bibinfo {author}
  {\bibfnamefont {M.}~\bibnamefont {Modugno}},\ }\href {\doibase
  10.1103/PhysRevA.102.043314} {\bibfield  {journal} {\bibinfo  {journal}
  {Phys. Rev. A}\ }\textbf {\bibinfo {volume} {102}},\ \bibinfo {pages}
  {043314} (\bibinfo {year} {2020})}\BibitemShut {NoStop}%
\bibitem [{\citenamefont {Bloch}\ and\ \citenamefont
  {Immanuel}(2005)}]{Bloch2005}%
  \BibitemOpen
  \bibfield  {author} {\bibinfo {author} {\bibfnamefont {I.}~\bibnamefont
  {Bloch}}\ and\ \bibinfo {author} {\bibnamefont {Immanuel}},\ }\href {\doibase
  10.1038/nphys138} {\bibfield  {journal} {\bibinfo  {journal} {Nat. Phys.}\
  }\textbf {\bibinfo {volume} {1}},\ \bibinfo {pages} {23} (\bibinfo {year}
  {2005})}\BibitemShut {NoStop}%
\bibitem [{\citenamefont {Bloch}\ \emph {et~al.}(2007)\citenamefont {Bloch},
  \citenamefont {Dalibard},\ and\ \citenamefont {Zwerger}}]{Bloch2007}%
  \BibitemOpen
  \bibfield  {author} {\bibinfo {author} {\bibfnamefont {I.}~\bibnamefont
  {Bloch}}, \bibinfo {author} {\bibfnamefont {J.}~\bibnamefont {Dalibard}}, \
  and\ \bibinfo {author} {\bibfnamefont {W.}~\bibnamefont {Zwerger}},\ }\href
  {\doibase 10.1103/RevModPhys.80.885} {\bibfield  {journal} {\bibinfo
  {journal} {Review of Modern Physics}\ }\textbf {\bibinfo {volume} {80}}
  (\bibinfo {year} {2007}),\ 10.1103/RevModPhys.80.885}\BibitemShut {NoStop}%
\bibitem [{\citenamefont {Greiner}\ \emph {et~al.}(2002)\citenamefont
  {Greiner}, \citenamefont {Mandel}, \citenamefont {Esslinger}, \citenamefont
  {Hänsch},\ and\ \citenamefont {Bloch}}]{Greiner_2002}%
  \BibitemOpen
  \bibfield  {author} {\bibinfo {author} {\bibfnamefont {M.}~\bibnamefont
  {Greiner}}, \bibinfo {author} {\bibfnamefont {O.}~\bibnamefont {Mandel}},
  \bibinfo {author} {\bibfnamefont {T.}~\bibnamefont {Esslinger}}, \bibinfo
  {author} {\bibfnamefont {T.~W.}\ \bibnamefont {Hänsch}}, \ and\ \bibinfo
  {author} {\bibfnamefont {I.}~\bibnamefont {Bloch}},\ }\href {\doibase
  10.1038/415039a} {\bibfield  {journal} {\bibinfo  {journal} {Nature}\
  }\textbf {\bibinfo {volume} {415}},\ \bibinfo {pages} {39–44} (\bibinfo
  {year} {2002})}\BibitemShut {NoStop}%
\bibitem [{\citenamefont {Kinoshita}\ \emph {et~al.}(2004)\citenamefont
  {Kinoshita}, \citenamefont {Wenger},\ and\ \citenamefont
  {Weiss}}]{Kinoshita2004}%
  \BibitemOpen
  \bibfield  {author} {\bibinfo {author} {\bibfnamefont {T.}~\bibnamefont
  {Kinoshita}}, \bibinfo {author} {\bibfnamefont {T.}~\bibnamefont {Wenger}}, \
  and\ \bibinfo {author} {\bibfnamefont {D.~S.}\ \bibnamefont {Weiss}},\ }\href
  {\doibase 10.1126/science.1100700} {\bibfield  {journal} {\bibinfo  {journal}
  {Science}\ }\textbf {\bibinfo {volume} {305}},\ \bibinfo {pages} {1125}
  (\bibinfo {year} {2004})},\ \Eprint
  {http://arxiv.org/abs/https://www.science.org/doi/pdf/10.1126/science.1100700}
  {https://www.science.org/doi/pdf/10.1126/science.1100700} \BibitemShut
  {NoStop}%
\bibitem [{\citenamefont {Fallani}\ \emph {et~al.}(2007)\citenamefont
  {Fallani}, \citenamefont {Lye}, \citenamefont {Guarrera}, \citenamefont
  {Fort},\ and\ \citenamefont {Inguscio}}]{Fallani2007}%
  \BibitemOpen
  \bibfield  {author} {\bibinfo {author} {\bibfnamefont {L.}~\bibnamefont
  {Fallani}}, \bibinfo {author} {\bibfnamefont {J.~E.}\ \bibnamefont {Lye}},
  \bibinfo {author} {\bibfnamefont {V.}~\bibnamefont {Guarrera}}, \bibinfo
  {author} {\bibfnamefont {C.}~\bibnamefont {Fort}}, \ and\ \bibinfo {author}
  {\bibfnamefont {M.}~\bibnamefont {Inguscio}},\ }\href {\doibase
  10.1103/PhysRevLett.98.130404} {\bibfield  {journal} {\bibinfo  {journal}
  {Phys. Rev. Lett.}\ }\textbf {\bibinfo {volume} {98}},\ \bibinfo {pages}
  {130404} (\bibinfo {year} {2007})}\BibitemShut {NoStop}%
\bibitem [{\citenamefont {Chin}\ \emph {et~al.}(2006)\citenamefont {Chin},
  \citenamefont {Miller}, \citenamefont {Liu}, \citenamefont {Stan},
  \citenamefont {Setiawan}, \citenamefont {Sanner}, \citenamefont {Xu},\ and\
  \citenamefont {Ketterle}}]{Chin_2006}%
  \BibitemOpen
  \bibfield  {author} {\bibinfo {author} {\bibfnamefont {J.~K.}\ \bibnamefont
  {Chin}}, \bibinfo {author} {\bibfnamefont {D.~E.}\ \bibnamefont {Miller}},
  \bibinfo {author} {\bibfnamefont {Y.}~\bibnamefont {Liu}}, \bibinfo {author}
  {\bibfnamefont {C.}~\bibnamefont {Stan}}, \bibinfo {author} {\bibfnamefont
  {W.}~\bibnamefont {Setiawan}}, \bibinfo {author} {\bibfnamefont
  {C.}~\bibnamefont {Sanner}}, \bibinfo {author} {\bibfnamefont
  {K.}~\bibnamefont {Xu}}, \ and\ \bibinfo {author} {\bibfnamefont
  {W.}~\bibnamefont {Ketterle}},\ }\href {\doibase 10.1038/nature05224}
  {\bibfield  {journal} {\bibinfo  {journal} {Nature}\ }\textbf {\bibinfo
  {volume} {443}},\ \bibinfo {pages} {961–964} (\bibinfo {year}
  {2006})}\BibitemShut {NoStop}%
\bibitem [{\citenamefont {Mehboudi}\ \emph {et~al.}(2015)\citenamefont
  {Mehboudi}, \citenamefont {Moreno-Cardoner}, \citenamefont {Chiara},\ and\
  \citenamefont {Sanpera}}]{Mehboudi_2015}%
  \BibitemOpen
  \bibfield  {author} {\bibinfo {author} {\bibfnamefont {M.}~\bibnamefont
  {Mehboudi}}, \bibinfo {author} {\bibfnamefont {M.}~\bibnamefont
  {Moreno-Cardoner}}, \bibinfo {author} {\bibfnamefont {G.~D.}\ \bibnamefont
  {Chiara}}, \ and\ \bibinfo {author} {\bibfnamefont {A.}~\bibnamefont
  {Sanpera}},\ }\href {\doibase 10.1088/1367-2630/17/5/055020} {\bibfield
  {journal} {\bibinfo  {journal} {New Journal of Physics}\ }\textbf {\bibinfo
  {volume} {17}},\ \bibinfo {pages} {055020} (\bibinfo {year}
  {2015})}\BibitemShut {NoStop}%
\bibitem [{\citenamefont {M\"uller}\ \emph {et~al.}(2025)\citenamefont
  {M\"uller}, \citenamefont {K\"ose}, \citenamefont {Meixner}, \citenamefont
  {Sch\"affer},\ and\ \citenamefont {Braun}}]{Muller2025}%
  \BibitemOpen
  \bibfield  {author} {\bibinfo {author} {\bibfnamefont {F.}~\bibnamefont
  {M\"uller}}, \bibinfo {author} {\bibfnamefont {E.}~\bibnamefont {K\"ose}},
  \bibinfo {author} {\bibfnamefont {A.~J.}\ \bibnamefont {Meixner}}, \bibinfo
  {author} {\bibfnamefont {E.}~\bibnamefont {Sch\"affer}}, \ and\ \bibinfo
  {author} {\bibfnamefont {D.}~\bibnamefont {Braun}},\ }\href {\doibase
  10.1103/PhysRevA.111.043501} {\bibfield  {journal} {\bibinfo  {journal}
  {Phys. Rev. A}\ }\textbf {\bibinfo {volume} {111}},\ \bibinfo {pages}
  {043501} (\bibinfo {year} {2025})}\BibitemShut {NoStop}%
\bibitem [{\citenamefont {Anderson}\ and\ \citenamefont
  {Kasevich}(1998)}]{Anderson1998}%
  \BibitemOpen
  \bibfield  {author} {\bibinfo {author} {\bibfnamefont {B.~P.}\ \bibnamefont
  {Anderson}}\ and\ \bibinfo {author} {\bibfnamefont {M.~A.}\ \bibnamefont
  {Kasevich}},\ }\href {\doibase 10.1126/science.282.5394.1686} {\bibfield
  {journal} {\bibinfo  {journal} {Science}\ }\textbf {\bibinfo {volume}
  {282}},\ \bibinfo {pages} {1686} (\bibinfo {year} {1998})}\BibitemShut
  {NoStop}%
\bibitem [{\citenamefont {Dimopoulos}\ and\ \citenamefont
  {Geraci}(2003)}]{Dimopoulos2003}%
  \BibitemOpen
  \bibfield  {author} {\bibinfo {author} {\bibfnamefont {S.}~\bibnamefont
  {Dimopoulos}}\ and\ \bibinfo {author} {\bibfnamefont {A.~A.}\ \bibnamefont
  {Geraci}},\ }\href {\doibase 10.1103/PhysRevD.68.124021} {\bibfield
  {journal} {\bibinfo  {journal} {Phys. Rev. D}\ }\textbf {\bibinfo {volume}
  {68}},\ \bibinfo {pages} {124021} (\bibinfo {year} {2003})}\BibitemShut
  {NoStop}%
\bibitem [{\citenamefont {Roati}\ \emph {et~al.}(2004)\citenamefont {Roati},
  \citenamefont {de~Mirandes}, \citenamefont {Ferlaino}, \citenamefont {Ott},
  \citenamefont {Modugno},\ and\ \citenamefont {Inguscio}}]{Roati2004}%
  \BibitemOpen
  \bibfield  {author} {\bibinfo {author} {\bibfnamefont {G.}~\bibnamefont
  {Roati}}, \bibinfo {author} {\bibfnamefont {E.}~\bibnamefont {de~Mirandes}},
  \bibinfo {author} {\bibfnamefont {F.}~\bibnamefont {Ferlaino}}, \bibinfo
  {author} {\bibfnamefont {H.}~\bibnamefont {Ott}}, \bibinfo {author}
  {\bibfnamefont {G.}~\bibnamefont {Modugno}}, \ and\ \bibinfo {author}
  {\bibfnamefont {M.}~\bibnamefont {Inguscio}},\ }\href {\doibase
  10.1103/PhysRevLett.92.230402} {\bibfield  {journal} {\bibinfo  {journal}
  {Phys. Rev. Lett.}\ }\textbf {\bibinfo {volume} {92}},\ \bibinfo {pages}
  {230402} (\bibinfo {year} {2004})}\BibitemShut {NoStop}%
\bibitem [{\citenamefont {Carusotto}\ \emph {et~al.}(2005)\citenamefont
  {Carusotto}, \citenamefont {Pitaevskii}, \citenamefont {Stringari},
  \citenamefont {Modugno},\ and\ \citenamefont {Inguscio}}]{Carusotto2005}%
  \BibitemOpen
  \bibfield  {author} {\bibinfo {author} {\bibfnamefont {I.}~\bibnamefont
  {Carusotto}}, \bibinfo {author} {\bibfnamefont {L.}~\bibnamefont
  {Pitaevskii}}, \bibinfo {author} {\bibfnamefont {S.}~\bibnamefont
  {Stringari}}, \bibinfo {author} {\bibfnamefont {G.}~\bibnamefont {Modugno}},
  \ and\ \bibinfo {author} {\bibfnamefont {M.}~\bibnamefont {Inguscio}},\
  }\href {\doibase 10.1103/PhysRevLett.95.093202} {\bibfield  {journal}
  {\bibinfo  {journal} {Phys. Rev. Lett.}\ }\textbf {\bibinfo {volume} {95}},\
  \bibinfo {pages} {093202} (\bibinfo {year} {2005})}\BibitemShut {NoStop}%
\bibitem [{\citenamefont {Wolf}\ \emph {et~al.}(2007)\citenamefont {Wolf},
  \citenamefont {Lemonde}, \citenamefont {Lambrecht}, \citenamefont {Bize},
  \citenamefont {Landragin},\ and\ \citenamefont {Clairon}}]{Wolf2007}%
  \BibitemOpen
  \bibfield  {author} {\bibinfo {author} {\bibfnamefont {P.}~\bibnamefont
  {Wolf}}, \bibinfo {author} {\bibfnamefont {P.}~\bibnamefont {Lemonde}},
  \bibinfo {author} {\bibfnamefont {A.}~\bibnamefont {Lambrecht}}, \bibinfo
  {author} {\bibfnamefont {S.}~\bibnamefont {Bize}}, \bibinfo {author}
  {\bibfnamefont {A.}~\bibnamefont {Landragin}}, \ and\ \bibinfo {author}
  {\bibfnamefont {A.}~\bibnamefont {Clairon}},\ }\href {\doibase
  10.1103/PhysRevA.75.063608} {\bibfield  {journal} {\bibinfo  {journal} {Phys.
  Rev. A}\ }\textbf {\bibinfo {volume} {75}},\ \bibinfo {pages} {063608}
  (\bibinfo {year} {2007})}\BibitemShut {NoStop}%
\bibitem [{\citenamefont {Ivanov}\ \emph {et~al.}(2008)\citenamefont {Ivanov},
  \citenamefont {Alberti}, \citenamefont {Schioppo}, \citenamefont {Ferrari},
  \citenamefont {Artoni}, \citenamefont {Chiofalo},\ and\ \citenamefont
  {Tino}}]{Ivanov2008}%
  \BibitemOpen
  \bibfield  {author} {\bibinfo {author} {\bibfnamefont {V.~V.}\ \bibnamefont
  {Ivanov}}, \bibinfo {author} {\bibfnamefont {A.}~\bibnamefont {Alberti}},
  \bibinfo {author} {\bibfnamefont {M.}~\bibnamefont {Schioppo}}, \bibinfo
  {author} {\bibfnamefont {G.}~\bibnamefont {Ferrari}}, \bibinfo {author}
  {\bibfnamefont {M.}~\bibnamefont {Artoni}}, \bibinfo {author} {\bibfnamefont
  {M.~L.}\ \bibnamefont {Chiofalo}}, \ and\ \bibinfo {author} {\bibfnamefont
  {G.~M.}\ \bibnamefont {Tino}},\ }\href {\doibase
  10.1103/PhysRevLett.100.043602} {\bibfield  {journal} {\bibinfo  {journal}
  {Phys. Rev. Lett.}\ }\textbf {\bibinfo {volume} {100}},\ \bibinfo {pages}
  {043602} (\bibinfo {year} {2008})}\BibitemShut {NoStop}%
\bibitem [{\citenamefont {Anderlini}\ \emph {et~al.}(2007)\citenamefont
  {Anderlini}, \citenamefont {Lee}, \citenamefont {Brown}, \citenamefont
  {Sebby-Strabley}, \citenamefont {Phillips},\ and\ \citenamefont
  {Porto}}]{Anderlini_2007}%
  \BibitemOpen
  \bibfield  {author} {\bibinfo {author} {\bibfnamefont {M.}~\bibnamefont
  {Anderlini}}, \bibinfo {author} {\bibfnamefont {P.~J.}\ \bibnamefont {Lee}},
  \bibinfo {author} {\bibfnamefont {B.~L.}\ \bibnamefont {Brown}}, \bibinfo
  {author} {\bibfnamefont {J.}~\bibnamefont {Sebby-Strabley}}, \bibinfo
  {author} {\bibfnamefont {W.~D.}\ \bibnamefont {Phillips}}, \ and\ \bibinfo
  {author} {\bibfnamefont {J.~V.}\ \bibnamefont {Porto}},\ }\href {\doibase
  10.1038/nature06011} {\bibfield  {journal} {\bibinfo  {journal} {Nature}\
  }\textbf {\bibinfo {volume} {448}},\ \bibinfo {pages} {452–456} (\bibinfo
  {year} {2007})}\BibitemShut {NoStop}%
\bibitem [{\citenamefont {{Bloch}}(2008)}]{Bloch2008}%
  \BibitemOpen
  \bibfield  {author} {\bibinfo {author} {\bibfnamefont {I.}~\bibnamefont
  {{Bloch}}},\ }\href {\doibase 10.1038/nature07126} {\bibfield  {journal}
  {\bibinfo  {journal} {\nat}\ }\textbf {\bibinfo {volume} {453}},\ \bibinfo
  {pages} {1016} (\bibinfo {year} {2008})}\BibitemShut {NoStop}%
\bibitem [{\citenamefont {Mandel}\ \emph
  {et~al.}(2003{\natexlab{a}})\citenamefont {Mandel}, \citenamefont {Greiner},
  \citenamefont {Widera}, \citenamefont {Rom}, \citenamefont {H\"ansch},\ and\
  \citenamefont {Bloch}}]{Mandel2003}%
  \BibitemOpen
  \bibfield  {author} {\bibinfo {author} {\bibfnamefont {O.}~\bibnamefont
  {Mandel}}, \bibinfo {author} {\bibfnamefont {M.}~\bibnamefont {Greiner}},
  \bibinfo {author} {\bibfnamefont {A.}~\bibnamefont {Widera}}, \bibinfo
  {author} {\bibfnamefont {T.}~\bibnamefont {Rom}}, \bibinfo {author}
  {\bibfnamefont {T.~W.}\ \bibnamefont {H\"ansch}}, \ and\ \bibinfo {author}
  {\bibfnamefont {I.}~\bibnamefont {Bloch}},\ }\href {\doibase
  10.1103/PhysRevLett.91.010407} {\bibfield  {journal} {\bibinfo  {journal}
  {Phys. Rev. Lett.}\ }\textbf {\bibinfo {volume} {91}},\ \bibinfo {pages}
  {010407} (\bibinfo {year} {2003}{\natexlab{a}})}\BibitemShut {NoStop}%
\bibitem [{\citenamefont {Mandel}\ \emph
  {et~al.}(2003{\natexlab{b}})\citenamefont {Mandel}, \citenamefont {Greiner},
  \citenamefont {Widera}, \citenamefont {Rom}, \citenamefont {Hänsch},\ and\
  \citenamefont {Bloch}}]{Mandel_2003}%
  \BibitemOpen
  \bibfield  {author} {\bibinfo {author} {\bibfnamefont {O.}~\bibnamefont
  {Mandel}}, \bibinfo {author} {\bibfnamefont {M.}~\bibnamefont {Greiner}},
  \bibinfo {author} {\bibfnamefont {A.}~\bibnamefont {Widera}}, \bibinfo
  {author} {\bibfnamefont {T.}~\bibnamefont {Rom}}, \bibinfo {author}
  {\bibfnamefont {T.~W.}\ \bibnamefont {Hänsch}}, \ and\ \bibinfo {author}
  {\bibfnamefont {I.}~\bibnamefont {Bloch}},\ }\href {\doibase
  10.1038/nature02008} {\bibfield  {journal} {\bibinfo  {journal} {Nature}\
  }\textbf {\bibinfo {volume} {425}},\ \bibinfo {pages} {937–940} (\bibinfo
  {year} {2003}{\natexlab{b}})}\BibitemShut {NoStop}%
\bibitem [{\citenamefont {Agarwal}\ \emph {et~al.}(2025)\citenamefont
  {Agarwal}, \citenamefont {Mondal}, \citenamefont {Sahoo}, \citenamefont
  {Rakshit}, \citenamefont {De},\ and\ \citenamefont {Sen}}]{Agarwal2025}%
  \BibitemOpen
  \bibfield  {author} {\bibinfo {author} {\bibfnamefont {K.~D.}\ \bibnamefont
  {Agarwal}}, \bibinfo {author} {\bibfnamefont {S.}~\bibnamefont {Mondal}},
  \bibinfo {author} {\bibfnamefont {A.}~\bibnamefont {Sahoo}}, \bibinfo
  {author} {\bibfnamefont {D.}~\bibnamefont {Rakshit}}, \bibinfo {author}
  {\bibfnamefont {A.~S.}\ \bibnamefont {De}}, \ and\ \bibinfo {author}
  {\bibfnamefont {U.}~\bibnamefont {Sen}},\ }\href {\doibase
  10.1142/S0129183125430065} {\  (\bibinfo {year} {2025}),\
  10.1142/S0129183125430065},\ \Eprint {http://arxiv.org/abs/2507.06348}
  {arXiv:2507.06348 [quant-ph]} \BibitemShut {NoStop}%
\bibitem [{\citenamefont {Sarkar}\ and\ \citenamefont
  {Sowi\ifmmode~\acute{n}\else \'{n}\fi{}ski}(2020)}]{Sarkar_2020}%
  \BibitemOpen
  \bibfield  {author} {\bibinfo {author} {\bibfnamefont {S.}~\bibnamefont
  {Sarkar}}\ and\ \bibinfo {author} {\bibfnamefont {T.}~\bibnamefont
  {Sowi\ifmmode~\acute{n}\else \'{n}\fi{}ski}},\ }\href {\doibase
  10.1103/PhysRevA.102.043326} {\bibfield  {journal} {\bibinfo  {journal}
  {Phys. Rev. A}\ }\textbf {\bibinfo {volume} {102}},\ \bibinfo {pages}
  {043326} (\bibinfo {year} {2020})}\BibitemShut {NoStop}%
\bibitem [{\citenamefont {Cl\'ement}\ \emph {et~al.}(2009)\citenamefont
  {Cl\'ement}, \citenamefont {Fabbri}, \citenamefont {Fallani}, \citenamefont
  {Fort},\ and\ \citenamefont {Inguscio}}]{Clement2009}%
  \BibitemOpen
  \bibfield  {author} {\bibinfo {author} {\bibfnamefont {D.}~\bibnamefont
  {Cl\'ement}}, \bibinfo {author} {\bibfnamefont {N.}~\bibnamefont {Fabbri}},
  \bibinfo {author} {\bibfnamefont {L.}~\bibnamefont {Fallani}}, \bibinfo
  {author} {\bibfnamefont {C.}~\bibnamefont {Fort}}, \ and\ \bibinfo {author}
  {\bibfnamefont {M.}~\bibnamefont {Inguscio}},\ }\href {\doibase
  10.1103/PhysRevLett.102.155301} {\bibfield  {journal} {\bibinfo  {journal}
  {Phys. Rev. Lett.}\ }\textbf {\bibinfo {volume} {102}},\ \bibinfo {pages}
  {155301} (\bibinfo {year} {2009})}\BibitemShut {NoStop}%
\bibitem [{\citenamefont {Gemelke}\ \emph {et~al.}(2009)\citenamefont
  {Gemelke}, \citenamefont {Zhang}, \citenamefont {Hung},\ and\ \citenamefont
  {Chin}}]{Gemelke2009}%
  \BibitemOpen
  \bibfield  {author} {\bibinfo {author} {\bibfnamefont {N.}~\bibnamefont
  {Gemelke}}, \bibinfo {author} {\bibfnamefont {X.}~\bibnamefont {Zhang}},
  \bibinfo {author} {\bibfnamefont {C.-L.}\ \bibnamefont {Hung}}, \ and\
  \bibinfo {author} {\bibfnamefont {C.}~\bibnamefont {Chin}},\ }\href {\doibase
  10.1038/nature08244} {\bibfield  {journal} {\bibinfo  {journal} {Nature}\
  }\textbf {\bibinfo {volume} {460}},\ \bibinfo {pages} {995} (\bibinfo {year}
  {2009})}\BibitemShut {NoStop}%
\bibitem [{\citenamefont {Will}\ \emph {et~al.}(2010)\citenamefont {Will},
  \citenamefont {Best}, \citenamefont {Schneider}, \citenamefont
  {Hackermueller}, \citenamefont {Lühmann},\ and\ \citenamefont
  {Bloch}}]{Will2010}%
  \BibitemOpen
  \bibfield  {author} {\bibinfo {author} {\bibfnamefont {S.}~\bibnamefont
  {Will}}, \bibinfo {author} {\bibfnamefont {T.}~\bibnamefont {Best}}, \bibinfo
  {author} {\bibfnamefont {U.}~\bibnamefont {Schneider}}, \bibinfo {author}
  {\bibfnamefont {L.}~\bibnamefont {Hackermueller}}, \bibinfo {author}
  {\bibfnamefont {D.-S.}\ \bibnamefont {Lühmann}}, \ and\ \bibinfo {author}
  {\bibfnamefont {I.}~\bibnamefont {Bloch}},\ }\href {\doibase
  10.1038/nature09036} {\bibfield  {journal} {\bibinfo  {journal} {Nature}\
  }\textbf {\bibinfo {volume} {465}},\ \bibinfo {pages} {197} (\bibinfo {year}
  {2010})}\BibitemShut {NoStop}%
\bibitem [{\citenamefont {Greiner}\ \emph {et~al.}(2003)\citenamefont
  {Greiner}, \citenamefont {Mandel}, \citenamefont {Rom}, \citenamefont
  {Altmeyer}, \citenamefont {Widera}, \citenamefont {Hänsch},\ and\
  \citenamefont {Bloch}}]{GREINER2003}%
  \BibitemOpen
  \bibfield  {author} {\bibinfo {author} {\bibfnamefont {M.}~\bibnamefont
  {Greiner}}, \bibinfo {author} {\bibfnamefont {O.}~\bibnamefont {Mandel}},
  \bibinfo {author} {\bibfnamefont {T.}~\bibnamefont {Rom}}, \bibinfo {author}
  {\bibfnamefont {A.}~\bibnamefont {Altmeyer}}, \bibinfo {author}
  {\bibfnamefont {A.}~\bibnamefont {Widera}}, \bibinfo {author} {\bibfnamefont
  {T.}~\bibnamefont {Hänsch}}, \ and\ \bibinfo {author} {\bibfnamefont
  {I.}~\bibnamefont {Bloch}},\ }\href {\doibase
  https://doi.org/10.1016/S0921-4526(02)01872-0} {\bibfield  {journal}
  {\bibinfo  {journal} {Physica B: Condensed Matter}\ }\textbf {\bibinfo
  {volume} {329-333}},\ \bibinfo {pages} {11} (\bibinfo {year} {2003})},\
  \bibinfo {note} {proceedings of the 23rd International Conference on Low
  Temperature Physics}\BibitemShut {NoStop}%
\bibitem [{\citenamefont {Sachdev}\ \emph {et~al.}(2002)\citenamefont
  {Sachdev}, \citenamefont {Sengupta},\ and\ \citenamefont
  {Girvin}}]{Sachdev_2002}%
  \BibitemOpen
  \bibfield  {author} {\bibinfo {author} {\bibfnamefont {S.}~\bibnamefont
  {Sachdev}}, \bibinfo {author} {\bibfnamefont {K.}~\bibnamefont {Sengupta}}, \
  and\ \bibinfo {author} {\bibfnamefont {S.~M.}\ \bibnamefont {Girvin}},\
  }\href {\doibase 10.1103/PhysRevB.66.075128} {\bibfield  {journal} {\bibinfo
  {journal} {Phys. Rev. B}\ }\textbf {\bibinfo {volume} {66}},\ \bibinfo
  {pages} {075128} (\bibinfo {year} {2002})}\BibitemShut {NoStop}%
\bibitem [{\citenamefont {Buyskikh}\ \emph {et~al.}(2019)\citenamefont
  {Buyskikh}, \citenamefont {Tagliacozzo}, \citenamefont {Schuricht},
  \citenamefont {Hooley}, \citenamefont {Pekker},\ and\ \citenamefont
  {Daley}}]{Buyskikh_2019}%
  \BibitemOpen
  \bibfield  {author} {\bibinfo {author} {\bibfnamefont {A.~S.}\ \bibnamefont
  {Buyskikh}}, \bibinfo {author} {\bibfnamefont {L.}~\bibnamefont
  {Tagliacozzo}}, \bibinfo {author} {\bibfnamefont {D.}~\bibnamefont
  {Schuricht}}, \bibinfo {author} {\bibfnamefont {C.~A.}\ \bibnamefont
  {Hooley}}, \bibinfo {author} {\bibfnamefont {D.}~\bibnamefont {Pekker}}, \
  and\ \bibinfo {author} {\bibfnamefont {A.~J.}\ \bibnamefont {Daley}},\ }\href
  {\doibase 10.1103/PhysRevLett.123.090401} {\bibfield  {journal} {\bibinfo
  {journal} {Phys. Rev. Lett.}\ }\textbf {\bibinfo {volume} {123}},\ \bibinfo
  {pages} {090401} (\bibinfo {year} {2019})}\BibitemShut {NoStop}%
\bibitem [{\citenamefont {{Inouye}}\ \emph {et~al.}(1998)\citenamefont
  {{Inouye}}, \citenamefont {{Andrews}}, \citenamefont {{Stenger}},
  \citenamefont {{Miesner}}, \citenamefont {{Stamper-Kurn}},\ and\
  \citenamefont {{Ketterle}}}]{Inouye1998}%
  \BibitemOpen
  \bibfield  {author} {\bibinfo {author} {\bibfnamefont {S.}~\bibnamefont
  {{Inouye}}}, \bibinfo {author} {\bibfnamefont {M.~R.}\ \bibnamefont
  {{Andrews}}}, \bibinfo {author} {\bibfnamefont {J.}~\bibnamefont
  {{Stenger}}}, \bibinfo {author} {\bibfnamefont {H.-J.}\ \bibnamefont
  {{Miesner}}}, \bibinfo {author} {\bibfnamefont {D.~M.}\ \bibnamefont
  {{Stamper-Kurn}}}, \ and\ \bibinfo {author} {\bibfnamefont {W.}~\bibnamefont
  {{Ketterle}}},\ }\href {\doibase 10.1038/32354} {\bibfield  {journal}
  {\bibinfo  {journal} {\nat}\ }\textbf {\bibinfo {volume} {392}},\ \bibinfo
  {pages} {151} (\bibinfo {year} {1998})}\BibitemShut {NoStop}%
\bibitem [{\citenamefont {Theis}\ \emph {et~al.}(2004)\citenamefont {Theis},
  \citenamefont {Thalhammer}, \citenamefont {Winkler}, \citenamefont {Hellwig},
  \citenamefont {Ruff}, \citenamefont {Grimm},\ and\ \citenamefont
  {Denschlag}}]{Theis2004}%
  \BibitemOpen
  \bibfield  {author} {\bibinfo {author} {\bibfnamefont {M.}~\bibnamefont
  {Theis}}, \bibinfo {author} {\bibfnamefont {G.}~\bibnamefont {Thalhammer}},
  \bibinfo {author} {\bibfnamefont {K.}~\bibnamefont {Winkler}}, \bibinfo
  {author} {\bibfnamefont {M.}~\bibnamefont {Hellwig}}, \bibinfo {author}
  {\bibfnamefont {G.}~\bibnamefont {Ruff}}, \bibinfo {author} {\bibfnamefont
  {R.}~\bibnamefont {Grimm}}, \ and\ \bibinfo {author} {\bibfnamefont {J.~H.}\
  \bibnamefont {Denschlag}},\ }\href {\doibase 10.1103/PhysRevLett.93.123001}
  {\bibfield  {journal} {\bibinfo  {journal} {Phys. Rev. Lett.}\ }\textbf
  {\bibinfo {volume} {93}},\ \bibinfo {pages} {123001} (\bibinfo {year}
  {2004})}\BibitemShut {NoStop}%
\bibitem [{\citenamefont {Chin}\ \emph {et~al.}(2010)\citenamefont {Chin},
  \citenamefont {Grimm}, \citenamefont {Julienne},\ and\ \citenamefont
  {Tiesinga}}]{Chin2010}%
  \BibitemOpen
  \bibfield  {author} {\bibinfo {author} {\bibfnamefont {C.}~\bibnamefont
  {Chin}}, \bibinfo {author} {\bibfnamefont {R.}~\bibnamefont {Grimm}},
  \bibinfo {author} {\bibfnamefont {P.}~\bibnamefont {Julienne}}, \ and\
  \bibinfo {author} {\bibfnamefont {E.}~\bibnamefont {Tiesinga}},\ }\href
  {\doibase 10.1103/RevModPhys.82.1225} {\bibfield  {journal} {\bibinfo
  {journal} {Rev. Mod. Phys.}\ }\textbf {\bibinfo {volume} {82}},\ \bibinfo
  {pages} {1225} (\bibinfo {year} {2010})}\BibitemShut {NoStop}%
\bibitem [{\citenamefont {He}\ \emph {et~al.}(2023)\citenamefont {He},
  \citenamefont {Yousefjani},\ and\ \citenamefont {Bayat}}]{he2023stark}%
  \BibitemOpen
  \bibfield  {author} {\bibinfo {author} {\bibfnamefont {X.}~\bibnamefont
  {He}}, \bibinfo {author} {\bibfnamefont {R.}~\bibnamefont {Yousefjani}}, \
  and\ \bibinfo {author} {\bibfnamefont {A.}~\bibnamefont {Bayat}},\
  }\href@noop {} {\bibfield  {journal} {\bibinfo  {journal} {Phys. Rev. Lett.}\
  }\textbf {\bibinfo {volume} {131}},\ \bibinfo {pages} {010801} (\bibinfo
  {year} {2023})}\BibitemShut {NoStop}%
\bibitem [{\citenamefont {Yousefjani}\ \emph {et~al.}(2023)\citenamefont
  {Yousefjani}, \citenamefont {He},\ and\ \citenamefont
  {Bayat}}]{Yousefjani2023}%
  \BibitemOpen
  \bibfield  {author} {\bibinfo {author} {\bibfnamefont {R.}~\bibnamefont
  {Yousefjani}}, \bibinfo {author} {\bibfnamefont {X.}~\bibnamefont {He}}, \
  and\ \bibinfo {author} {\bibfnamefont {A.}~\bibnamefont {Bayat}},\
  }\href@noop {} {\bibfield  {journal} {\bibinfo  {journal} {Chin. Phys. B}\
  }\textbf {\bibinfo {volume} {32}},\ \bibinfo {pages} {100313} (\bibinfo
  {year} {2023})}\BibitemShut {NoStop}%
\bibitem [{\citenamefont {Sarkar}\ and\ \citenamefont
  {Bayat}(2025)}]{Sarkar2025}%
  \BibitemOpen
  \bibfield  {author} {\bibinfo {author} {\bibfnamefont {S.}~\bibnamefont
  {Sarkar}}\ and\ \bibinfo {author} {\bibfnamefont {A.}~\bibnamefont {Bayat}},\
  }\href {\doibase 10.1103/PhysRevA.111.062602} {\bibfield  {journal} {\bibinfo
   {journal} {Phys. Rev. A}\ }\textbf {\bibinfo {volume} {111}},\ \bibinfo
  {pages} {062602} (\bibinfo {year} {2025})}\BibitemShut {NoStop}%
\bibitem [{\citenamefont {Manshouri}\ \emph {et~al.}(2025)\citenamefont
  {Manshouri}, \citenamefont {Zarei}, \citenamefont {Abdi}, \citenamefont
  {Bose},\ and\ \citenamefont {Bayat}}]{Manshouri_2025}%
  \BibitemOpen
  \bibfield  {author} {\bibinfo {author} {\bibfnamefont {H.}~\bibnamefont
  {Manshouri}}, \bibinfo {author} {\bibfnamefont {M.}~\bibnamefont {Zarei}},
  \bibinfo {author} {\bibfnamefont {M.}~\bibnamefont {Abdi}}, \bibinfo {author}
  {\bibfnamefont {S.}~\bibnamefont {Bose}}, \ and\ \bibinfo {author}
  {\bibfnamefont {A.}~\bibnamefont {Bayat}},\ }\href {\doibase
  10.22331/q-2025-07-11-1793} {\bibfield  {journal} {\bibinfo  {journal}
  {Quantum}\ }\textbf {\bibinfo {volume} {9}},\ \bibinfo {pages} {1793}
  (\bibinfo {year} {2025})}\BibitemShut {NoStop}%
\bibitem [{\citenamefont {Yousefjani}\ and\ \citenamefont
  {Al-Kuwari}(2026)}]{yousefjani2026exponentially}%
  \BibitemOpen
  \bibfield  {author} {\bibinfo {author} {\bibfnamefont {R.}~\bibnamefont
  {Yousefjani}}\ and\ \bibinfo {author} {\bibfnamefont {S.}~\bibnamefont
  {Al-Kuwari}},\ }\href {https://arxiv.org/abs/2604.17262} {\enquote {\bibinfo
  {title} {Exponentially-enhanced weak-field sensing with quantum stark
  localization},}\ } (\bibinfo {year} {2026}),\ \Eprint
  {http://arxiv.org/abs/2604.17262} {arXiv:2604.17262 [quant-ph]} \BibitemShut
  {NoStop}%
\bibitem [{\citenamefont {Li}\ \emph {et~al.}(2025)\citenamefont {Li},
  \citenamefont {Yang}, \citenamefont {Shi}, \citenamefont {Wang},
  \citenamefont {Wang}, \citenamefont {Li}, \citenamefont {Zhang},
  \citenamefont {Zhao}, \citenamefont {Xu}, \citenamefont {Deng}, \citenamefont
  {Liu}, \citenamefont {Ma}, \citenamefont {Li}, \citenamefont {Zhang},
  \citenamefont {Fang}, \citenamefont {Song}, \citenamefont {Liu},
  \citenamefont {Zhou}, \citenamefont {Liu}, \citenamefont {Chen},
  \citenamefont {Liang}, \citenamefont {Song}, \citenamefont {Xiang},
  \citenamefont {Xu}, \citenamefont {Huang}, \citenamefont {Bayat},\ and\
  \citenamefont {Fan}}]{Li2025nonequilibrium}%
  \BibitemOpen
  \bibfield  {author} {\bibinfo {author} {\bibfnamefont {H.}~\bibnamefont
  {Li}}, \bibinfo {author} {\bibfnamefont {Y.}~\bibnamefont {Yang}}, \bibinfo
  {author} {\bibfnamefont {Y.-H.}\ \bibnamefont {Shi}}, \bibinfo {author}
  {\bibfnamefont {Z.-A.}\ \bibnamefont {Wang}}, \bibinfo {author}
  {\bibfnamefont {Z.}~\bibnamefont {Wang}}, \bibinfo {author} {\bibfnamefont
  {J.}~\bibnamefont {Li}}, \bibinfo {author} {\bibfnamefont {Y.}~\bibnamefont
  {Zhang}}, \bibinfo {author} {\bibfnamefont {K.}~\bibnamefont {Zhao}},
  \bibinfo {author} {\bibfnamefont {Y.-S.}\ \bibnamefont {Xu}}, \bibinfo
  {author} {\bibfnamefont {C.-L.}\ \bibnamefont {Deng}}, \bibinfo {author}
  {\bibfnamefont {Y.}~\bibnamefont {Liu}}, \bibinfo {author} {\bibfnamefont
  {W.-G.}\ \bibnamefont {Ma}}, \bibinfo {author} {\bibfnamefont {T.-M.}\
  \bibnamefont {Li}}, \bibinfo {author} {\bibfnamefont {J.-C.}\ \bibnamefont
  {Zhang}}, \bibinfo {author} {\bibfnamefont {C.-P.}\ \bibnamefont {Fang}},
  \bibinfo {author} {\bibfnamefont {J.-C.}\ \bibnamefont {Song}}, \bibinfo
  {author} {\bibfnamefont {H.-T.}\ \bibnamefont {Liu}}, \bibinfo {author}
  {\bibfnamefont {S.-Y.}\ \bibnamefont {Zhou}}, \bibinfo {author}
  {\bibfnamefont {Z.-H.}\ \bibnamefont {Liu}}, \bibinfo {author} {\bibfnamefont
  {B.-J.}\ \bibnamefont {Chen}}, \bibinfo {author} {\bibfnamefont {G.-H.}\
  \bibnamefont {Liang}}, \bibinfo {author} {\bibfnamefont {X.}~\bibnamefont
  {Song}}, \bibinfo {author} {\bibfnamefont {Z.}~\bibnamefont {Xiang}},
  \bibinfo {author} {\bibfnamefont {K.}~\bibnamefont {Xu}}, \bibinfo {author}
  {\bibfnamefont {K.}~\bibnamefont {Huang}}, \bibinfo {author} {\bibfnamefont
  {A.}~\bibnamefont {Bayat}}, \ and\ \bibinfo {author} {\bibfnamefont
  {H.}~\bibnamefont {Fan}},\ }\href {https://arxiv.org/abs/2508.14409}
  {\enquote {\bibinfo {title} {Non-equilibrium criticality-enhanced quantum
  sensing with superconducting qubits},}\ } (\bibinfo {year} {2025}),\ \Eprint
  {http://arxiv.org/abs/2508.14409} {arXiv:2508.14409 [quant-ph]} \BibitemShut
  {NoStop}%
\bibitem [{\citenamefont {Fisher}(1922)}]{fisher1922mathematical}%
  \BibitemOpen
  \bibfield  {author} {\bibinfo {author} {\bibfnamefont {R.~A.}\ \bibnamefont
  {Fisher}},\ }\href {\doibase https://doi.org/10.1098/rsta.1922.0009}
  {\bibfield  {journal} {\bibinfo  {journal} {Philos. Trans. Royal Soc.}\
  }\textbf {\bibinfo {volume} {222}},\ \bibinfo {pages} {309} (\bibinfo {year}
  {1922})}\BibitemShut {NoStop}%
\bibitem [{\citenamefont {Liu}\ \emph {et~al.}(2019)\citenamefont {Liu},
  \citenamefont {Yuan}, \citenamefont {Lu},\ and\ \citenamefont
  {Wang}}]{liu2019quantum}%
  \BibitemOpen
  \bibfield  {author} {\bibinfo {author} {\bibfnamefont {J.}~\bibnamefont
  {Liu}}, \bibinfo {author} {\bibfnamefont {H.}~\bibnamefont {Yuan}}, \bibinfo
  {author} {\bibfnamefont {X.-M.}\ \bibnamefont {Lu}}, \ and\ \bibinfo {author}
  {\bibfnamefont {X.}~\bibnamefont {Wang}},\ }\href@noop {} {\bibfield
  {journal} {\bibinfo  {journal} {J. Phys. A: Math. Theor.}\ }\textbf {\bibinfo
  {volume} {53}},\ \bibinfo {pages} {023001} (\bibinfo {year}
  {2019})}\BibitemShut {NoStop}%
\bibitem [{\citenamefont {Montenegro}\ \emph {et~al.}(2025)\citenamefont
  {Montenegro}, \citenamefont {Mukhopadhyay}, \citenamefont {Yousefjani},
  \citenamefont {Sarkar}, \citenamefont {Mishra}, \citenamefont {Paris},\ and\
  \citenamefont {Bayat}}]{MONTENEGRO2025}%
  \BibitemOpen
  \bibfield  {author} {\bibinfo {author} {\bibfnamefont {V.}~\bibnamefont
  {Montenegro}}, \bibinfo {author} {\bibfnamefont {C.}~\bibnamefont
  {Mukhopadhyay}}, \bibinfo {author} {\bibfnamefont {R.}~\bibnamefont
  {Yousefjani}}, \bibinfo {author} {\bibfnamefont {S.}~\bibnamefont {Sarkar}},
  \bibinfo {author} {\bibfnamefont {U.}~\bibnamefont {Mishra}}, \bibinfo
  {author} {\bibfnamefont {M.~G.}\ \bibnamefont {Paris}}, \ and\ \bibinfo
  {author} {\bibfnamefont {A.}~\bibnamefont {Bayat}},\ }\href {\doibase
  https://doi.org/10.1016/j.physrep.2025.05.005} {\bibfield  {journal}
  {\bibinfo  {journal} {Physics Reports}\ }\textbf {\bibinfo {volume} {1134}},\
  \bibinfo {pages} {1} (\bibinfo {year} {2025})},\ \bibinfo {note} {review:
  Quantum metrology and sensing with many-body systems}\BibitemShut {NoStop}%
\bibitem [{\citenamefont {Alberti}\ \emph {et~al.}(2010)\citenamefont
  {Alberti}, \citenamefont {Ferrari}, \citenamefont {Ivanov}, \citenamefont
  {Chiofalo},\ and\ \citenamefont {Tino}}]{Alberti_2010}%
  \BibitemOpen
  \bibfield  {author} {\bibinfo {author} {\bibfnamefont {A.}~\bibnamefont
  {Alberti}}, \bibinfo {author} {\bibfnamefont {G.}~\bibnamefont {Ferrari}},
  \bibinfo {author} {\bibfnamefont {V.~V.}\ \bibnamefont {Ivanov}}, \bibinfo
  {author} {\bibfnamefont {M.~L.}\ \bibnamefont {Chiofalo}}, \ and\ \bibinfo
  {author} {\bibfnamefont {G.~M.}\ \bibnamefont {Tino}},\ }\href {\doibase
  10.1088/1367-2630/12/6/065037} {\bibfield  {journal} {\bibinfo  {journal}
  {New Journal of Physics}\ }\textbf {\bibinfo {volume} {12}},\ \bibinfo
  {pages} {065037} (\bibinfo {year} {2010})}\BibitemShut {NoStop}%
\bibitem [{\citenamefont {Thommen}\ \emph {et~al.}(2004)\citenamefont
  {Thommen}, \citenamefont {Garreau},\ and\ \citenamefont
  {Zehnlé}}]{Thommen_2004}%
  \BibitemOpen
  \bibfield  {author} {\bibinfo {author} {\bibfnamefont {Q.}~\bibnamefont
  {Thommen}}, \bibinfo {author} {\bibfnamefont {J.~C.}\ \bibnamefont
  {Garreau}}, \ and\ \bibinfo {author} {\bibfnamefont {V.}~\bibnamefont
  {Zehnlé}},\ }\href {\doibase 10.1088/1464-4266/6/7/007} {\bibfield
  {journal} {\bibinfo  {journal} {Journal of Optics B: Quantum and
  Semiclassical Optics}\ }\textbf {\bibinfo {volume} {6}},\ \bibinfo {pages}
  {301} (\bibinfo {year} {2004})}\BibitemShut {NoStop}%
\bibitem [{\citenamefont {Dalfovo}\ \emph {et~al.}(1999)\citenamefont
  {Dalfovo}, \citenamefont {Giorgini}, \citenamefont {Pitaevskii},\ and\
  \citenamefont {Stringari}}]{Dalfovo1999}%
  \BibitemOpen
  \bibfield  {author} {\bibinfo {author} {\bibfnamefont {F.}~\bibnamefont
  {Dalfovo}}, \bibinfo {author} {\bibfnamefont {S.}~\bibnamefont {Giorgini}},
  \bibinfo {author} {\bibfnamefont {L.~P.}\ \bibnamefont {Pitaevskii}}, \ and\
  \bibinfo {author} {\bibfnamefont {S.}~\bibnamefont {Stringari}},\ }\href
  {\doibase 10.1103/RevModPhys.71.463} {\bibfield  {journal} {\bibinfo
  {journal} {Rev. Mod. Phys.}\ }\textbf {\bibinfo {volume} {71}},\ \bibinfo
  {pages} {463} (\bibinfo {year} {1999})}\BibitemShut {NoStop}%
\bibitem [{\citenamefont {Dalibard}(2013)}]{dalibard_2013}%
  \BibitemOpen
  \bibfield  {author} {\bibinfo {author} {\bibfnamefont {J.}~\bibnamefont
  {Dalibard}},\ }\href {https://hal.science/hal-04010525} {\enquote {\bibinfo
  {title} {{College de France lecture 2013: Optical traps and optical
  lattices}},}\ } (\bibinfo {year} {2013}),\ \bibinfo {note}
  {lecture}\BibitemShut {NoStop}%
\bibitem [{\citenamefont {Kohn}(1959)}]{Kohn_1959}%
  \BibitemOpen
  \bibfield  {author} {\bibinfo {author} {\bibfnamefont {W.}~\bibnamefont
  {Kohn}},\ }\href {\doibase 10.1103/PhysRev.115.809} {\bibfield  {journal}
  {\bibinfo  {journal} {Phys. Rev.}\ }\textbf {\bibinfo {volume} {115}},\
  \bibinfo {pages} {809} (\bibinfo {year} {1959})}\BibitemShut {NoStop}%
\bibitem [{\citenamefont {Krutitsky}(2016)}]{Krutitsky_2016}%
  \BibitemOpen
  \bibfield  {author} {\bibinfo {author} {\bibfnamefont {K.~V.}\ \bibnamefont
  {Krutitsky}},\ }\href {\doibase 10.1016/j.physrep.2015.10.004} {\bibfield
  {journal} {\bibinfo  {journal} {Physics Reports}\ }\textbf {\bibinfo {volume}
  {607}},\ \bibinfo {pages} {1–101} (\bibinfo {year} {2016})}\BibitemShut
  {NoStop}%
\bibitem [{\citenamefont {Moseley}\ \emph {et~al.}(2008)\citenamefont
  {Moseley}, \citenamefont {Fialko},\ and\ \citenamefont
  {Ziegler}}]{Moseley_2008}%
  \BibitemOpen
  \bibfield  {author} {\bibinfo {author} {\bibfnamefont {C.}~\bibnamefont
  {Moseley}}, \bibinfo {author} {\bibfnamefont {O.}~\bibnamefont {Fialko}}, \
  and\ \bibinfo {author} {\bibfnamefont {K.}~\bibnamefont {Ziegler}},\ }\href
  {\doibase 10.1002/andp.20085200804} {\bibfield  {journal} {\bibinfo
  {journal} {Annalen der Physik}\ }\textbf {\bibinfo {volume} {520}},\ \bibinfo
  {pages} {561–608} (\bibinfo {year} {2008})}\BibitemShut {NoStop}%
\bibitem [{\citenamefont {Zwerger}(2002)}]{Zwerger}%
  \BibitemOpen
  \bibfield  {author} {\bibinfo {author} {\bibfnamefont {W.}~\bibnamefont
  {Zwerger}},\ }\href {\doibase 10.1088/1464-4266/5/2/352} {\bibfield
  {journal} {\bibinfo  {journal} {Journal of Optics B: Quantum and
  Semiclassical Optics}\ }\textbf {\bibinfo {volume} {5}} (\bibinfo {year}
  {2002}),\ 10.1088/1464-4266/5/2/352}\BibitemShut {NoStop}%
\bibitem [{\citenamefont {Wannier}(1960)}]{Wannier_1960}%
  \BibitemOpen
  \bibfield  {author} {\bibinfo {author} {\bibfnamefont {G.~H.}\ \bibnamefont
  {Wannier}},\ }\href {\doibase 10.1103/PhysRev.117.432} {\bibfield  {journal}
  {\bibinfo  {journal} {Phys. Rev.}\ }\textbf {\bibinfo {volume} {117}},\
  \bibinfo {pages} {432} (\bibinfo {year} {1960})}\BibitemShut {NoStop}%
\bibitem [{\citenamefont {Yuan}\ and\ \citenamefont {Fung}(2015)}]{Yuan_2015}%
  \BibitemOpen
  \bibfield  {author} {\bibinfo {author} {\bibfnamefont {H.}~\bibnamefont
  {Yuan}}\ and\ \bibinfo {author} {\bibfnamefont {C.-H.~F.}\ \bibnamefont
  {Fung}},\ }\href {\doibase 10.1103/physrevlett.115.110401} {\bibfield
  {journal} {\bibinfo  {journal} {Physical Review Letters}\ }\textbf {\bibinfo
  {volume} {115}} (\bibinfo {year} {2015}),\
  10.1103/physrevlett.115.110401}\BibitemShut {NoStop}%
\bibitem [{\citenamefont {Boixo}\ \emph {et~al.}(2007)\citenamefont {Boixo},
  \citenamefont {Flammia}, \citenamefont {Caves},\ and\ \citenamefont
  {Geremia}}]{Boixo2007Generalized}%
  \BibitemOpen
  \bibfield  {author} {\bibinfo {author} {\bibfnamefont {S.}~\bibnamefont
  {Boixo}}, \bibinfo {author} {\bibfnamefont {S.~T.}\ \bibnamefont {Flammia}},
  \bibinfo {author} {\bibfnamefont {C.~M.}\ \bibnamefont {Caves}}, \ and\
  \bibinfo {author} {\bibfnamefont {J.}~\bibnamefont {Geremia}},\ }\href
  {\doibase 10.1103/PhysRevLett.98.090401} {\bibfield  {journal} {\bibinfo
  {journal} {Phys. Rev. Lett.}\ }\textbf {\bibinfo {volume} {98}},\ \bibinfo
  {pages} {090401} (\bibinfo {year} {2007})}\BibitemShut {NoStop}%
\end{thebibliography}%

\clearpage
\appendix

\section{}\label{appendixB}
\begin{figure}[h]
	\includegraphics[width=.49\linewidth]{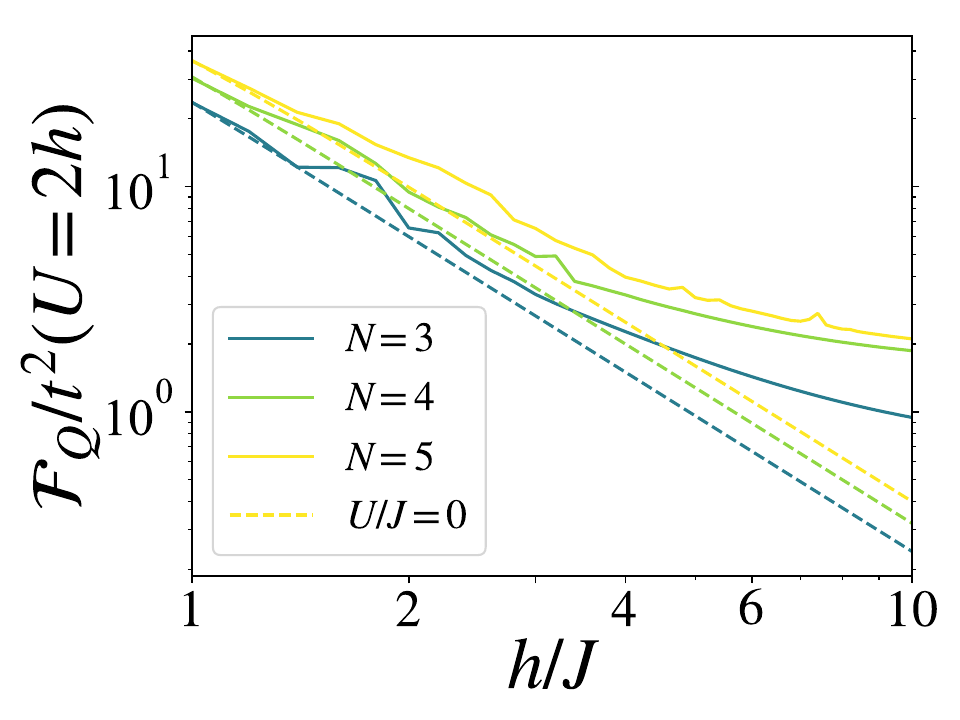}
	\put(-20,77){\color{black}(a)}
	\includegraphics[width=.49\linewidth]{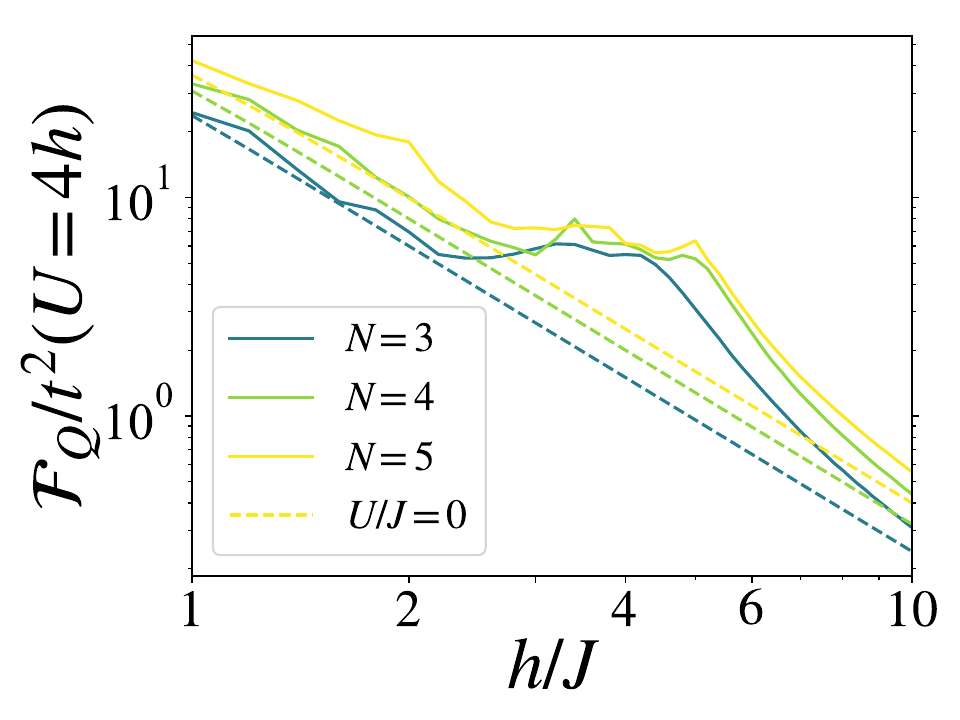}
	\put(-20,77){\color{black}(b)}\\
	\includegraphics[width=.49\linewidth]{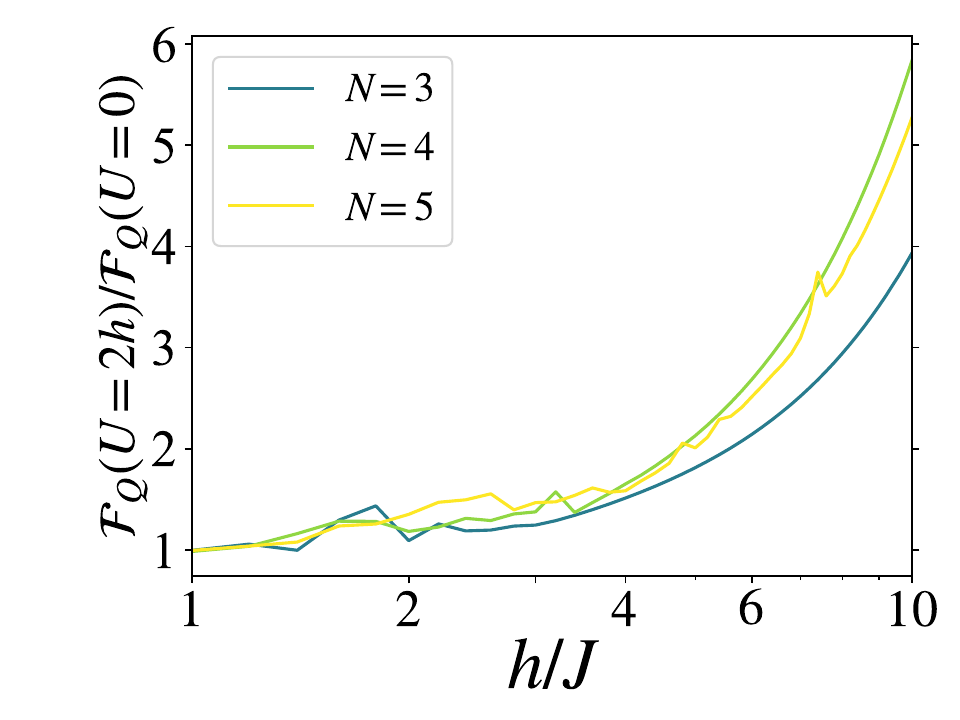}
	\put(-20,77){\color{black}(c)}
	\includegraphics[width=.49\linewidth]{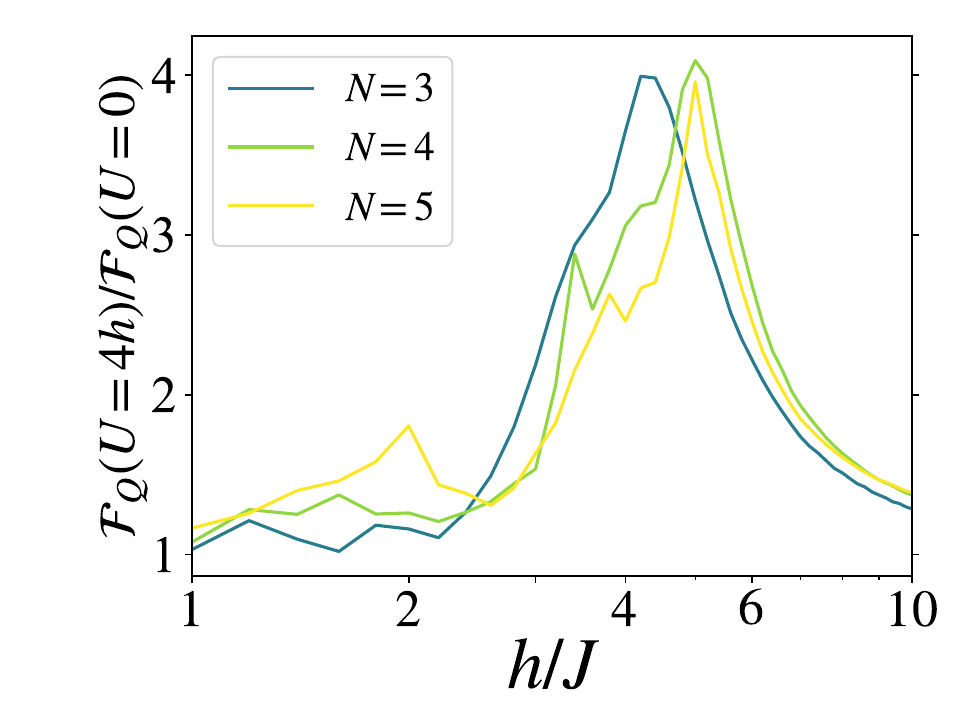}
	\put(-20,77){\color{black}(d)}
	\caption{For an specific value of system size $L=11$, the normalized QFI as a function of $h$ for different $N$ is depicted at the resonance condition (a) $m=2$ and (b) $m=4$. The corresponding values of $A_r(U,m){=} \mathcal{F}_{Q}(U{=}mh)/\mathcal{F}_{Q}(U{=}0)$ are plotted in (c) and (d) respectively.}
	\label{fig:QFImhN}
\end{figure}

In this appendix, we evaluate the behavior of the coefficient $A_r(U,m)$ in Eq.~\eqref{eq:QFI_scaling} for different values of $m$. To this end, Fig.~\ref{fig:QFImhN} for an specific value of system size $L$ and different particle numbers $N$, the normalized QFI as a function of $h$ is plotted under two different resonance conditions: $m=2$ and $m=4$. The resonance coefficient is defined as $A_r(U,m){=} \mathcal{F}_{Q}(U{=}mh)/\mathcal{F}_{Q}(U{=}0)$, which quantifies the enhancement of the QFI beyond the scaling behavior observed at $U=0$. One can find in Fig.~\ref{fig:QFImhN}(c) and (d), $A_r(U,m)$ is independent of particle number but depends on resonance condition $m$. 

\clearpage


\end{document}